\documentclass[twocolumn,dvipsnames]{aastex63}
\usepackage{soul} % added by FW
\RequirePackage{lineno}
\usepackage{graphicx}	% Including figure files
\usepackage{multirow,makecell}

\RequirePackage{bm,amsfonts,hhline,amsmath,amssymb,microtype,float,eurosym,
latexsym,epsf,mathtools,cuted,times,makecell,array,natbib,cancel}
\usepackage{tabularx}
\usepackage{epstopdf}

\usepackage{gensymb}
\usepackage{xcolor}

\usepackage{pifont}% http://ctan.org/pkg/pifont

\usepackage{hyperref}
\hypersetup{colorlinks=true, linkcolor=red, filecolor=magenta, urlcolor=Sepia, citecolor=blue}

\definecolor{ForestGreen}{HTML}{228B22}

\newcommand{\plb}{PhLB}

\received{XXX}
\revised{YYY}
\accepted{ZZZ}

\submitjournal{ApJ}

\shorttitle{Anisotropic magnetized compact stars }
\shortauthors{Deb, Mukhopadhyay \& Weber}

\begin{document}

\title{Effects of anisotropy on strongly magnetized neutron and strange quark stars in general relativity}

\author[0000-0003-4067-5283]{Debabrata Deb$^{\dagger}$}
\email{debabratadeb@iisc.ac.in$^{\dagger}$}
\affiliation{Department of Physics, Indian Institute of Science, Bangalore 560012, India}

\author[0000-0002-3020-9513]{Banibrata Mukhopadhyay$^{\ddagger}$}
\email{bm@iisc.ac.in$^{\ddagger}$}
\affiliation{Department of Physics, Indian Institute of Science, Bangalore 560012, India}

\author[0000-0002-5020-1906]{Fridolin Weber$^{\star}$}
\email{fweber@sdsu.edu$^{\star}$}
\affiliation{Department of Physics, San Diego State University, San Diego, CA 92182, USA}
\affiliation{Center for Astrophysics and Space Sciences, University of California at San Diego, La Jolla, CA 92093, USA}

%\author{xyz}
%\email{xyz}
%\affiliation{xyz}

\begin{abstract}
  
We investigate the properties of anisotropic, spherically symmetric
compact stars, especially neutron stars and strange quark stars, made
of strongly magnetized matter.  The neutron stars are described by 
SLy equation of state, the strange quark stars by an equation
of state based on the MIT Bag model. The stellar models are based on
an a priori assumed density dependence of the magnetic field and thus
anisotropy. Our study shows that not only the presence of a strong
magnetic field and anisotropy, but also the orientation of the
magnetic field itself, have an important influence on the physical
properties of stars. Two possible magnetic field orientations are considered, a
radial orientation, where the local magnetic fields point in the
radial direction, and a transverse orientation, where the local
magnetic fields are perpendicular to the radial
direction. Interestingly, we find that for a transverse orientation of
the magnetic field, the stars become more massive with increasing
anisotropy and magnetic field strength and increase in size, since the
repulsive, effective anisotropic force increases in this case. In the
case of a radially orientated magnetic field, however, the masses and
radii of the stars decrease with increasing magnetic field strength,
because of the decreasing effective anisotropic force. Importantly, we
also show that in order to achieve hydrostatic equilibrium
configurations of magnetized matter, it is essential to account for
both the local anisotropy effects as well as the anisotropy effects
caused by a strong magnetic field. Otherwise, hydrostatic equilibrium
is not achieved for magnetized stellar models.

\end{abstract}

\keywords{Gravitation – Stars: general - stars: fundamental parameters
  – stars: magnetic field – stars: massive – stars: neutron }

%%%%%%%%%%%%%%%%%%%%%%%%%%%%%%%%%%%%%%%%%%%%%%%%%%%%%%%%%%%%%%%%%%%%%%%%%%%%%%%%%%%%%%%%%%%%%%%%%%%%%%%%%%%%%%%%%%%%%%

%%%%%%%%%%%%%%%%%%%%%%%%%%%%%%%%

\section{Introduction}\label{sec:int}

Compact stars present unique astrophysical laboratories to study the
nature of matter (and several astrophysical phenomena) under extreme
physical conditions~\citep{Glendenning2000,Weber1999}. Neutron stars
(NSs) and Strange Quark stars (SQSs) represent the ultra-dense classes
of compact stars. Magnetic flux conservation during stellar collapse
leads to the presence of ultra-strong magnetic fields inside of compact
stars. Some researchers have found that at the center of
inhomogeneous, ultradense and gravitationally bound compact stars, the
magnetic field may be as high as $\sim10^{19}$
G~\citep{Yuan1998,Tatsumi2000,Ferrer2010}. However, from the available
observational evidence, it is difficult to confirm the strength of the
magnetic field inside of compact stars, which urges researchers to
develop suitable theoretical models that help to investigate
appropriately the effects of high magnetic fields on the physical
parameters of compact stellar objects. Evidently, to study the effects
that high magnetic fields have on compact stars, it is essential to
carry out such a study in the realm of general relativity
(GR). Following the pioneering works
of~\cite{Tolman1939,Oppenheimer1939} (TOV), many researchers have
investigated the properties of NSs and SQSs based on the GR
hydrostatic equilibrium equation derived by
TOV~\citep{Bowers1974,Hillebrandt1976,Mak2002,Weber2005,Negreiros2009,Weber2014,Arbanil2016,Deb2017,Deb2018}.

Since the ground breaking observation of radio pulsars
by~\cite{Hewish1968}, NSs remain to be one of the most studied
astrophysical objects, which are assumed to be the possible sources of
high-energy emission. The typical values of the surface magnetic field
as inferred from simple magnetic dipole models and spin-down rates are
in the range $10^{8}-10^{13}$~G~\citep{Taylor1993,Alpar1982}. Note
that among the radio pulsars, PSR J1847$-$0130 exhibits a strong
magnetic field of $B=9.4\times {14}^{13}$ G~\citep{McLaughlin2003}. On
the other hand, besides the X-ray luminosities observed from the
anomalous X-ray pulsars (AXPs), the inferred periods of AXPs and
soft-$\gamma$ repeaters (SGRs) suggest that such NSs have even larger
surface magnetic fields of $10^{14}-10^{15}$
G~\citep{Paczynski1992,Thompson1992,Thompson1996,Melatos1999}. Neutron
stars with such high surface magnetic fields are popularly known as
\emph{magnetars}. Further on, several interesting
studies~\citep{Usov1992,Kluzniak1998,Wheeler2000,Starling2009,Cenko2010}
have predicted that magnetars are the probable source of $\gamma$-ray
bursts and they require higher magnetic field such as
$10^{16}-10^{17}$ G to initiate the Poynting flux-dominated
jets. Although till now only around 30 magnetars have been detected,
it is speculated that these astrophysical objects may account for
$10\%$ of the NSs population~\citep{Kouveliotou1998}. For NSs, the
effects of a strong magnetic field on the ultra-dense electron gases
in their interiors have been studied in several papers
~\citep{Canuto1977,Fushiki1989,Abrahams1991,Fushiki1992,Roegnvaldsson1993}. Studies
of dense and strongly magnetized nuclear matter have also been carried
out
by~\cite{Chakrabarty1997,Bandyopadhyay1998,Broderick2000,Suh2001,Harding2006,Chen2007,Rabhi2008}
and references therein. Finally we mention that studies of NSs
with different magnetic field configurations, viz., toroidal,
poloidal, or mixed were carried out
by~\cite{Bocquet1995,Cardall2001,Pili2014}.
 
Because of its profound significance for strong interaction physics
and astrophysics, the possible existence of SQS has attracted great
scientific interest over the past three decades. SQSs are hypothetical
compact stellar objects made completely of strange quark matter
(SQM). Such matter consists entirely of deconfined up ($u$), down
($d$) and strange ($s$) quarks, which, according to the SQM hypothesis, could be lower in energy than nuclear matter and
thus be the true ground state of the strong interaction
\citep{Bodmer1971,Witten1984,Terazawa1979}. Various researchers have
studied the properties of SQSs (see, for instance,
~\citealt{Itoh1970,Alcock1986,Haensel1986,Alcock1988,Madsen1999,Bombaci2004,Weber2005,Staff2007,Herzog2011}).  Furthermore,
theoretical studies have shown that the birth of SQSs could occur
through the conversion of NSs to SQSs within a few milliseconds via a
strong deflagration process, which leads to the emission of a powerful
neutrino
signal~\citep{Martemyanov1994,Bombaci2004,Staff2007,Herzog2011}. A distinguishing feature between NSs and SQSs is that
the radii of the latter become monotonically smaller with decreasing
star mass, which is not the case for NSs \citep{Alcock1986,Alcock1988,Kapoor2001}. In the
past, it has been speculated that compact stars such as 4U~1728-34,
4U~1820-30, SAX~J1808.4C3658, Her X-1 and RX~J1856.5C3754 could be SQS
candidates \citep{Weber2005}.  Hence, it will be interesting to
investigate the GR effects on strongly magnetized SQSs. Important
studies which have examined the effects of strong magnetic fields on
SQSs have been carried out
by~\cite{Chakrabarty1996,Chaichian2000,Felipe2008,Menezes2009a,Menezes2009b,Rabhi2009}.

\cite{Ruderman1972} has shown that when the
nuclear matter in the stellar interior reaches a density beyond
${10}^{15}~{\rm g/cm}^3$, interactions become relativistic, and the
presence of a type-P superfluid leads to a pressure anisotropy in the
stars. However,~\cite{Bowers1974} in their study strongly argued
against the over-simplistic assumption that compact stars are
composed of only an isotropic perfect fluid. They presented the
non-negligible effects of a local anisotropy on the physical
parameters of compact stars, such as maximum equilibrium mass and
surface redshift, by generalizing the TOV equations in terms of a
local anisotropy.~\cite{Letelier1980} and~\cite{Bayin1982} strongly
argued that the presence of two (or more) fluids or a mixture thereof
in  compact stars, may be the possible reason for pressure
anisotropy, which~\cite{Herrera1997} confirmed later. Further,
anisotropy may be caused by phase
transitions~\citep{Sokolov1980,Carter1998} in the interiors of compact
stars, when the matter forms superfluid or superconducting
states. Some works~\citep{Barreto1992,Barreto1993} also showed that
the presence of viscosity might be the possible source of local
anisotropy within dense compact stars. Other investigations revealed
additional reasons for the existence of local anisotropy, such as pion
condensation~\citep{Sawyer1972,Dev2002}, the existence of a solid core
at densities $10^{14-15} ~{\rm
  g/cm}^3$~\citep{Cameron1973,Canuto1974,Canuto1977a}, and the
presence of a type-3A superfluid~\citep{Kippenhahn1990}, which are
considered to offer a more realistic view of the structure of the
ultra-dense cores of compact stellar objects. For a further detailed
understanding of the mechanisms that produce anisotropies, one may see
the seminal articles~\citep{Herrera1997,Dev2002} and references
therein. Further, to emphasize the relevance of local anisotropy,
several recent
articles~\citep{Corchero2001,Ivanov2002,Mak2003,Schunck2003,Usov2004,Chaisi2005,Varela2010,Rahaman2010,Rahaman2011,Rahaman2012,Silva2015,Arbanil2016,Deb2017,Deb2018}
may also be recalled, where the effects of local anisotropy on
spherically symmetric compact stars were studied in
detail.~\cite{Ferrer2010} showed that the presence of a strong
magnetic field may also lead to anisotropy in compact stars by breaking
the spatial rotational $[{\mathcal O}(3)]$ symmetry, which
later~\cite{Isayev2011,Isayev2012} confirmed in their articles.
 
Interestingly,
researchers~\citep{Bandyopadhyay1997a,Bandyopadhyay1998a,Broderick2000,Cardall2001,Menezes2009a,Menezes2009b,Ryu2010,Paulucci2011,Ryu2012,Dexheimer2014,Casali2014,Hou2015,Kayanikhoo12020}
still do not agree unanimously on whether the maximum mass of a
compact stellar object increases or decreases due to the presence of a
strong magnetic field and it remains still an important open issue
that needs to be resolved.~\cite{Chu2014} tried to address this issue
by introducing the idea of magnetic field orientation. When the local
magnetic fields are directed towards the radial direction, they are
termed radially oriented (RO), and when the magnetic fields are
randomly oriented in the direction perpendicular to the radial
direction (say along $\theta$ direction), they are referred to as
transversely orientated (TO).~\cite{Chu2014} showed in
  their study that not only the strength of the magnetic field but
  also its orientation has a significant effect on the maximum mass of
  a compact stellar object. However, it is important to point out that
  in their study, neither the effect of the magnetic field nor of the
  magnetic field orientation on the TOV equation has been taken into
  account.  Hence, no effect of the orientation of the magnetic field
  was observed, with the exception of a change in the total stellar
  mass.  This is not expected in reality. Hence, it will be
interesting to investigate the properties of anisotropic compact stars
by considering the effects of magnetic field orientations and its
spatial distribution in the TOV equation. Therefore, in the present study,
we consider the presence of the effective anisotropy
that is arising due to (i) the local anisotropy of the fluid, and (ii)
the presence of a strong magnetic field.
Note that the magnetic field strength at the surface of magnetars is
  relatively weak, and it gradually increases up to several orders to
  reach its maximum value at the center~(see~\citealt{Melatos1999,Makishima2014,Dexheimer2017}).  Although the
  pressure anisotropy inside the magnetars may not be very large, the
  present study shows that the consideration of anisotropy offers a
  more generalized TOV equation to calculate the properties of
  magnetized anisotropic compact stars and also the effective
  anisotropy due to both the fields and matter plays a crucial role to
  ensure stability at the stellar center.

The present article is arranged as follows. In the next section, we
discuss the basic formalism and the modified hydrostatic equilibrium
equations of highly magnetized compact stellar objects. To close the
system of equations, we consider an ansatz for the anisotropy which is
introduced in subsection~\ref{subsec1.2}. The equation of state (EOS)
is discussed in subsection~\ref{subsec1.1} and the functional form of
the density-dependent magnetic field is considered in
subsection~\ref{subsec1.3}. In Section~\ref{secII} we discuss the
achieved results. Further, possible future directions of our study are
presented in Section~\ref{secIII}. We conclude this work with a brief
discussion in Section~\ref{secIV}.

\section{Basic Formalism and Structure equations for magnetized compact stars}\label{sec1}

To describe the interior spacetime of static, spherically symmetric
compact stellar objects, we consider the
metric\begin{equation}\label{1.1} \mathrm{d}s^2=e^{\nu(r)}
\mathrm{d}t^2 - e^{\lambda(r)}
\mathrm{d}r^2-r^2(\mathrm{d}\theta^2+\sin^2\theta \mathrm{d}\phi^2),
\end{equation}
 where $\nu(r)$ and $\lambda(r)$ are the metric potentials. 

%%%%%%%%%%%%%%%%%%%%%%%%%%%%%%%%%%%%%%%%%%%%%%%%%%%%%%%%%%%%%%%%%%%%%%%%%%%%%%%%%%%%%%%%%%%%%%%%%%%%%%%%%%%%%%%%%%%

\begin{figure*}[!htpb]
\centering 
\includegraphics[width=0.45\textwidth]{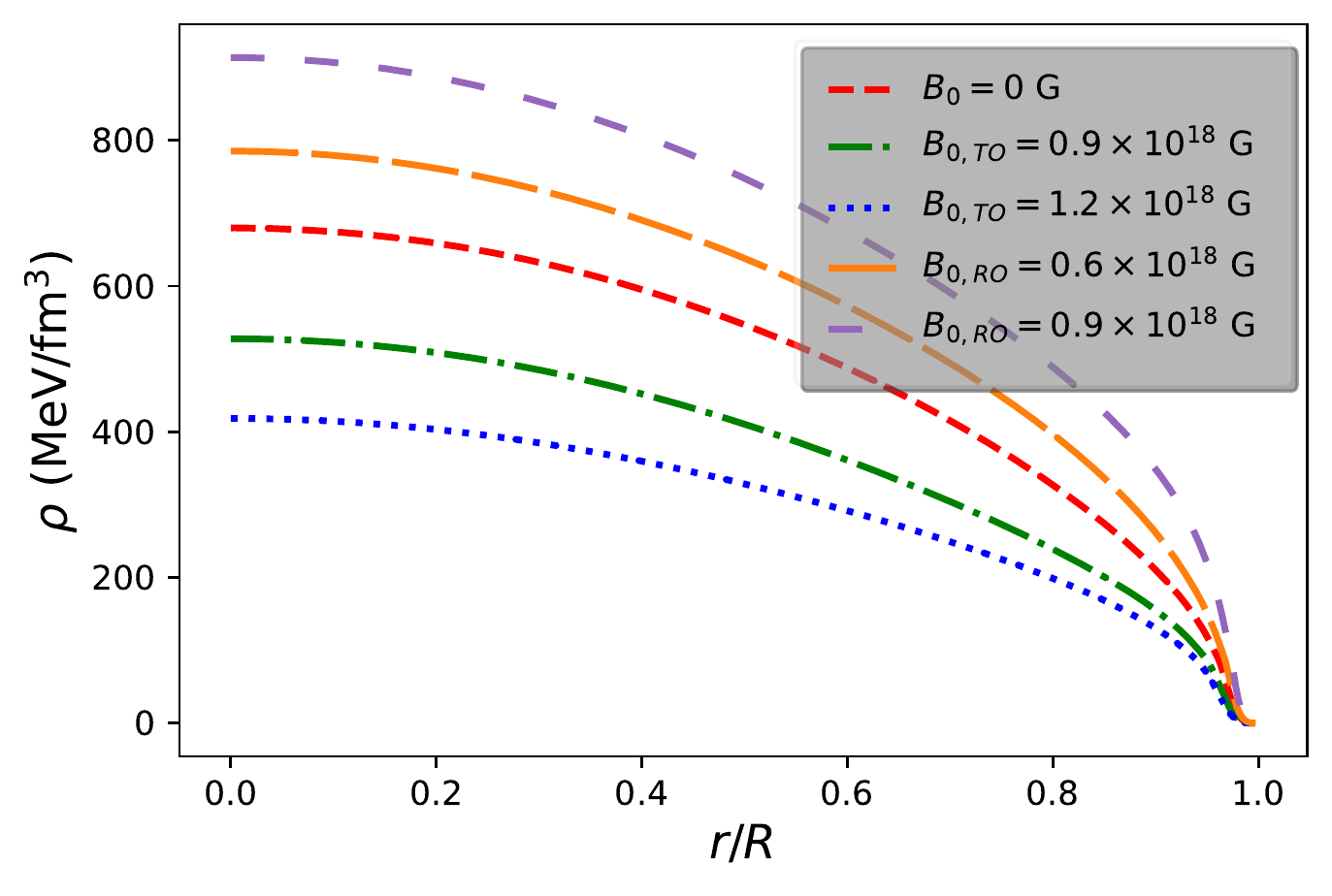} 
\includegraphics[width=0.45\textwidth]{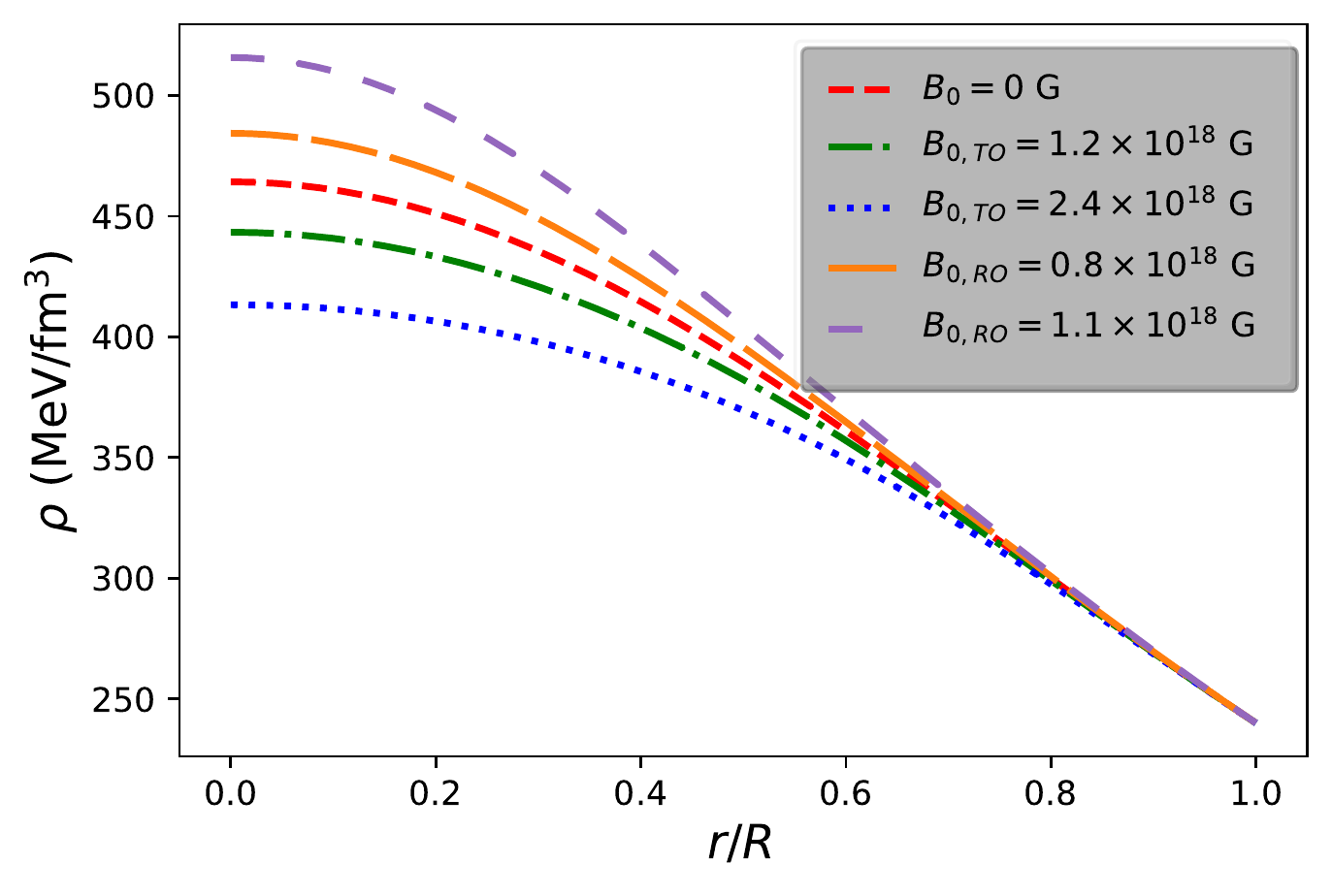} 
\includegraphics[width=0.45\textwidth]{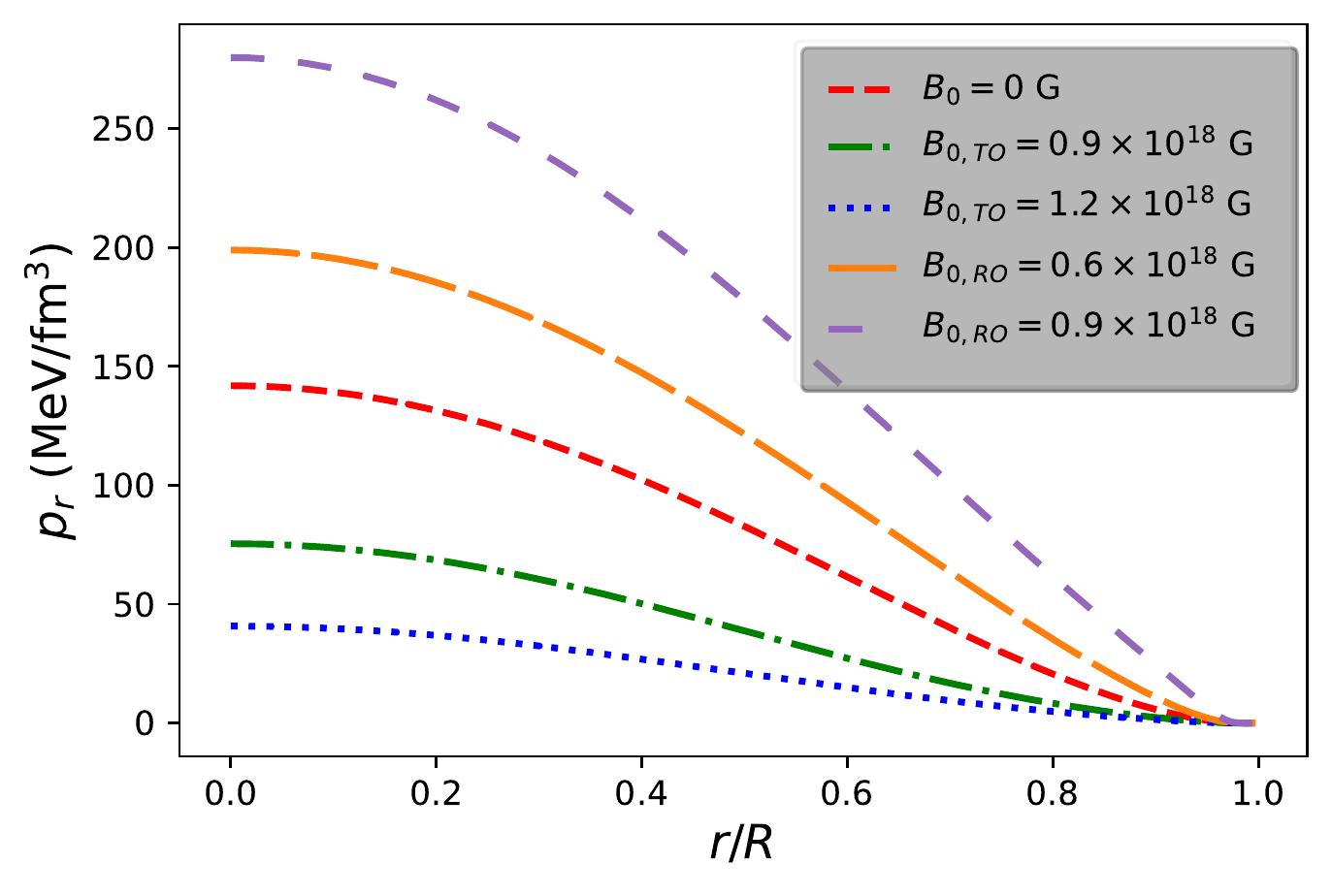} 
\includegraphics[width=0.45\textwidth]{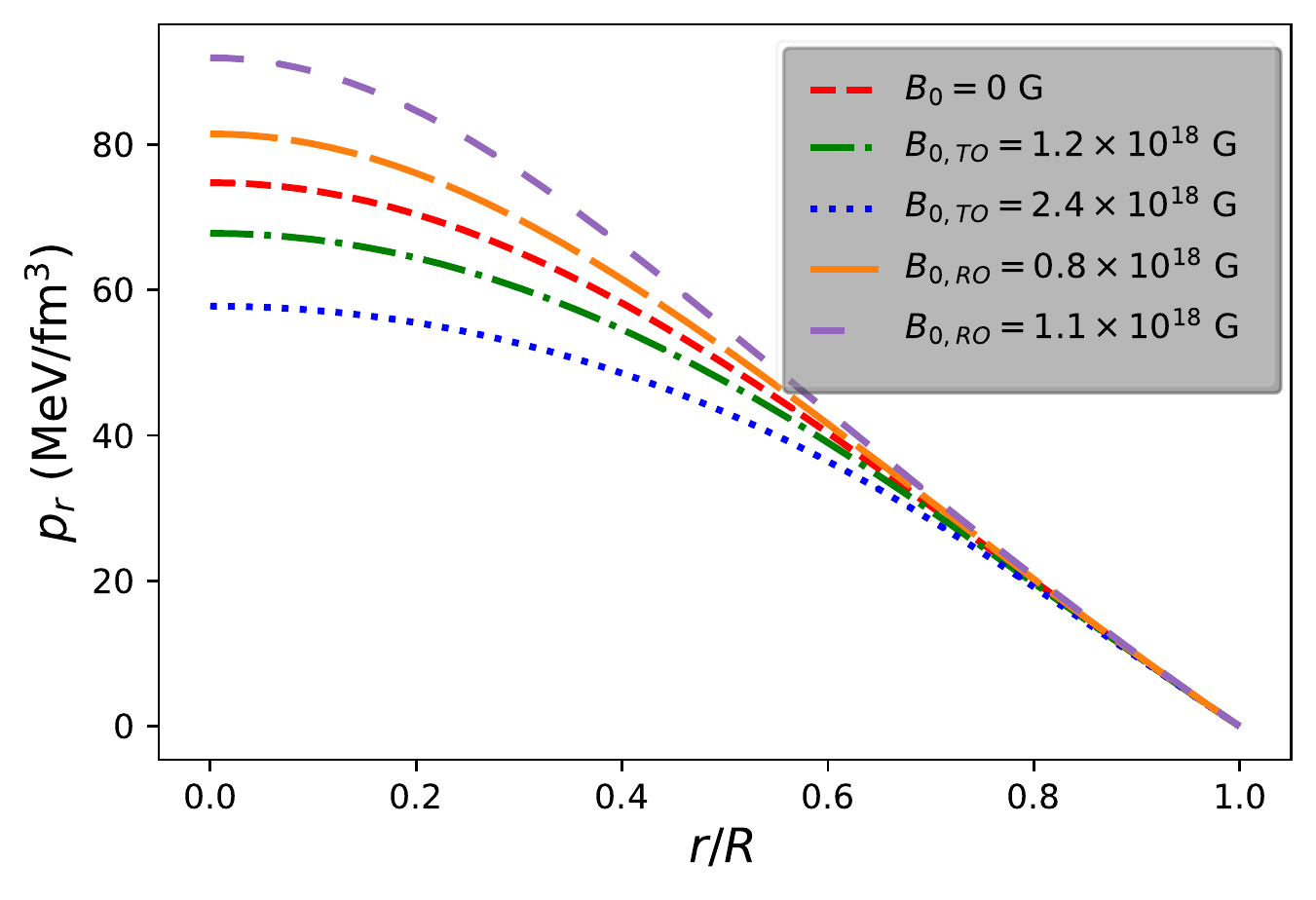} 
\includegraphics[width=0.45\textwidth]{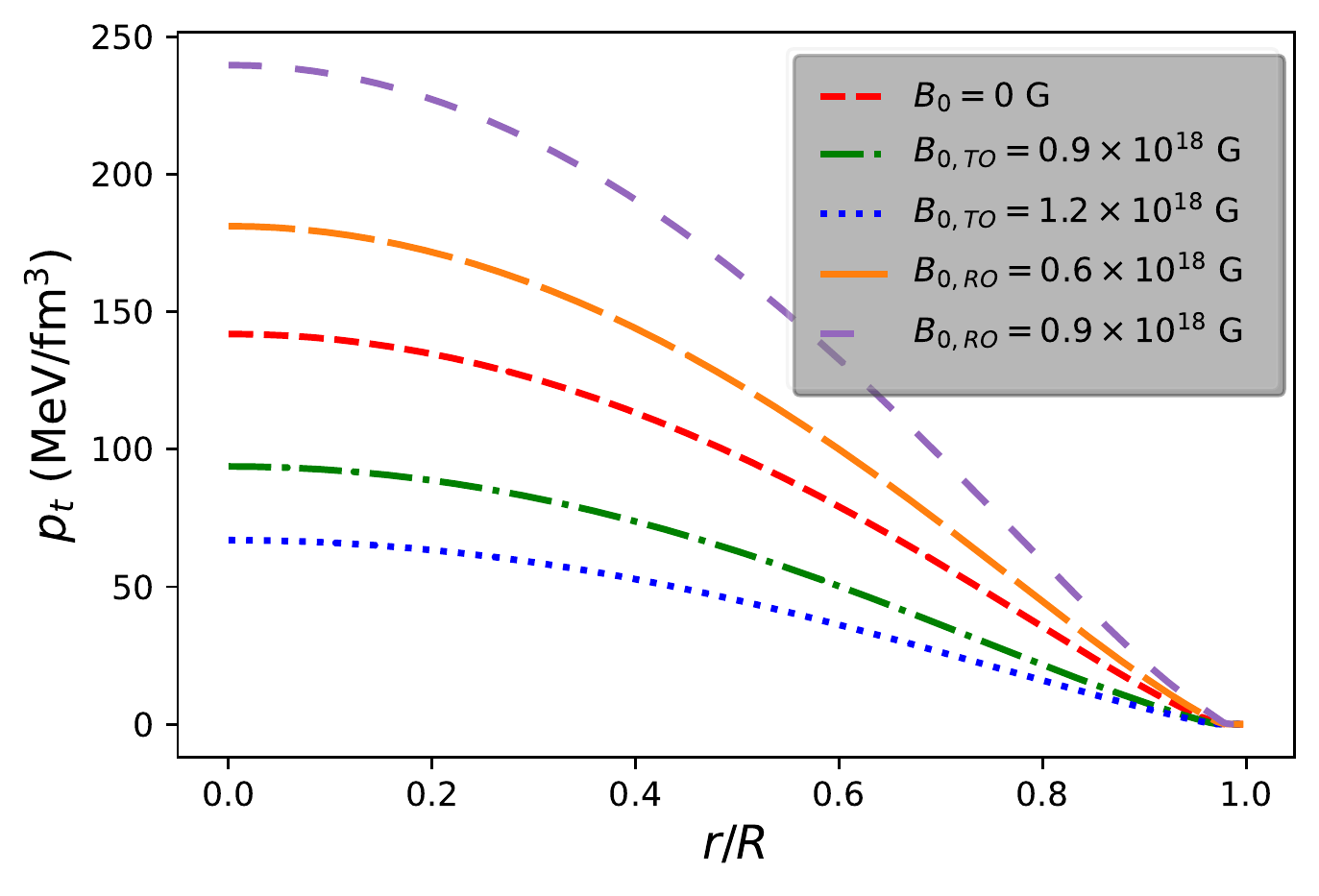} 
\includegraphics[width=0.45\textwidth]{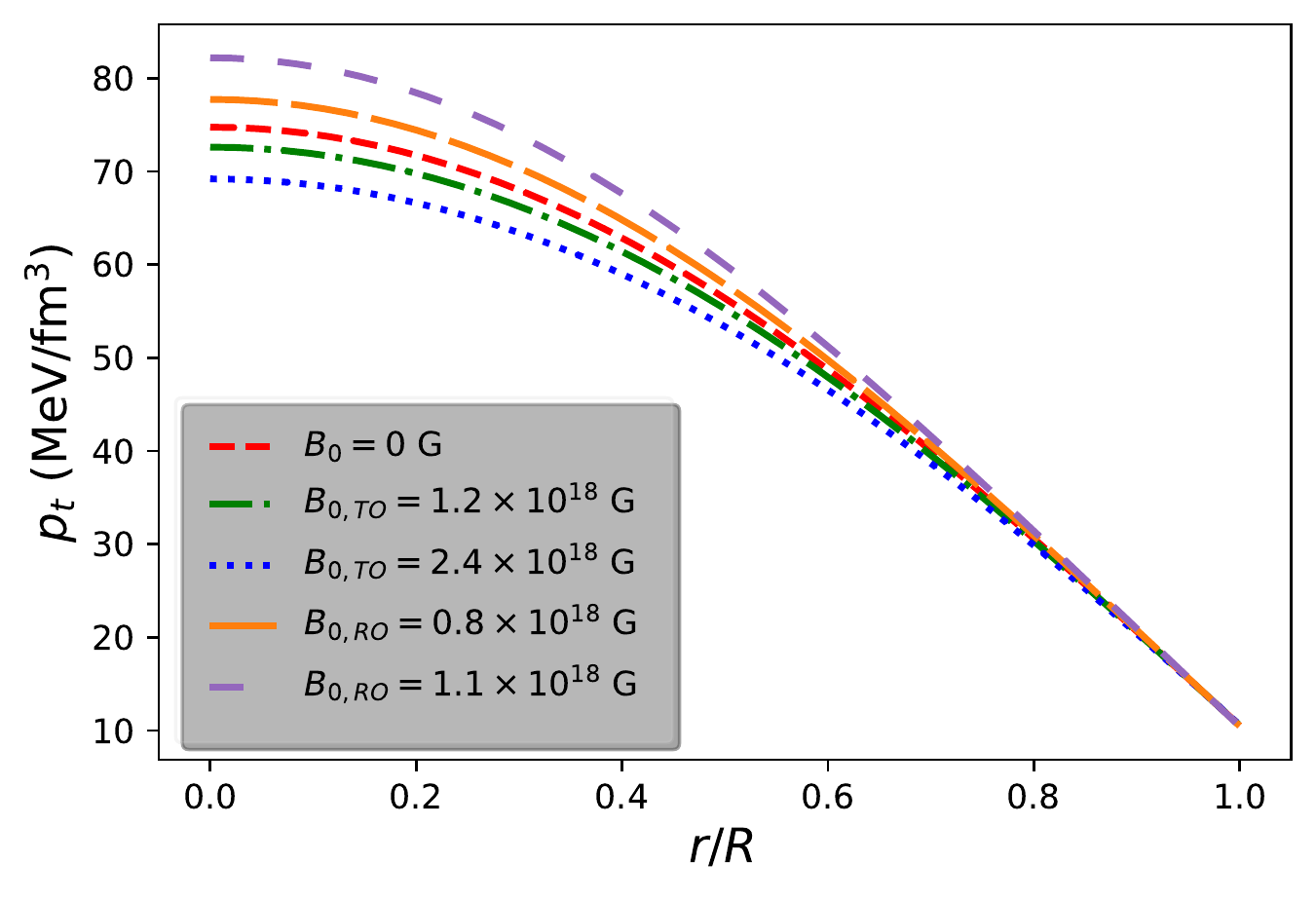}
\caption{Variation of (i) matter density ($\rho$), (ii)
  radial pressure $(p_r)$, and (iii) tangential pressure $(p_t)$ with
  radial coordinate $r/R$ for
  $2.01\pm0.04~M_\odot$~\citep{Antoniadis2013} NS candidate
  PSR~J0348+0432 (panels on the left) and
  $1.97\pm0.04~M_\odot$~\citep{Demorest2010} SQS candidate
  PSR~J1614$-$2230 (panels on the right).  Here and in what follows $\kappa=0.5$, $\eta=0.2$, $\gamma=2$ and $\mathcal{B}=60~\mathrm{MeV/{fm^3}}$.} \label{pressure}
\end{figure*}

%%%%%%%%%%%%%%%%%%%%%%%%%%%%%%%%%%%%%%%%%%%%%%%%%%%%%%%%%%%%%%%%%%%%%%%%%%%%%%%%%%%%%%%%%%%%%%%%%%%%%%%%%%%%%%%%%%%

The energy-momentum tensor of the system is given by
\begin{equation}\label{1.2}
T^{\mu\nu} = T^{\mu\nu}_m + T^{\mu\nu}_f,
\end{equation}
where $T^{\mu\nu}_m$ and $T^{\mu\nu}_f$ represent the contributions
due to the matter and field, respectively, which are given by
\begin{eqnarray}\label{1.3}
& \hspace{-1cm} T^{\mu\nu}_m = (\rho+{p_t})u^\mu
  u^\nu-{p_t}g^{\mu\nu}+\left({p_r}-{p_t}\right)v^\mu v^\nu
  \nonumber\\ & \hspace{3.5cm} +
  \frac{1}{2}\left(\mathcal{M}^{\mu\alpha}F^{\nu}_{\alpha}+\mathcal{M}^{\nu\alpha}F^{\mu}_{\alpha}\right),\\ \label{1.4}
  & \hspace{-2.2cm} T^{\mu\nu}_f =
  -\frac{1}{4\pi}F^{\mu\alpha}F^{\nu}_{\alpha} + \frac{1}{16\pi}
  g^{\mu\nu} F^{\beta\sigma}F_{\beta\sigma},
\end{eqnarray}
where $u^{\mu} =\delta^\mu_0 e^{-\nu(r)/2}$ which is the time-like
unit vector denoting the fluid 4-velocity of matter, whereas
$v^{\mu}=\delta^\mu_1 e^{-\lambda(r)/2}$ represents the space-like
unit vector in the radial direction. They satisfy $u^\mu
u_{\mu}=-v^{\mu}v_{\mu}=1$ and $u^{\mu}v_{\mu}=0$. The quantities
$\rho$, $p_r$ and $p_t$ represent the energy density of matter, the
radial pressure in the direction of $v^\mu$ and the tangential
pressure orthogonal to $v_\mu$, respectively. The quantities
$\mathcal{M}^{\mu\nu}$ and $F^{\mu\nu}$ represent the magnetization
tensor and the Maxwell tensor, respectively, and $g_{\mu\nu}$ is the
metric tensor. Now considering that in the bulk matter there are no
macroscopic charges, we can neglect the effects due to the electric
field and immediately obtain from the Eqs.~\eqref{1.3} and \eqref{1.4}
\begin{eqnarray}\label{1.5}
& \hspace{-1cm} T^{\mu\nu}_m = (\rho+{p_t})u^\mu
  u^\nu-{p_t}g^{\mu\nu}+\left({p_r}-{p_t}\right)v^\mu v^\nu\nonumber
  \\ & \hspace{3cm} +
  \mathcal{M}B\left(g^{\mu\nu}-u^{\mu}u^{\nu}+\frac{B^{\mu}B^{\nu}}{B^2}\right),\\ \label{1.6}
  & \hspace{-2.6cm} T^{\mu\nu}_f =
  \frac{B^2}{4\pi}\left(u^{\mu}u^{\nu}-\frac{1}{2}g^{\mu\nu}\right)-\frac{B^{\mu}B^{\nu}}{4\pi},
\end{eqnarray}
where $\mathcal{M}$ is the magnetization per unit volume and $B^\mu
B_\mu = -B^2$. \cite{Ferrer2010} and~\cite{Sinha2013} found in their
works that the magnetization is at least one order of
  magnitude smaller than the magnetic pressure and that magnetization
has no effect on the physical properties of magnetized
matter.  Hence the magnetization effect is very small and
will therefore be neglected in the numerical results of
  our study.  Importantly, we assume the field strengths to be such
that they do not or only minimally effect the spherical shape of a
compact star. Moreover, toroidally dominated magnetized compact stars
do not deviate much from spherical
symmetry~\citep{Das2015b,Subramanian2015,Kalita2019}. We therefore
make use of the standard form of the TOV equation for
the description of magnetized compact stars in the present work.

The system density $(\widetilde{\rho})$, which is the sum of the
contribution from the matter and field, is given by
\begin{eqnarray}\label{1.7}
\widetilde{\rho} = \rho + \frac{B^2}{8\pi}.
\end{eqnarray}
Depending on the magnetic field orientation, the system's parallel
pressure along the magnetic field reads
 \begin{eqnarray}\label{1.8}
 p_{\parallel} = \begin{cases}
 p_r - \frac{B^2}{8\pi}, \hspace{1cm} &\textrm{for~ RO}\\
 p_t - \frac{B^2}{8\pi}. \hspace{1cm} &\textrm{for~ TO}
 \end{cases}
 \end{eqnarray}
Similarly, the system transverse pressure perpendicular to the magnetic field is given by
\begin{eqnarray}\label{1.9}
 p_{\bot} = \begin{cases}
 p_t + \frac{B^2}{8\pi}, \hspace{1cm} &\textrm{for~ RO}\\
 p_r + \frac{B^2}{8\pi}. \hspace{1cm} &\textrm{for~ TO}
 \end{cases}
 \end{eqnarray}
Note that $p_t$ stands for the pressure either in the polar or in the
azimuthal directions in spherical symmetry.

The mass function of a star in the presence of a magnetic filed is defined as
\begin{eqnarray}\label{ece1}
m\left(r\right)= \int_0^r  4\pi r^2 \widetilde{\rho} \;\mathrm{d}r.
\end{eqnarray}

Finally, the conservation of the energy momentum tensor is expressed as
\begin{equation}\label{ece2}
\nabla_\mu T^{\mu\nu}=0.
\end{equation}

Following Eqs.~\eqref{1.1},~\eqref{1.2},~\eqref{1.5}-\eqref{ece2}, the
essential stellar structure equations needed to describe static,
anisotropic, spherically symmetric compact objects in the presence of
a strong magnetic field take the form
\begin{eqnarray}\label{1.10}
& \hspace{-4.5cm} \frac{\mathrm{d}m}{\mathrm{d}r} = 4\pi
  \left(\rho+\frac{B^2}{8\pi}\right) r^2,\\ \label{1.11}
  & \hspace{-0.7cm} \begin{cases} {\frac {{\rm d}p_r}{{\rm
          d}r}}=\frac{-\left(\rho+p_{{r}}\right)\frac{4\pi{r}^{3}
        \left( p_{{r}}-{\frac {{B}^{2}}{8\pi}}
        \right)+m}{r\left(r-2m\right)}+\frac{2}{r}\Delta}{\left[1-\frac{\rm
          d}{{\rm d}\rho}\left(\frac{B^2}{8\pi}\right)\frac{{\rm
            d}\rho}{{\rm d}{p_r}}\right]}, \hspace{0.94cm}\textrm{for
      RO}\\ {\frac {{\rm d}p_r}{{\rm d}r}}=
    \frac{-\left(\rho+p_{{r}}+\frac{B^2}{4\pi}\right)\frac{4\pi{r}^{3}
        \left( p_{{r}}+{\frac {{B}^{2}}{8\pi}}
        \right)+m}{r\left(r-2m\right)}+\frac{2}{r}\Delta}{\left[1+\frac{\rm
          d}{{\rm d}\rho}\left(\frac{B^2}{8\pi}\right)\frac{{\rm
            d}\rho}{{\rm d}{p_r}}\right]}, \hspace{0.2cm}\textrm{for
      TO}
\end{cases} 
\end{eqnarray}
where $\Delta=\left(p_t-p_r+\frac{B^2}{4\pi}\right)$ or
$\left(p_t-p_r-\frac{B^2}{8\pi}\right)$, which denote the effective
anisotropy of stellar structures for RO or TO, respectively.

For the non-magnetized case, i.e., $B=0$, Eq.~\eqref{1.11} reduces
to the standard form of the TOV
equation~\citep{Bowers1974,Herrera2013}. It is important to mention
that throughout the present investigation, we consider field
magnitudes $<3\times {10^{18}}$~G, hence the effects of Landau
quantization are negligible. In fact, the effects due to Landau
quantization become significant only for fields larger than
${10}^{19}$~G~\citep{Sinha2013}. Therefore,
following~\cite{Sinha2013}, we consider Landau quantization effects to
be negligible. The anisotropic contribution due to the magnetic field,
however, will still be there since the difference between the parallel
and transverse pressures is proportional to the square of the field
magnitude. Besides, one should not forget the local anisotropy of the
fluid.

%%%%%%%%%%%%%%%%%%%%%%%%%%%%%%%%%%%%%%%%%%%%%%%%%%%%%%%%%%%%%%%%%%%%%%%%%%%%%%%%%%%%%%%%%%%%%%%%%%%%%%%%%%%%%%%%%%%

\begin{figure}[!htpb]
\centering 
\includegraphics[width=0.45\textwidth]{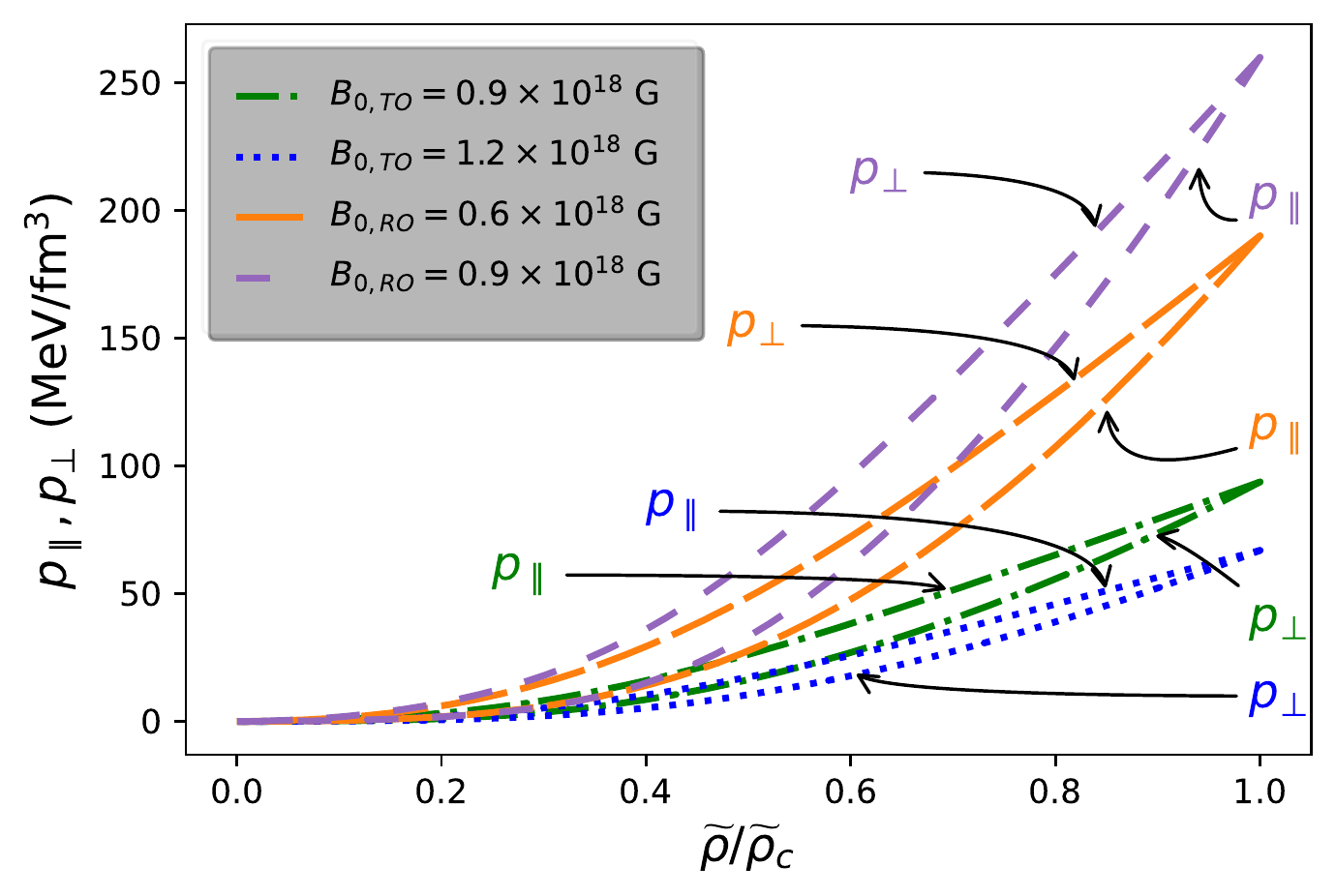}
\includegraphics[width=0.45\textwidth]{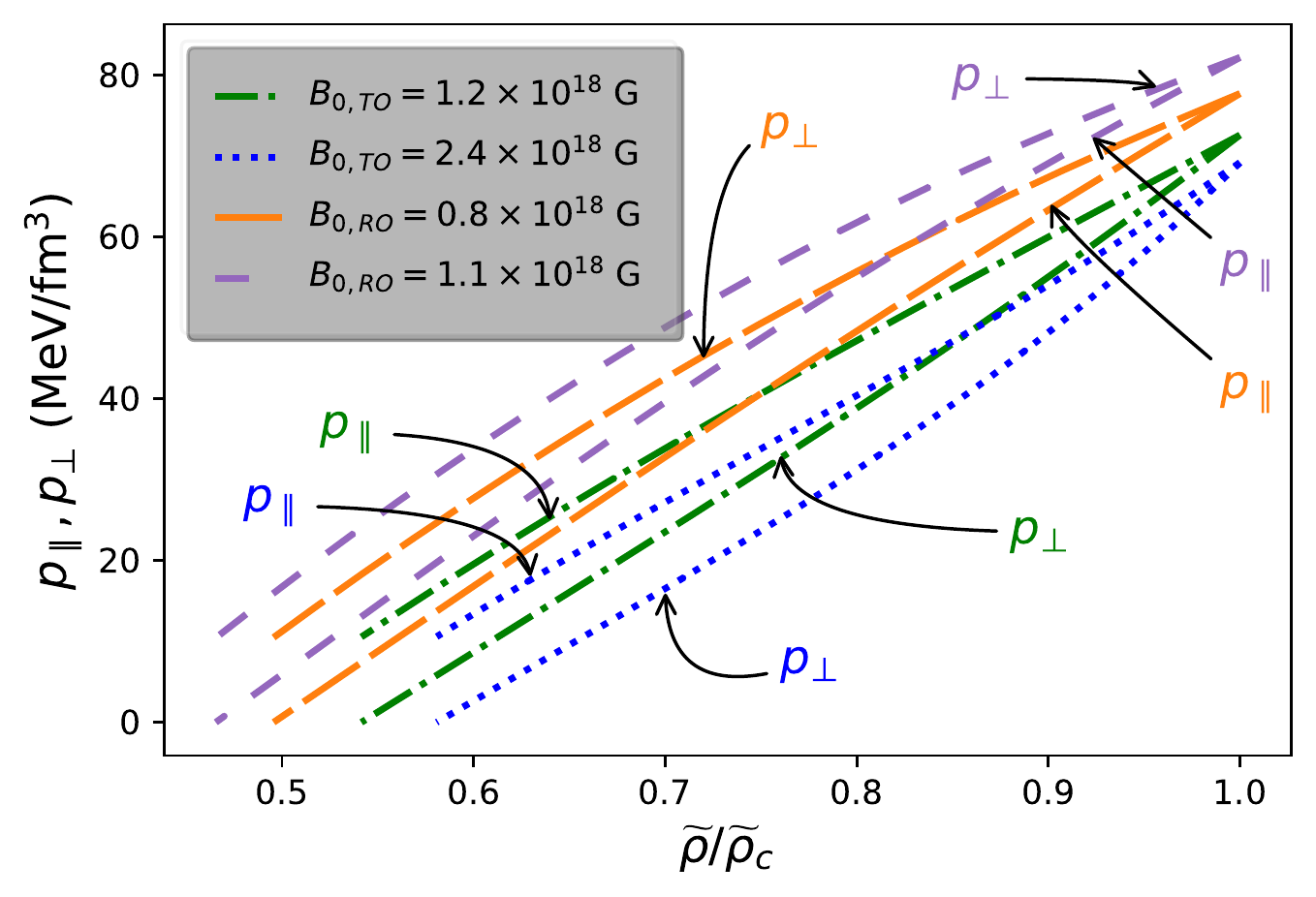}
\caption{Variation of parallel pressure $(p_\parallel)$
  and transverse pressure $(p_\bot)$ with the system
  density~$(\widetilde{\rho})$, normalized to the central system
  density~$(\widetilde{\rho}_c)$, for
  $2.01\pm0.04~M_\odot$~\citep{Antoniadis2013} NS candidate
  PSR~J0348+0432 and $1.97\pm0.04~M_\odot$~\citep{Demorest2010} SQS
  candidate PSR~J1614$-$2230.  The upper panel features NS pressure
  profiles, whereas the lower panel presents SQS pressure
  profiles. The dotted, dash-dotted, long-dashed and short-dashed
  curves correspond to $B_0=2.4\times{10}^{18}$~G (TO),
  $B_0=1.2\times{10}^{18}$~G (TO), $B_0=0.8\times{10}^{18}$~G (RO) and
  $B_0=1.1\times{10}^{18}$~G (RO), respectively.} \label{EOS}
\end{figure}

%%%%%%%%%%%%%%%%%%%%%%%%%%%%%%%%%%%%%%%%%%%%%%%%%%%%%%%%%%%%%%%%%%%%%%%%%%%%%%%%%%%%%%%%%%%%%%%%%%%%%%%%%%%%%%%%%%%

\subsection{Ansatz for Anisotropy}\label{subsec1.2}

Further, we require a functional form for the anisotropy $(\Delta)$ to
close the system of equation in such a way that we may include the
anisotropic effect due to both the local anisotropy of the fluid and
the presence of a strong magnetic field. Unfortunately, there is no
available explicit form of anisotropy in the existing literature
derived directly from the microscopic theory, which can explain the
combined anisotropic effects due to both the fluid and magnetic
field. To overcome this delicate issue, we consider a phenomenological
approach based on the essential assumptions given bellow.

(i) At the stellar center the hydrodynamic force $F_h$ and gravitational force $F_g$ are zero. To maintain the stability of the system via equilibrium of the forces (non-diverging nature), the anisotropic force essentially should be zero at the center, which implies that the anisotropy must vanish quadratically at the center.

(ii) The anisotropy should vary with position inside the system and
also depend non-linearly on $p_r$~\citep{Bowers1974,Silva2015}.

(iii) Based on the present study, the functional form of the
anisotropy should include the anisotropic effects due to both the
local anisotropy of the fluid and the presence of a strong magnetic
field. It is also important to include the effects due to magnetic
field orientation.

\cite{Bowers1974} derived a general parametric form for $\Delta$ in
general relativity for a spherically symmetric star, which is
consistent with the above mentioned essential assumptions (i) to (iii). In the
years following the Bowers and Liang paper, hundreds of articles have
investigated the effects of anisotropy for compact stars using this
parametric form of anisotropy, which has become widely accepted within
the community. To include the effects of the magnetic field and its
orientation, here we modify the Bowers-Liang anisotropic form, which
reads
\begin{eqnarray}\label{1.14}
 \Delta = \begin{cases} \kappa
   \frac{\left(\rho+p_r\right)\left(\rho+3\,p_r-\frac{B^2}{4\pi}\right)}{\left(1-\frac{2m}{r}\right)}r^2, \hspace{1.5cm}
   \textrm{for RO}\\ \kappa
   \frac{\left(\rho+p_r+\frac{B^2}{4\pi}\right)\left(\rho+3\,p_r+\frac{B^2}{2\pi}\right)}{\left(1-\frac{2m}{r}\right)}r^2, \hspace{0.8cm}
   \textrm{for TO}
 \end{cases}
\end{eqnarray}
where the dimensionless constant $\kappa$ controls the strength of the
anisotropy in the system. Note that we consider the parametric values
of $\kappa$ well within its limiting values given by
$\left[-\frac{2}{3},\frac{2}{3}\right]$~\citep{Silva2015}. Note
that~\cite{Ferrer2010} introduced ``\emph{anisotropy, which leads to
the distinction between longitudinal- and transverse-to-the-field
pressures}". To this end, our study has focused on the important fact
that the anisotropy necessarily should be zero at the center in the
case of magnetized compact stars and the stellar models with non-zero
anisotropies at the center would have unstable cores [see
  Eq.~\eqref{1.10}], which would eliminate such theoretical models.

 The chosen parametric form for the
  anisotropy based on a phenomenological approach is consistent with
  the essential physical assumptions (i) to (iii) and has been widely
  accepted by the community. It includes the effects of magnetic
  fields and their orientations and constitutes currently the best
  possible physically viable way of solving the hydrostatic
  equilibrium equations of magnetized compact stars.

%%%%%%%%%%%%%%%%%%%%%%%%%%%%%%%%%%%%%%%%%%%%%%%%%%%%%%%%%%%%%%%%%%%%%%%%%%%%%%%%%%%%%%%%%%%%%%%%%%%%%%%%%%%%%%%%%%%

\begin{figure}[!htpb]
\centering 
\includegraphics[width=0.45\textwidth]{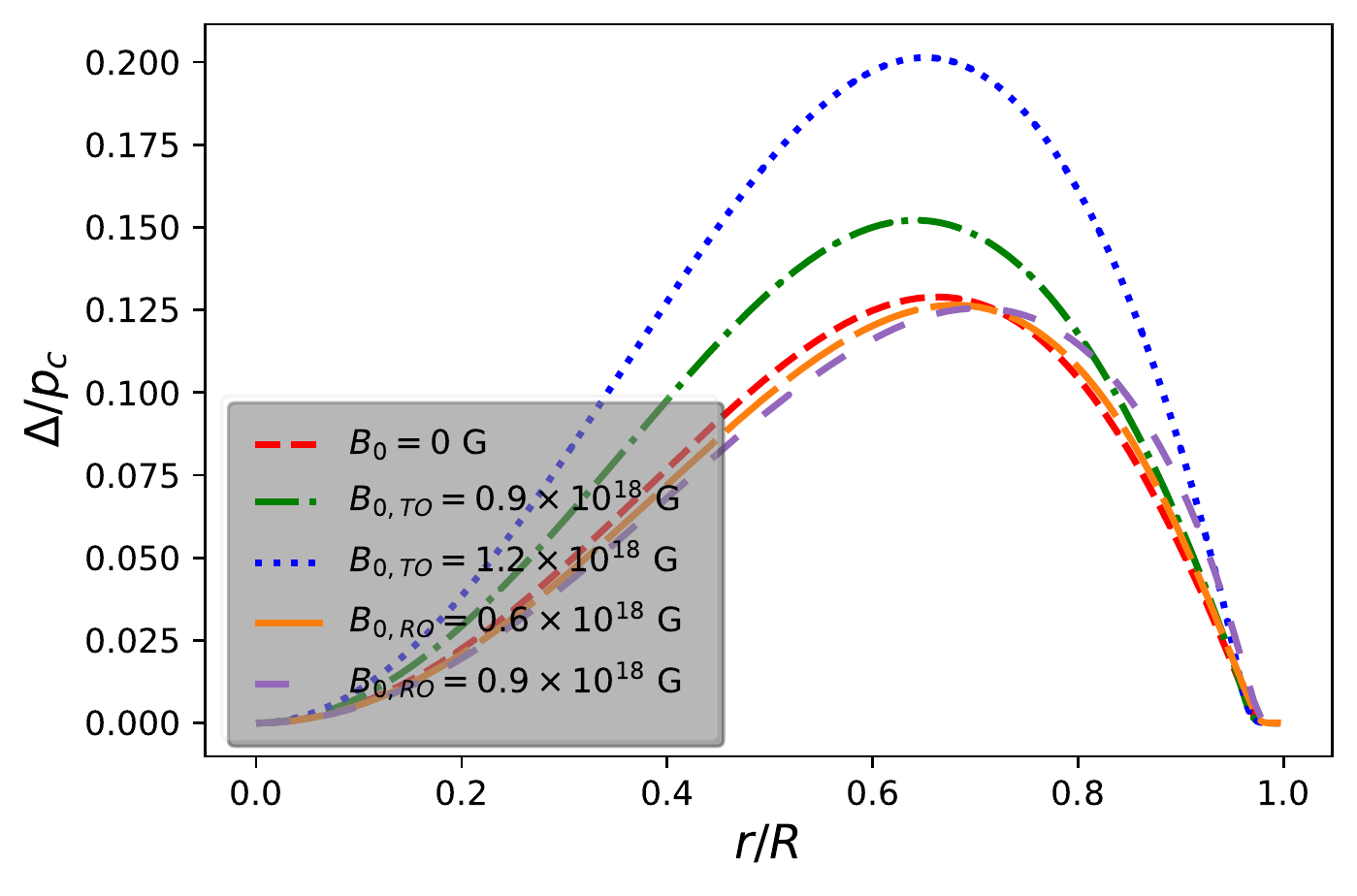}
\includegraphics[width=0.45\textwidth]{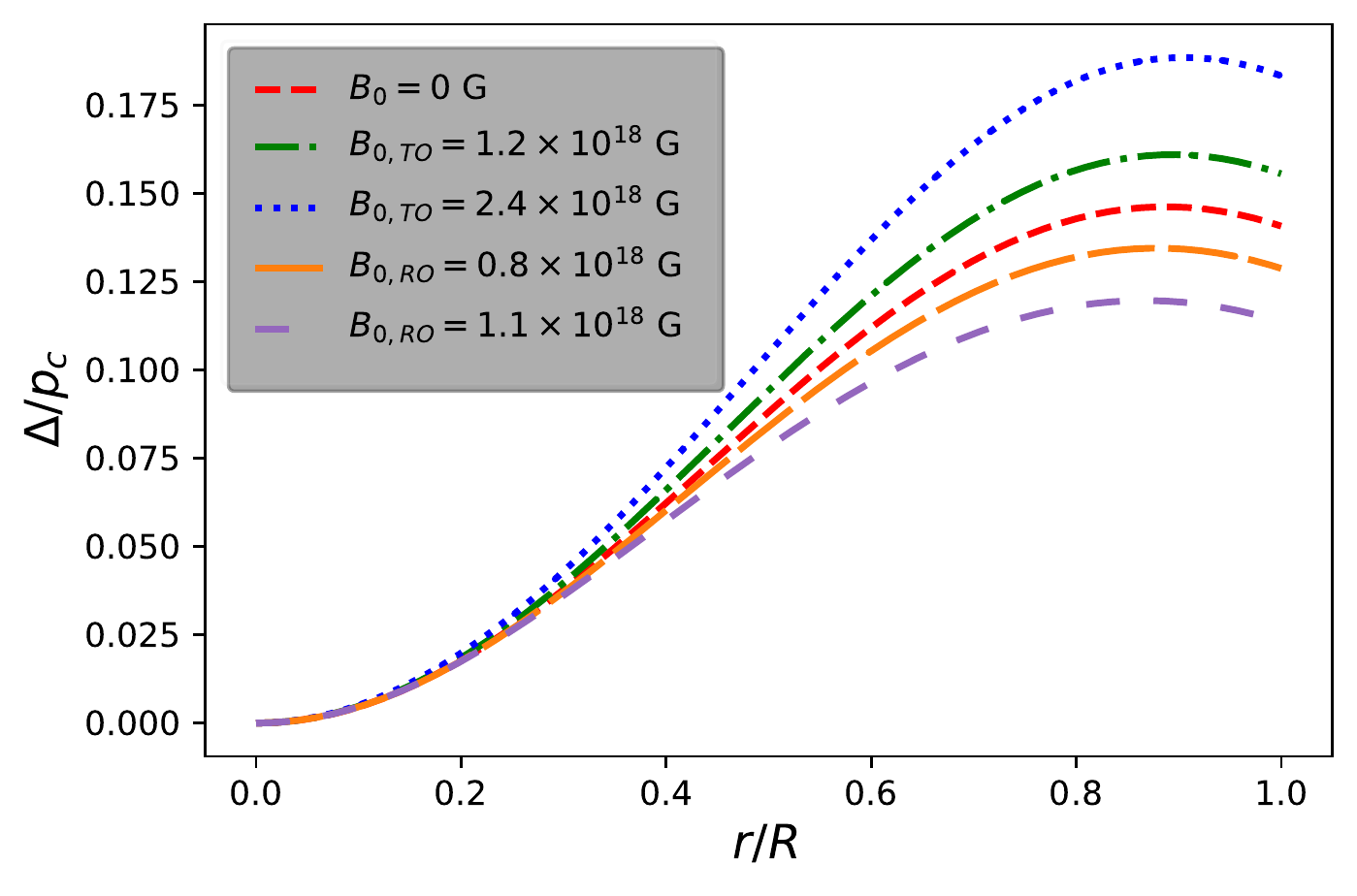}
\caption{Anisotropy profiles $\Delta$, normalized to
  central pressure $(p_c)$, for $2.01\pm0.04~M_\odot$~\citep{Antoniadis2013} NS candidate
  PSR~J0348+0432 and $1.97\pm0.04~M_\odot$~\citep{Demorest2010} SQS
  candidate PSR~J1614$-$2230. The top panel is for NSs, the bottom
  panel is for SQSs.} \label{anisotropy}
\end{figure}

%%%%%%%%%%%%%%%%%%%%%%%%%%%%%%%%%%%%%%%%%%%%%%%%%%%%%%%%%%%%%%%%%%%%%%%%%%%%%%%%%%%%%%%%%%%%%%%%%%%%%%%%%%%%%%%%%%%

\subsection{Equation of state}\label{subsec1.1}

Next we consider the relation between $\rho$ and $p_r$, known
as EOS, to close our system of equations. By providing the EOS of
matter together with the functional form for the anisotropy of the
system, the stellar structure equations~\eqref{1.10} and \eqref{1.11}
can be then solved numerically. In order to obtain solutions of the 
coupled stellar structure equations, it is required to integrate
Eqs.~\eqref{1.10} and \eqref{1.11} simultaneously, from the stellar center to
the surface.

We consider SLy EOS, which is moderately stiff in classification.  Based on Skyrme-type energy density functional, \cite{Douchin2001} proposed SLy EOS, which is widely used in the literature to discuss NSs.  Note that SLy EOS is equally consistent in the NS core and the crust.

The phenomenological MIT bag model EOS was introduced
by~\cite{Chodos1974} to study strongly interacting particles, viz.,
hadrons.  In our work, we use the MIT bag
  model EOS to describe (absolutely stable) SQM and
  to compute the properties of SQSs.  The $u$, $d$, and $s$ quarks are treated  as massless but
  relativistic  particles confined inside a spherical bag, in which case  the
  EOS of SQM is  given by
\begin{eqnarray}\label{1.13} 
p_r = \frac{1}{3} \left(\rho - 4\,\mathcal{B}\right),
\end{eqnarray}
where $\mathcal{B}$ denotes the MIT bag constant.  Our
  numerical results are computed for a bag constant value of
  $\mathcal{B}=60~\mathrm{MeV/{fm^3}}$ $(\mathcal{B}^{1/4} =
  146~\mathrm{MeV})$, which corresponds to SQM that is strongly bound 
  (of strange quark mass $\sim 100$~MeV) with respect to ordinary nuclear matter and
  $^{56}{\text Fe}$ \citep{Farhi1984,Weber2005}. As required by 
  SQM hypothesis
  \citep{Bodmer1971,Witten1984,Terazawa1979}, the energy per baryon
  of 2-flavor ($u$, $d$) quark matter for this value of the bag
  constant is higher than the energy per baryon of nuclear matter and
  $^{56}{\text Fe}$.  We also note that this $\mathcal{B}$ value lies
  within the range of $57-94~\mathrm{MeV/{fm^3}}$, which is
  $(145 \lesssim \mathcal{B}^{1/4} \lesssim 164~\mathrm{MeV})$
  frequently studied in the literature dealing with absolutely stable
  strange quark
  matter~\citep{Farhi1984,Alcock1986,Burgio2002,Jaikumar2006,Bordbar2012,Maharaj2014,Arbanil2016,Moraes2016,Alaverdyan2017,Lugones2017,Deb2019}.

%%%%%%%%%%%%%%%%%%%%%%%%%%%%%%%%%%%%%%%%%%%%%%%%%%%%%%%%%%%%%%%%%%%%%%%%%%%%%%%%%%%%%%%%%%%%%%%%%%%%%%%%%%%%%%%%%%%

\begin{figure}[!htpb]
\centering
\includegraphics[width=0.45\textwidth]{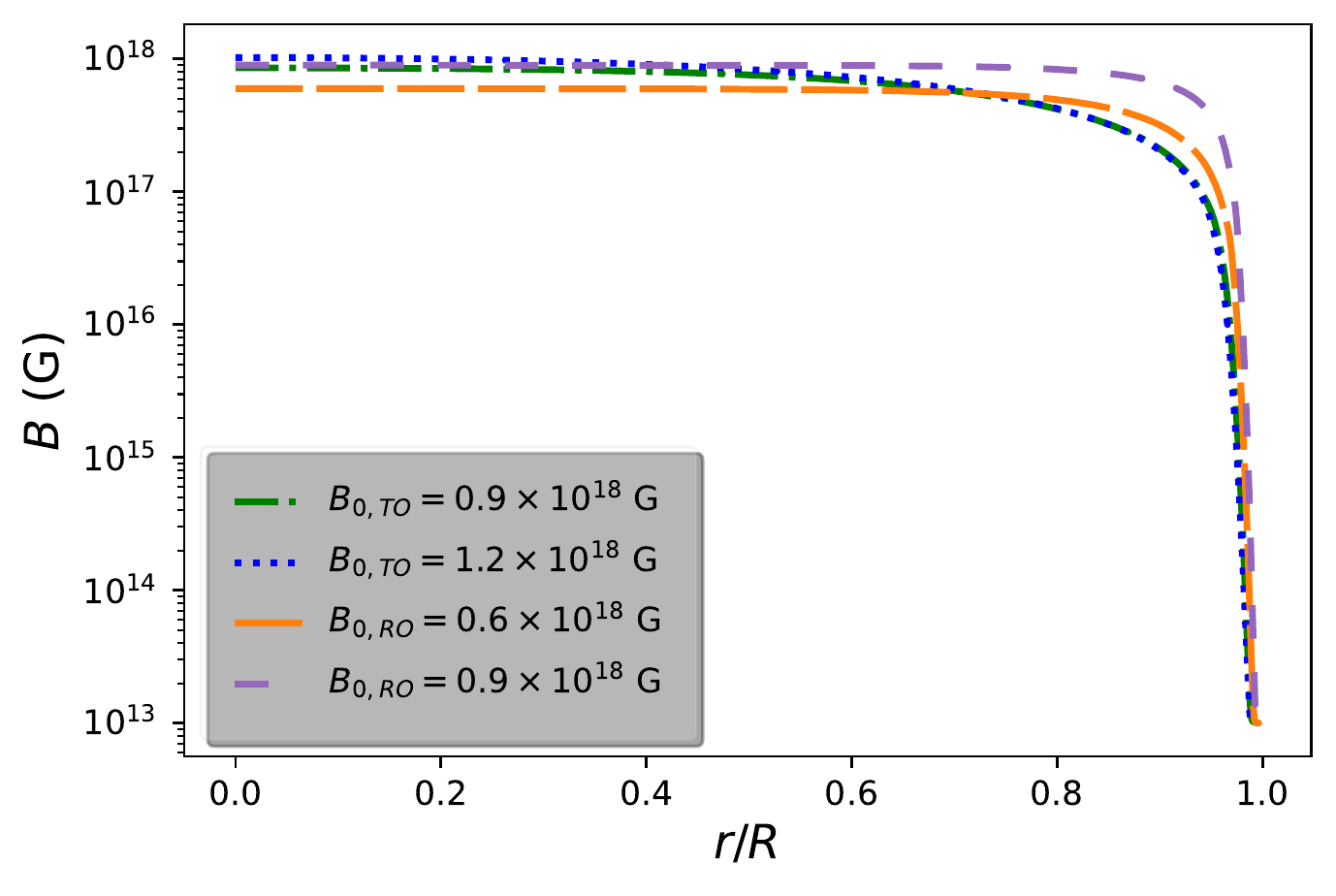}
\includegraphics[width=0.45\textwidth]{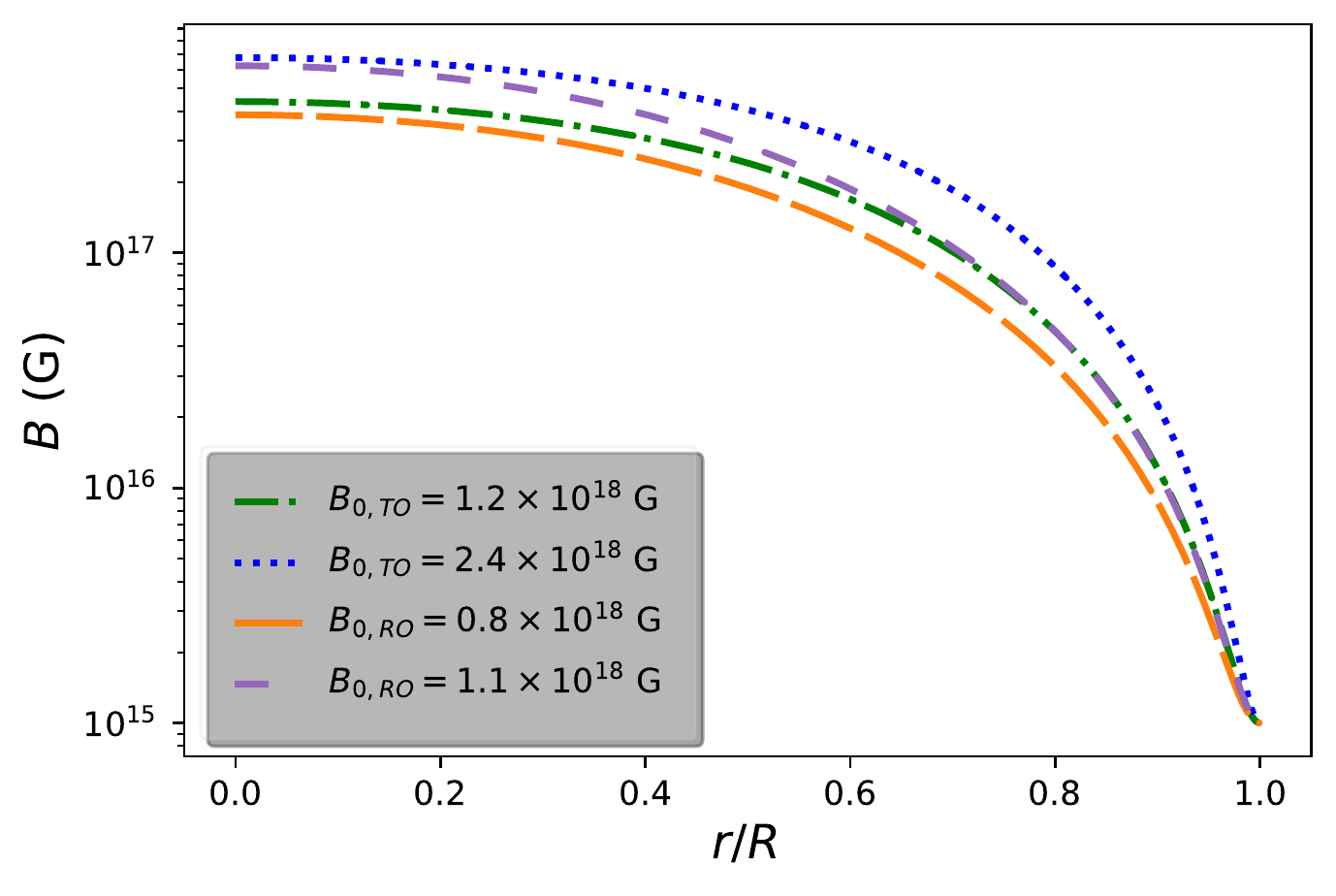}
\caption{Variation of magnetic field $B$ with
  radial coordinate $r/R$, for
  $2.01\pm0.04~M_\odot$~\citep{Antoniadis2013} NS candidate
  PSR~J0348+0432 and $1.97\pm0.04~M_\odot$~\citep{Demorest2010} SQS
  candidate PSR~J1614$-$2230. The upper and lower panels represent NS
  and SQS magnetic field profiles, respectively.} \label{mag_field}
\end{figure}

%%%%%%%%%%%%%%%%%%%%%%%%%%%%%%%%%%%%%%%%%%%%%%%%%%%%%%%%%%%%%%%%%%%%%%%%%%%%%%%%%%%%%%%%%%%%%%%%%%%%%%%%%%%%%%%%%%%

\subsection{Profile for density-dependent magnetic fields}\label{subsec1.3}
To solve Eqs.~\eqref{1.10} and \eqref{1.11} simultaneously, one needs
to close the system of equations by specifying a parametric form of
the magnetic field strength. To mimic the spatial dependence of the
magnetic field strength, which varies from the stellar center to the
surface, in the present study we consider a density-dependent
parametric form for the magnetic field strength which was conceptualized
by~\cite{Bandyopadhyay1997a,Bandyopadhyay1998a} and later was widely
applied in
literature~\citep{Menezes2009a,Ryu2010,Ryu2012,Sinha2013,Chu2014,Chu2015,Isayev2018,Roy2019,Aguirre2020,Thapa2020}.

Therefore, following~\cite{Bandyopadhyay1997a,Bandyopadhyay1998a}, we
choose the profile for the density-dependent magnetic field in such a
way that the magnetic field at the stellar core, $B_c$, complies with
the virial theorem and the surface magnetic field, $B_s$, fits
observed values. This profile is given by
\begin{eqnarray}\label{1.15}
B(\rho)= B_s +
B_0\left[1-\exp\left\lbrace-\eta\left(\frac{\rho}{\rho_0}\right)^\gamma\right\rbrace\right],
\end{eqnarray}
where $B_0$ is a parameter that has the same dimension as the magnetic
field strength, and the dimensionless parameters $\eta$ and $\gamma$
control how the magnetic field decays from its maximum value at the
center to the minimum value at the surface. More precisely, $\eta$
controls the field decay at the saturation density, and the width of
the transition is controlled by $\gamma$.  Here, $\rho_0$ denotes the
normal nuclear matter density. However, one may note that although
Eq.~\eqref{1.15} is applicable to NSs, for SQSs, where the surface
density is $\rho_s \neq 0$, a modification of the magnetic field
profile is required to ensure that the asymptotic value for $B_s$ is
obtained at the surface, given by
\begin{eqnarray}\label{1.16}
B(\rho)= B_s +
B_0\left[1-\exp\left\lbrace-\eta\left(\frac{\rho-\rho_s}{\rho_0}\right)^\gamma\right\rbrace\right].
\end{eqnarray} 

In the present study, we shall consider values of $B_s$ given by
$10^{13}$ and $10^{15}$ G for NSs and SQSs, respectively. However, we
found that our results are not sensitive to the particular choice of
the value of $B_s$.

\subsection{Consistency of Maxwell's equation with the magnetic field orientations} 

\cite{Chu2014,Chu2015} in their works showed that magnetic field
orientations have a significant effect on spherically symmetric
compact stars. Following their work, we offer a more general model by
considering the same magnetic field orientations, such as ``radial
orientation" when the local magnetic fields orient themselves along
the radial direction and the ``transverse orientation" when the local
magnetic fields are oriented perpendicularly to the radial
direction. Now we show in a straightforward way that there is no
violation of $\overrightarrow{\nabla} \cdot \overrightarrow{B}=0$ for
the present interest of magnetic field orientations:

(i) \textbf{Radial orientation:}
For a radial orientation, the magnetic field takes the form
\begin{eqnarray}\label{pf1}
\overrightarrow{B}=\left(B_r,0,0\right).
\end{eqnarray}
Now, 
\begin{eqnarray}\label{pf2}
\overrightarrow{\nabla} \cdot \overrightarrow{B}=0
\Rightarrow \frac{1}{r^2} \frac{\partial}{\partial r}\left(r^2 B_r\right)=0
\Rightarrow B_r = \frac{K}{r^2}.
\end{eqnarray}
Here $K$ cannot be a pure constant in order to avoid 
absurd possibility of magnetic monopole. Hence, $K$ could be $K\left(\theta,\phi\right)$. Hence, $B_r$ could
be thought of as $B_r={K {\rm sign}(\cos\theta)}/r^2$, i.e., with upper
hemisphere $+K$ and lower hemisphere $-K$. This physically implies that the field
lines coming out of the upper hemisphere and entering through the lower 
hemisphere of the star, hence having split monopole type in nature. Of course, this
is an approximate modeling of the magnetic field in a star assuming to be spherical in
shape. However, for the present purpose, this will not pose any practical hindrance 
in order to understand the physics.

For the ease of understanding, let us choose Minkowski space, which
leads the spatial components to
\begin{eqnarray}\label{pf3}
M^{ij}=\frac{B^2}{8\pi} \delta^{ij} - \frac{B^i B^j}{4\pi}.
\end{eqnarray}

From Eq.~\eqref{pf3} one can see that
$M^{rr}=\frac{B^2}{8\pi}-\frac{B^2}{4\pi}=-\frac{B^2}{8\pi}$,
$M^{\theta \theta}=\frac{B^2}{8\pi}=M^{\phi \phi}$.

Finally, we have
\begin{equation}\label{pf4}
T^{\mu\nu}_f = \left(\begin{array}{cccc}    \frac{B^2}{8\pi}    &      0               &     0              &           0\\
                                               0             &  -\frac{B^2}{8\pi}   &     0              &           0\\
                                               0             &      0               &  \frac{B^2}{8\pi}  &           0\\
                                               0             &      0               &     0              &  \frac{B^2}{8\pi}  \end{array} \right),
\end{equation}
where $B^2={B_r}^2=B(r)^2$ (as assumed). Equation (\ref{pf4}) confirms that for RO,
the assumption of spherical symmetry is quite valid and the basic idea
proposed by~\cite{Bowers1974} can be implemented for magnetized
stars.\\

(ii) \textbf{Transverse orientation:}

For transverse orientation, the magnetic field takes the form 
\begin{eqnarray}\label{pf5}
\overrightarrow{B}=\left(0,B_\theta,B_\phi \right).
\end{eqnarray}
Now
\begin{eqnarray}\label{pf6}
& \overrightarrow{\nabla} \cdot \overrightarrow{B}=0 \Rightarrow
  \frac{1}{r \sin\theta}\frac{\partial}{\partial \theta}(B_\theta
  \sin\theta)+\frac{1}{r \sin\theta}\frac{\partial B_\phi}{\partial
    \phi}=0 \nonumber \\ & \Rightarrow B_\theta=
  \frac{\tilde{K}(r)}{\sin \theta},
\end{eqnarray}
where the system is axisymmetric, which leads to $\frac{\partial
  B_\phi}{\partial \phi}=0$.

Furthermore, we have $B^2 = \frac{{\tilde{K}(r)}^2}{\sin^2 \theta} +
{B_{\phi}}^2={B(r)}^2$ (as assumed).

Therefore, ${B_{\phi}}^2={B(r)}^2-\frac{{\tilde{K(r)}}^2}{\sin^2
  \theta}$; $\frac{\partial B_\phi}{\partial \phi}=0$ are satisfied.

This leads to
\begin{eqnarray}\label{pf7}
	\overrightarrow{B}=\left(0,\frac{{\tilde{K}(r)}}{\sin \theta}
        , \sqrt{{B(r)}^2-\frac{{\tilde{K}(r)}^2}{\sin^2
            \theta}}\right).
\end{eqnarray}

Now, similarly using Eq.~\eqref{pf3}, we have

$M^{rr}=\frac{B^2}{8\pi}$, $M^{\theta\theta} = \frac{B^2}{8\pi} - \frac{{B_\theta}^2}{4\pi}=\frac{{B_\phi}^2-{B_\theta}^2}{8\pi}$, $M^{\phi\phi} = \frac{B^2}{8\pi} - \frac{{B_\phi}^2}{4\pi}=-\frac{{B_\phi}^2-{B_\theta}^2}{8\pi}$, $M_{\theta\phi}=M_{\phi\theta}= -\frac{B_{\theta}B_{\phi}}{4\pi}$.

Finally, we have
\begin{equation}\label{pf8}
T^{\mu\nu}_f = \left(\begin{array}{cccc}    \frac{B^2}{8\pi}    &      0               &     0              &           0\\
                                               0             &  -\frac{B^2}{8\pi}   &     0              &           0\\
                                               0             &      0               &  \frac{{B_\phi}^2-{B_\theta}^2}{8\pi}  &           -\frac{B_{\theta}B_{\phi}}{4\pi}\\
                                               0             &      0               &     -\frac{B_{\theta}B_{\phi}}{4\pi}             &  -\frac{{B_\phi}^2-{B_\theta}^2}{8\pi} \end{array} \right).
\end{equation}

But if $\overrightarrow{B}$ is only along the $\theta$ direction (say), then one has
\begin{equation}\label{pf44}
T^{\mu\nu}_f = \left(\begin{array}{cccc}    \frac{B^2}{8\pi}    &      0               &     0              &           0\\
                                               0             &  \frac{B^2}{8\pi}   &     0              &           0\\
                                               0             &      0               &  -\frac{B^2}{8\pi}  &           0\\
                                               0             &      0               &     0              &  \frac{B^2}{8\pi}  \end{array} \right).
\end{equation}

If $\overrightarrow{B}$ is only along the $\phi$ direction, $T^{\theta\theta}_f$ and $T^{\phi\phi}_f$ in the above equation are interchanged. Hence, following~\cite{Chu2014}, the assumptions about the orientation of the  magnetic field for spherically symmetric, anisotropic
compact stars is consistent with Maxwell's equations.

%%%%%%%%%%%%%%%%%%%%%%%%%%%%%%%%%%%%%%%%%%%%%%%%%%%%%%%%%%%%%%%%%%%%%%%%%%%%%%%%%%%%%%%%%%%%%%%%%%%%%%%%%%%%%%%%%%%

\begin{figure*}[!htpb]
\centering
\includegraphics[width=0.33\textwidth]{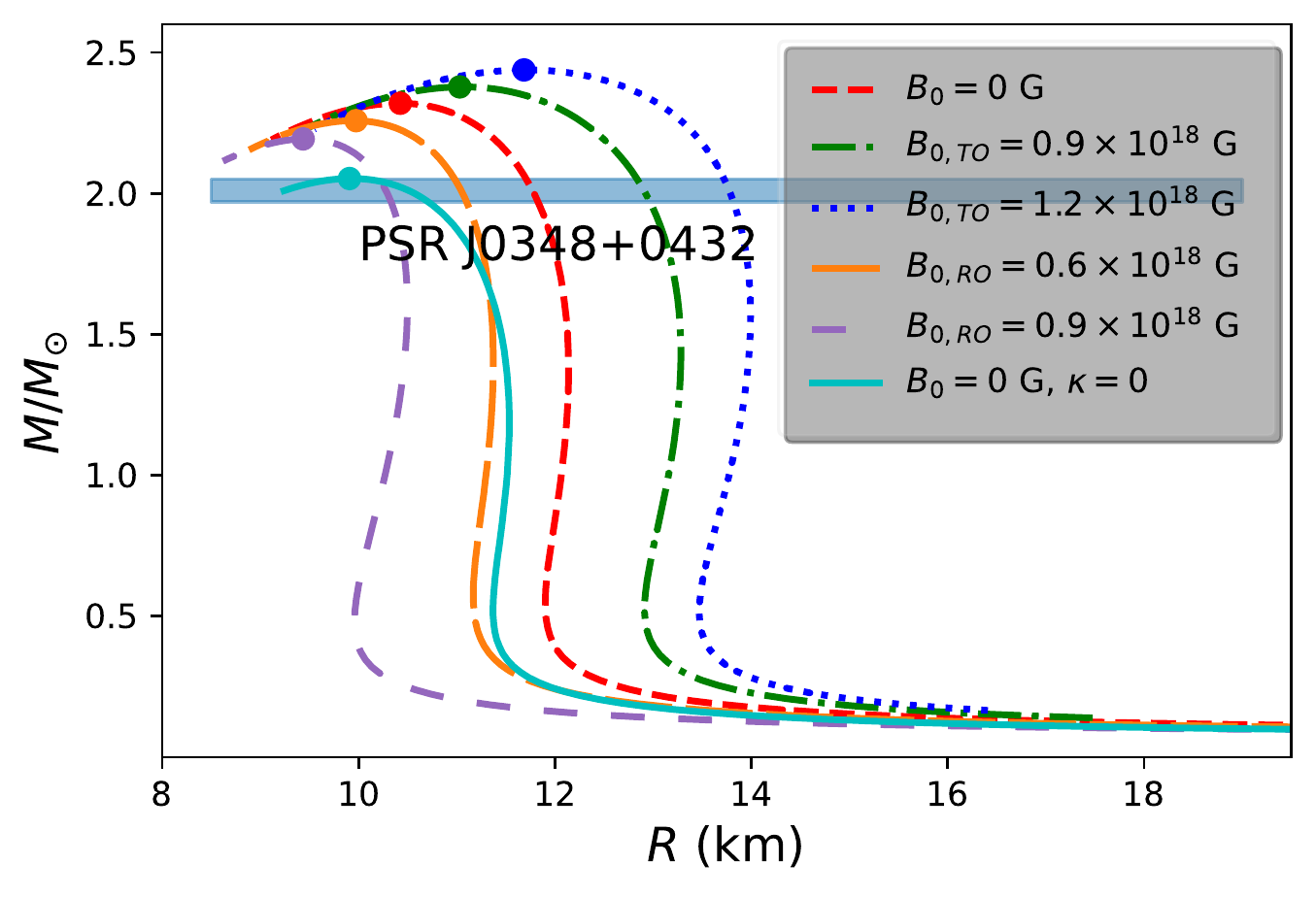} 
\includegraphics[width=0.33\textwidth]{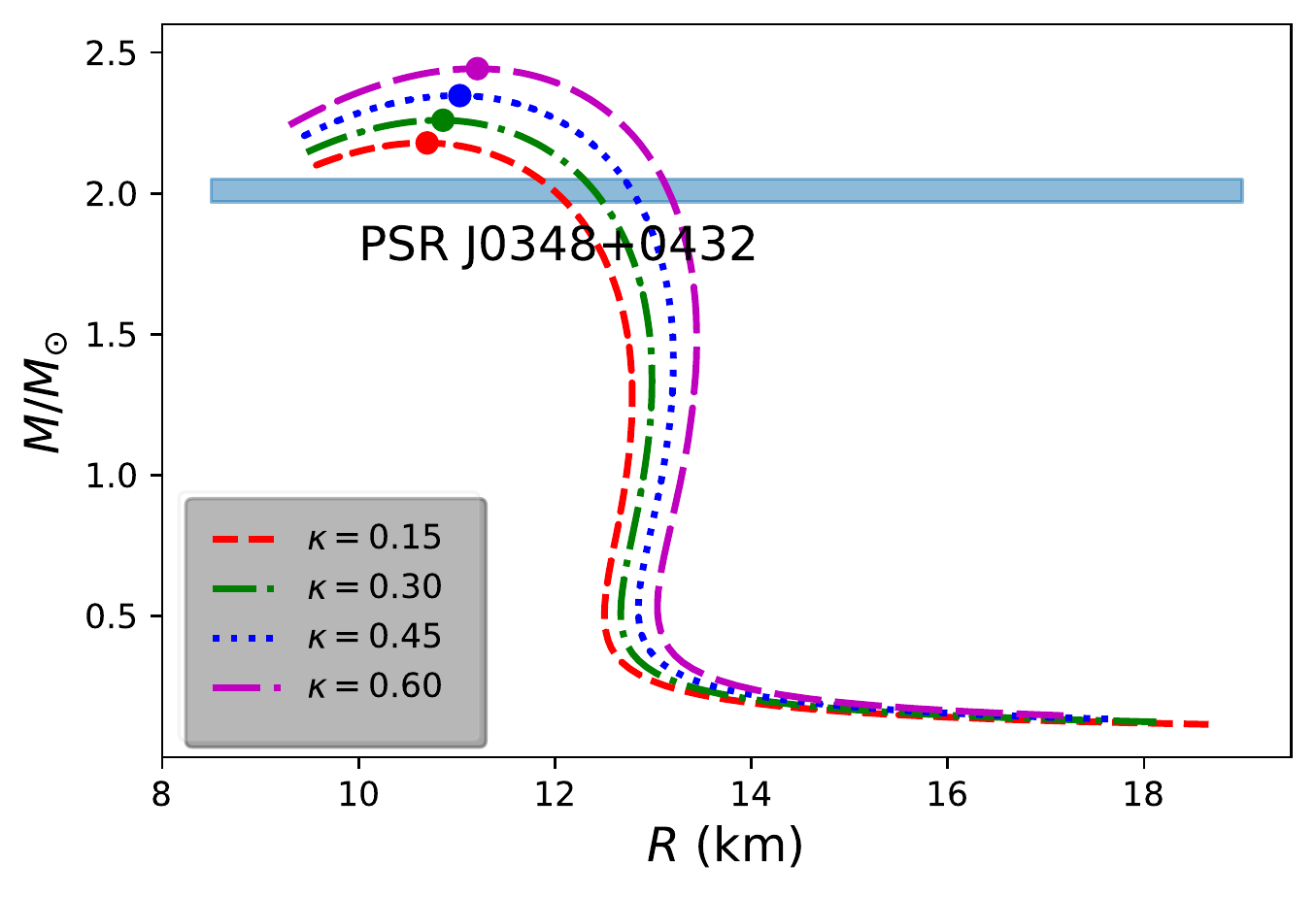} 
\includegraphics[width=0.33\textwidth]{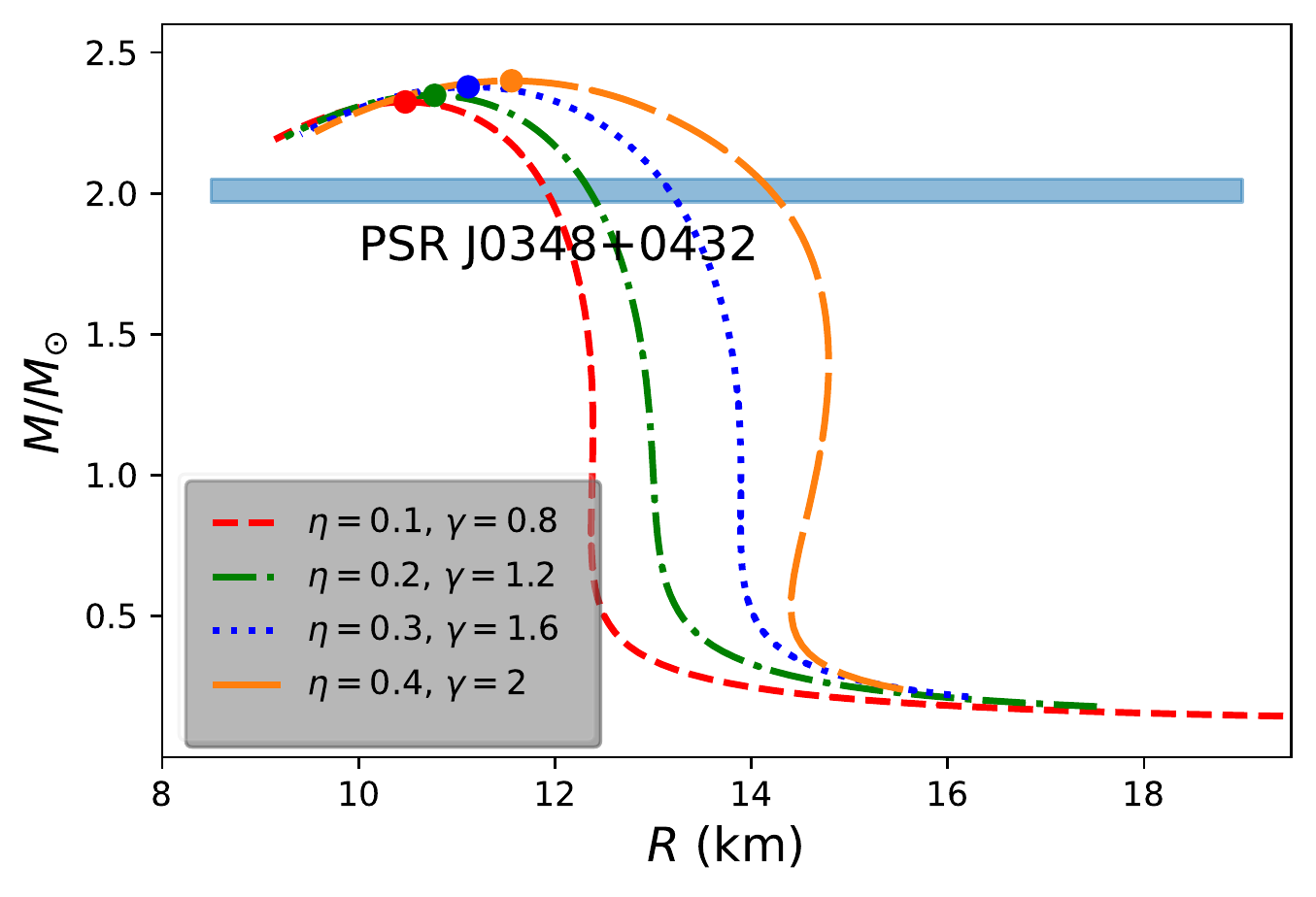} 
\includegraphics[width=0.33\textwidth]{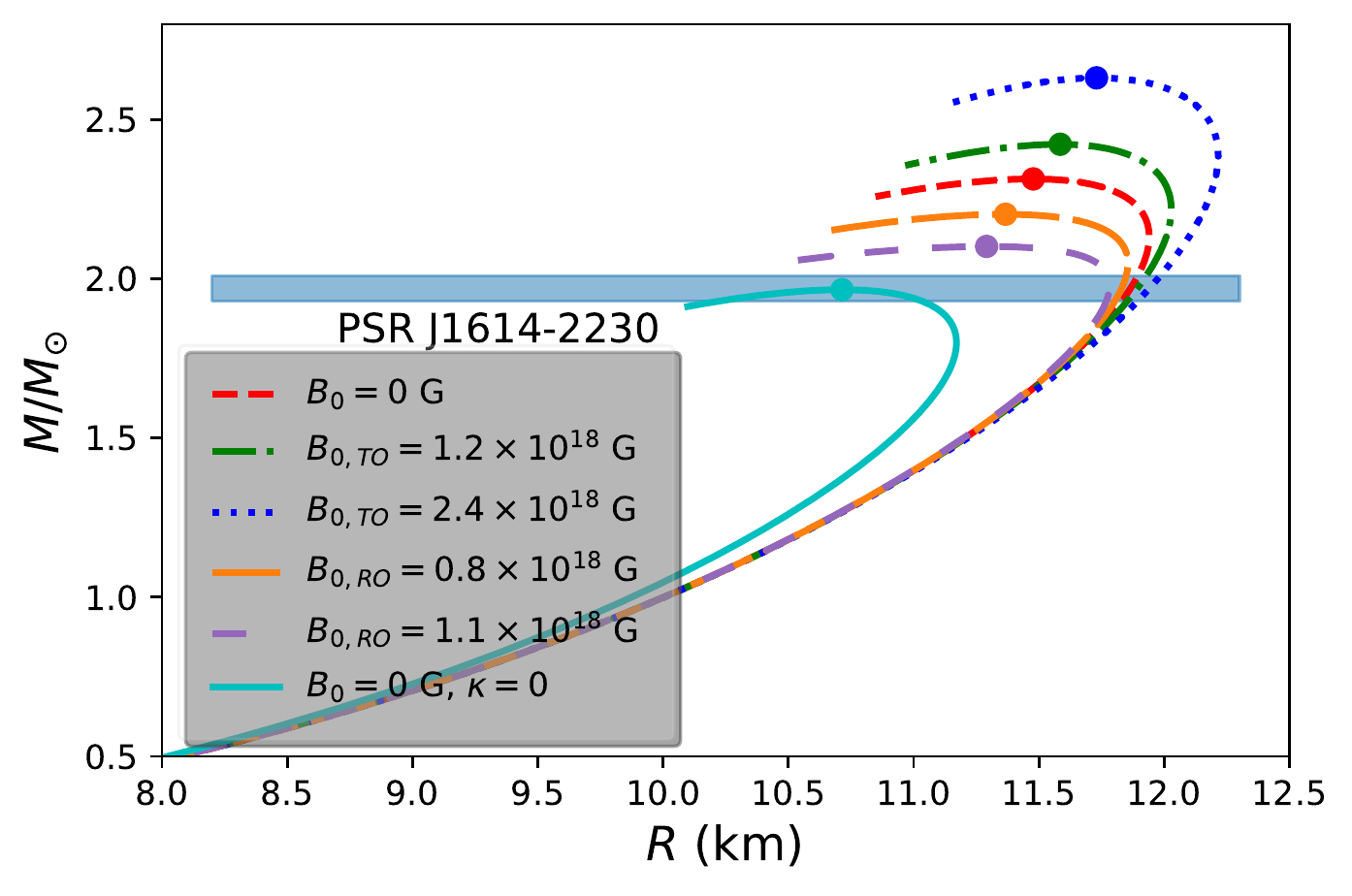} 
\includegraphics[width=0.33\textwidth]{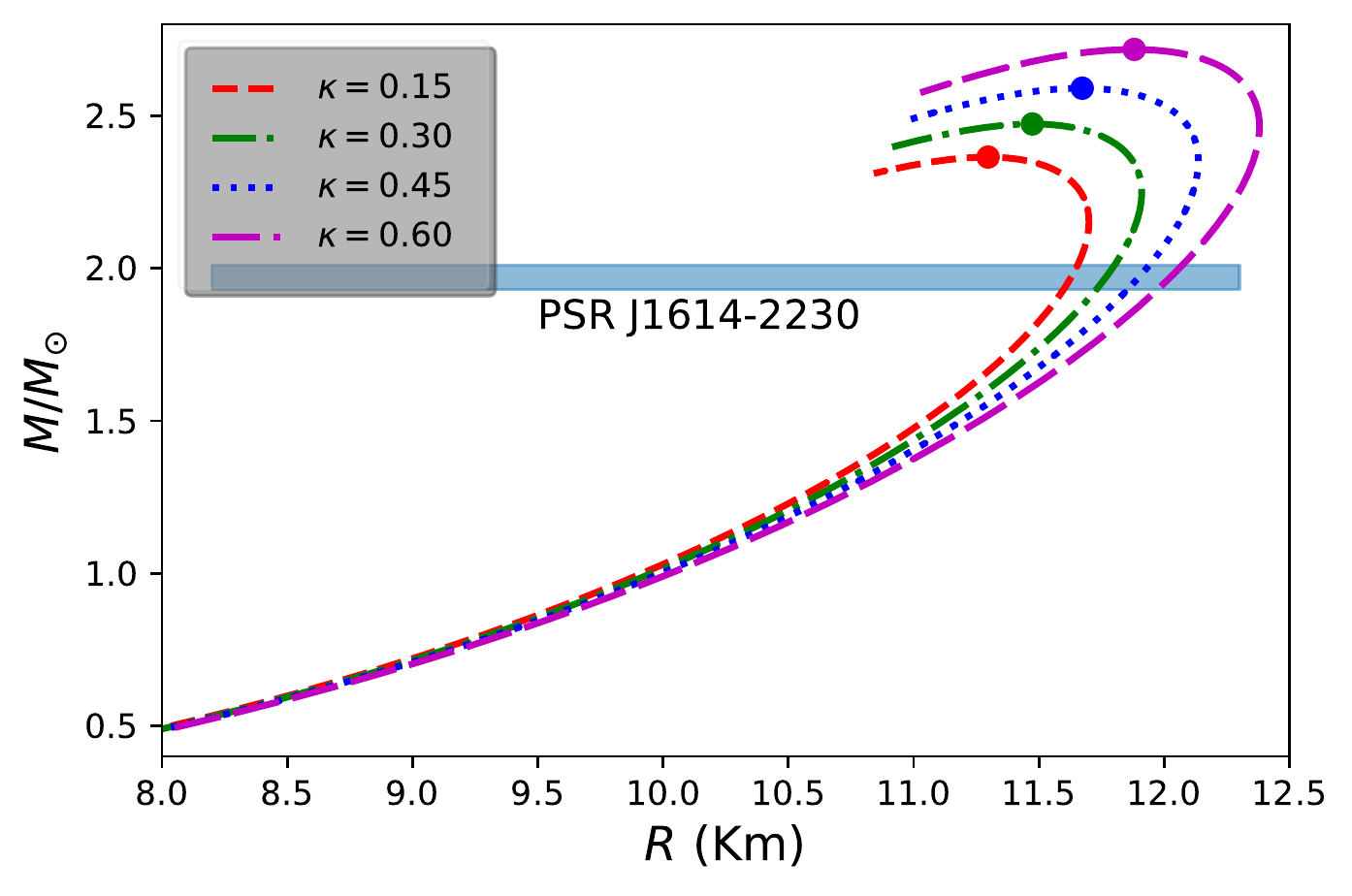} 
\includegraphics[width=0.33\textwidth]{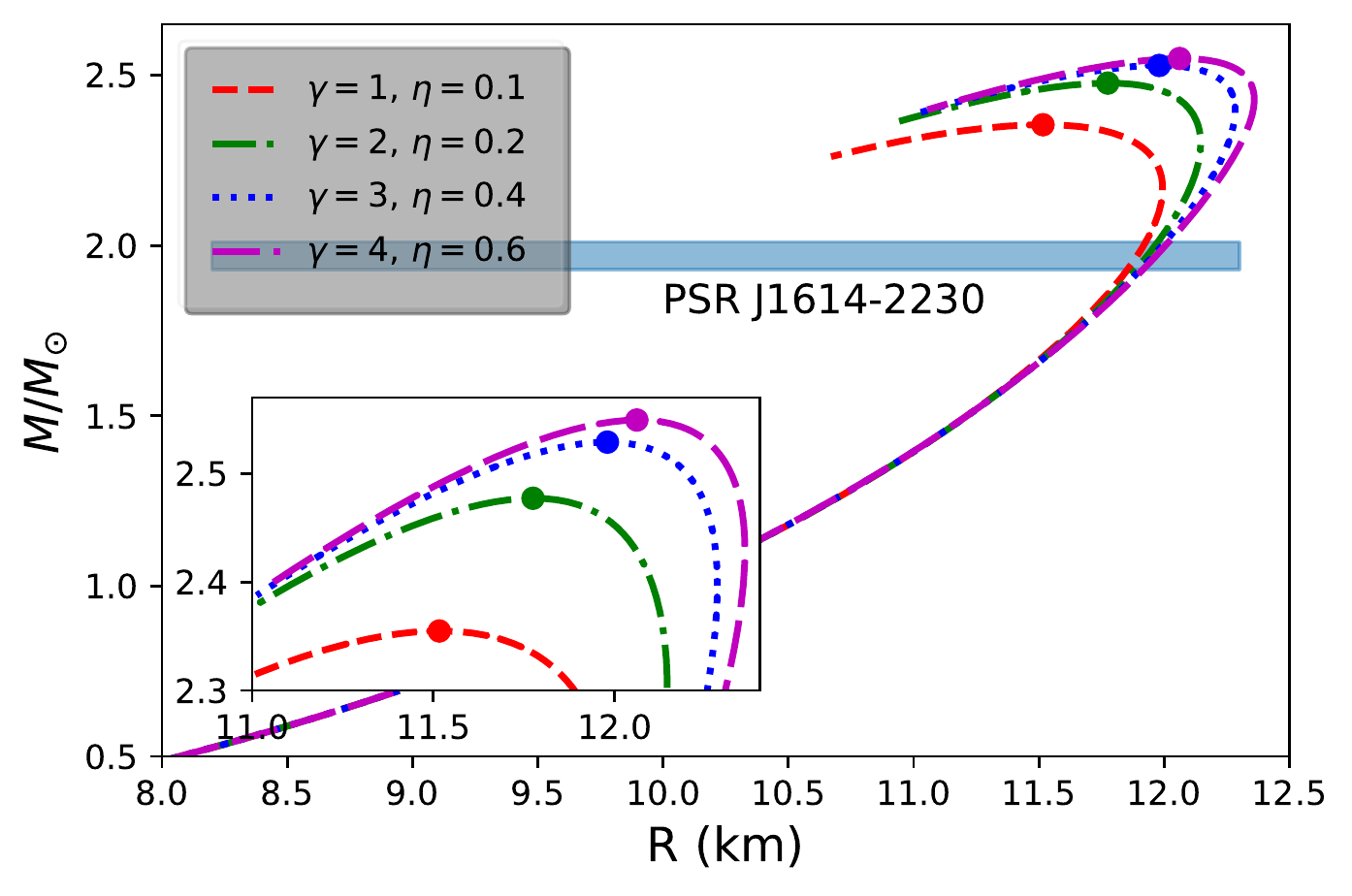}
\caption{Variation of stellar mass $M/M_\odot$ with
  stellar radius $R$. Solid circles represent the maximum-mass star of
  each stellar sequence. Here, upper panels feature NS $M/M_\odot$
  vs. $R$ for (i) varying $B_0$ and $\kappa=0.5$ (upper left), (ii)
  varying $\kappa$, where $B_0=6\times{10}^{17}$~G (upper middle) and
  (iii) varying $\eta$ and $\gamma$, where $B_0=6\times{10}^{17}$~G
  and $\kappa=0.5$ (upper right). However, lower panels feature SQS
  $M/M_\odot$ vs. $R$ for (i) varying $B_0$, where $\kappa=0.5$ (lower
  left), (ii) varying $\kappa$, where $B_0=2.4\times{10}^{18}$~G
  (lower middle) and (iii) varying $\eta$ and $\gamma$, where
  $B_0=1.2\times{10}^{18}$~G and $\kappa=0.5$ (lower
  right). }\label{MR}
\end{figure*}

%%%%%%%%%%%%%%%%%%%%%%%%%%%%%%%%%%%%%%%%%%%%%%%%%%%%%%%%%%%%%%%%%%%%%%%%%%%%%%%%%%%%%%%%%%%%%%%%%%%%%%%%%%%%%%%%%%%

\section{Results and discussions}\label{secII}
In the present article, we study compact stellar objects, viz., NSs
and SQSs with strong magnetic fields, assuming that they are
approximately spherically symmetric. Importantly, we show that not
only the magnetic field strength and anisotropy have a significant
effect on the stellar configurations, but that also the orientation of
the magnetic field (viz., RO or TO) has a pronounced, non-negligible
influence on the stars, too.

To describe NSs and SQSs we assume that their interiors can 
reasonably well be described by SLy EOS and the MIT bag model
EOS, respectively. 
Further, to close the system of equations, we assume a functional form for the
anisotropy that is shown in Eq.~\eqref{1.14}. Finally,
following~\cite{Bandyopadhyay1997a,Bandyopadhyay1998a} we consider
density dependent magnetic field strength profiles for NSs and SQSs
which are given by Eqs.~\eqref{1.15} and \eqref{1.16}. When presenting
the results of our study, we have chosen pulsars PSR J0348+0432 and
PSR J1614$-$2230 as reference stars. The observed masses of these
stars are $2.01 \pm 0.04~M_\odot$~\citep{Antoniadis2013} and $1.97 \pm
0.04~M_\odot$~\citep{Demorest2010}, respectively. To study NSs we have
chosen $B_s=10^{13}$~G and $B_0$ as $0.9\times{10}^{18}$~G and
$1.2\times{10}^{18}$~G for TO, and $6\times{10}^{17}$~G and
$9\times{10}^{17}$~G for RO. For the SQSs, we choose $B_s=10^{15}$~G
and $B_0$ as $1.2\times{10}^{17}$~G and $2.4\times{10}^{17}$~G for TO,
and $0.8\times{10}^{17}$~G and $1.1\times{10}^{17}$~G for RO. Note
that the choice of the values of $B_0$ is not same for TO and RO as
well as those between NS and SQS. This is because the effects on
stellar mass by the magnetic field are different between TO and RO
magnetic fields. For both types of stars we assume $\kappa=0.5$ as a
reference value. For SQSs we assume a bag constant value of
$\mathcal{B}=60~\mathrm{MeV/{fm^3}}$,  which describes SQM that is 
strongly bound with respect to nuclear matter.

%%%%%%%%%%%%%%%%%%%%%%%%%%%%%%%%%%%%%%%%%%%%%%%%%%%%%%%%%%%%%%%%%%%%%%%%%%%%%%%%%%%%%%%%%%%%%%%%%%%%%%%%%%%%%%%%%%%

\begin{figure}
\centering 
\includegraphics[width=0.45\textwidth]{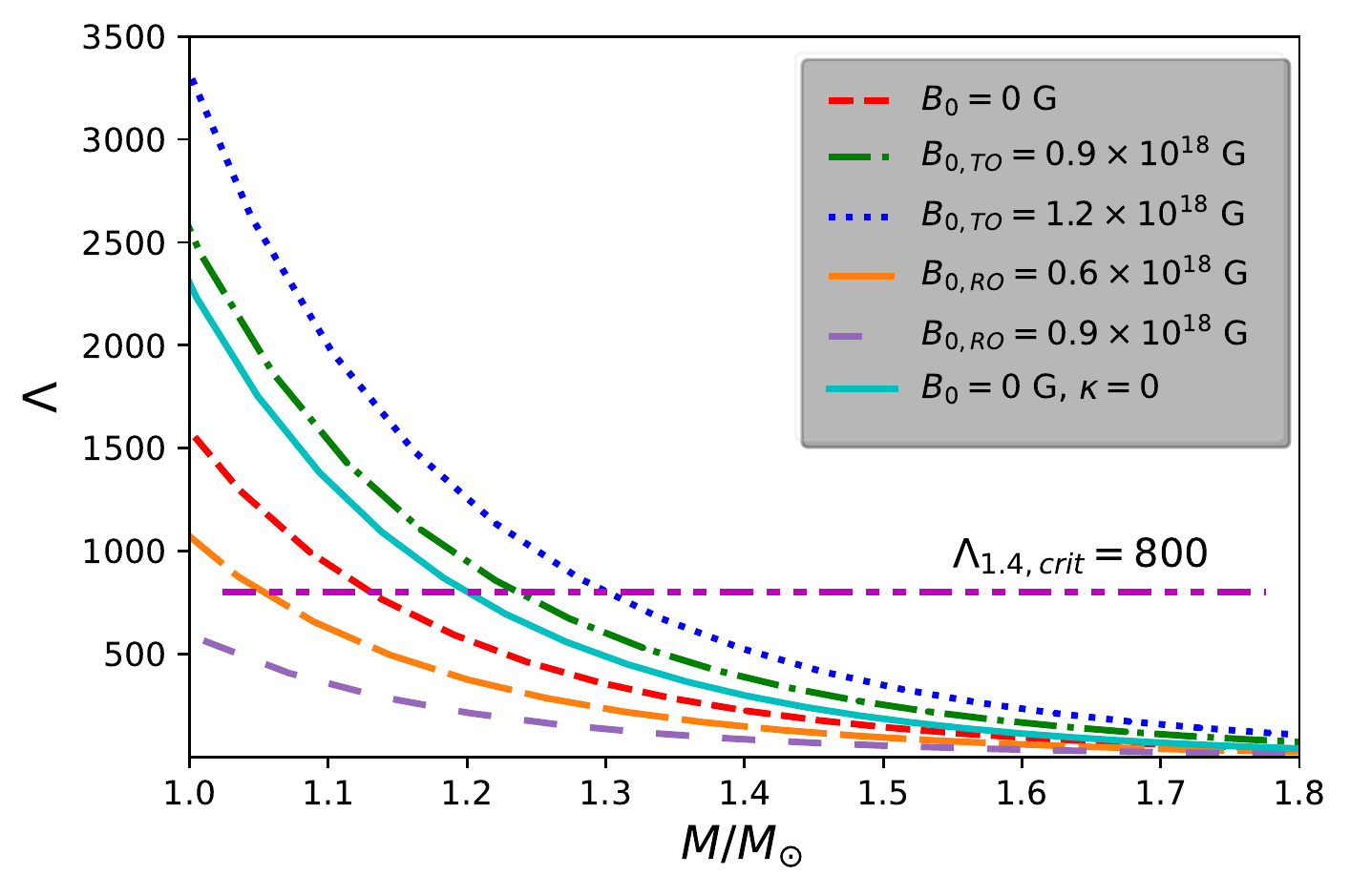} 
\includegraphics[width=0.45\textwidth]{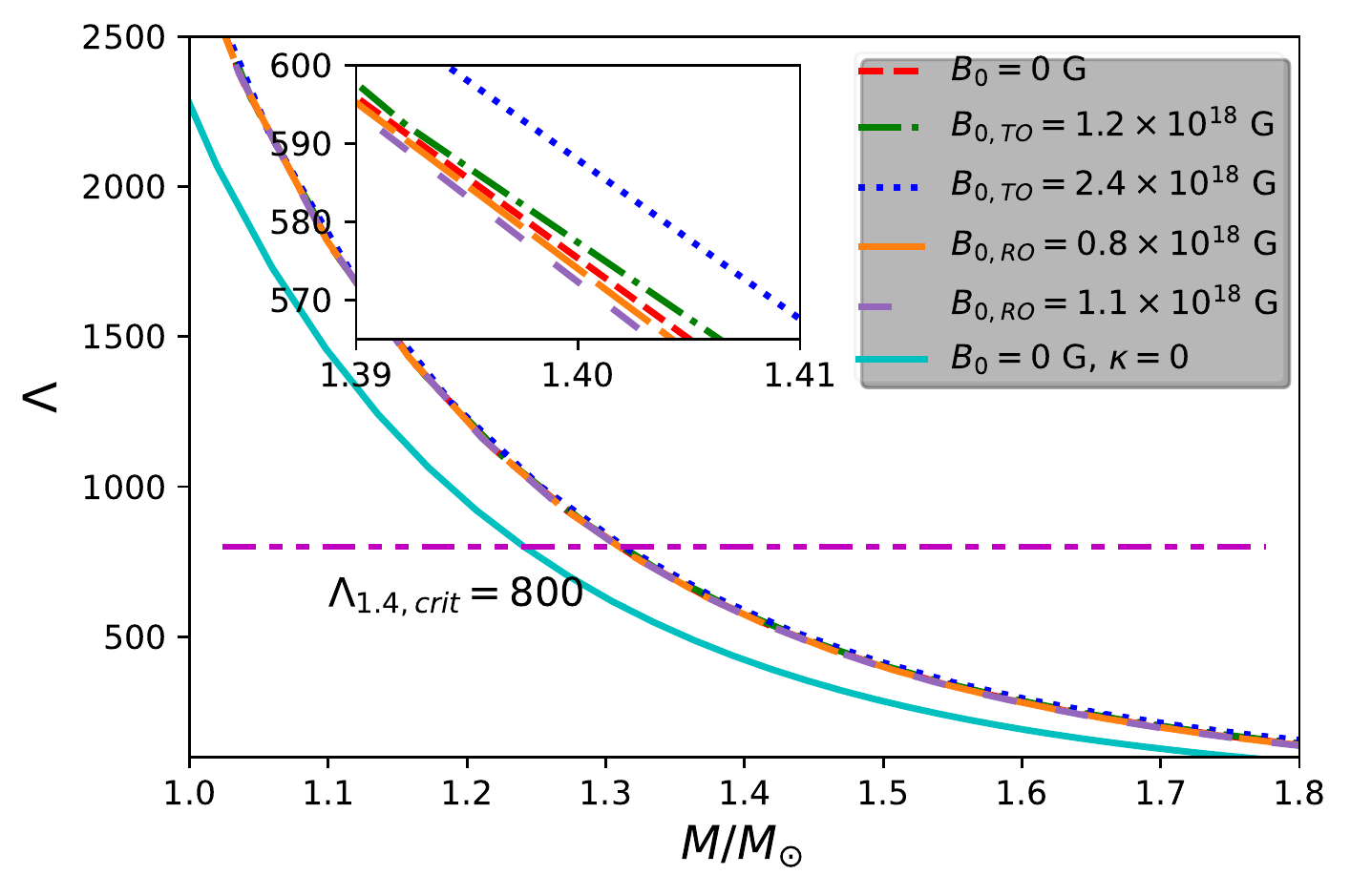} 
\caption{Variation of tidal deformability $\Lambda$ with respect to stellar mass $M/M_\odot$ 
	for various $B_0$ and $\kappa=0.5$.  In the upper and lower panels we present cases for NSs and SQSs respectively. } \label{tidal}
\end{figure}

%%%%%%%%%%%%%%%%%%%%%%%%%%%%%%%%%%%%%%%%%%%%%%%%%%%%%%%%%%%%%%%%%%%%%%%%%%%%%%%%%%%%%%%%%%%%%%%%%%%%%%%%%%%%%%%%%%%   

%%%%%%%%%%%%%%%%%%%%%%%%%%%%%%%%%%%%%%%%%%%%%%%%%%%%%%%%%%%%%%%%%%%%%%%%%%%%%%%%%%%%%%%%%%%%%%%%%%%%%%%%%%%%%%%%%%%

\begin{figure*}[t]
\centering 
\includegraphics[width=0.33\textwidth]{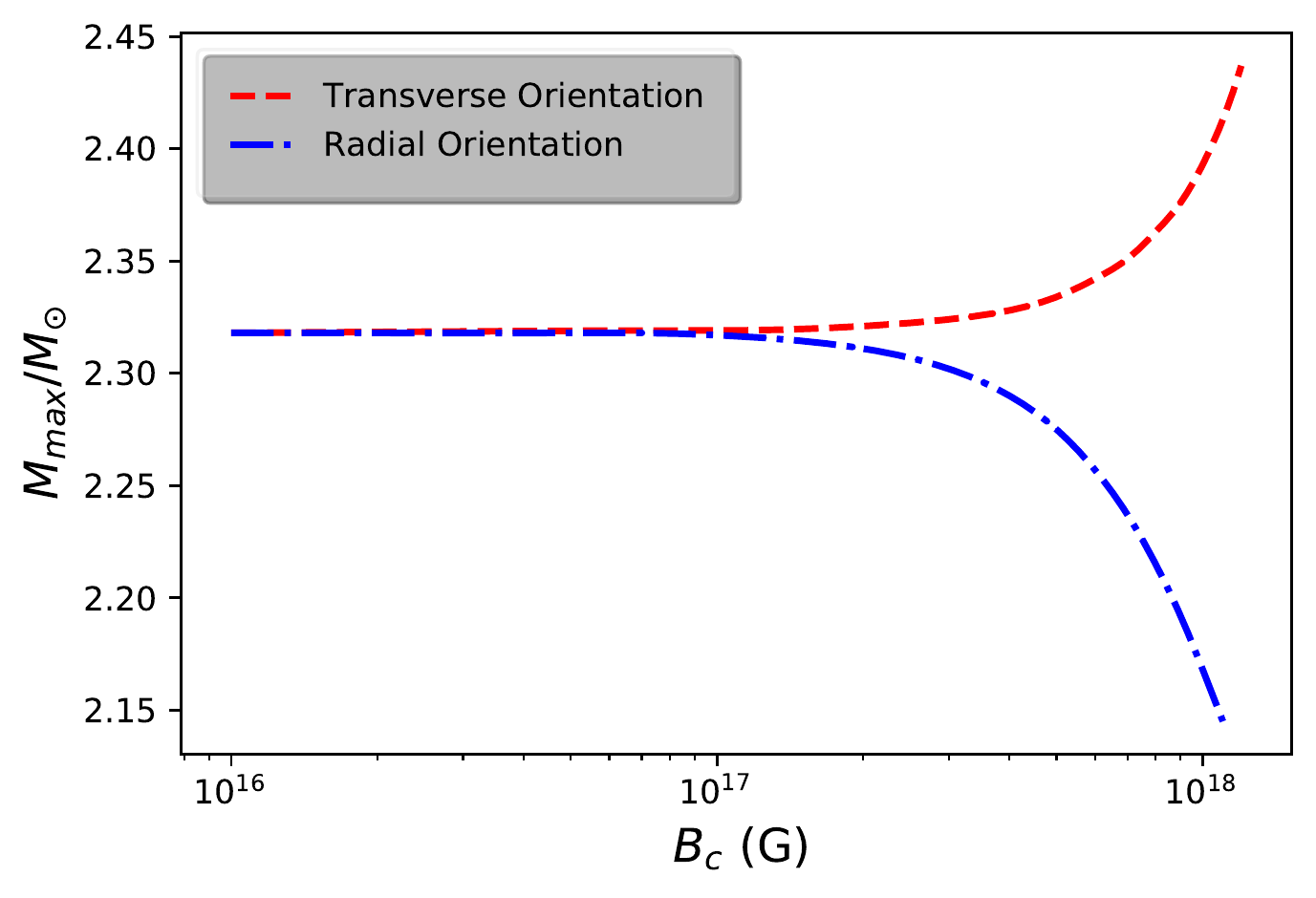} 
\includegraphics[width=0.33\textwidth]{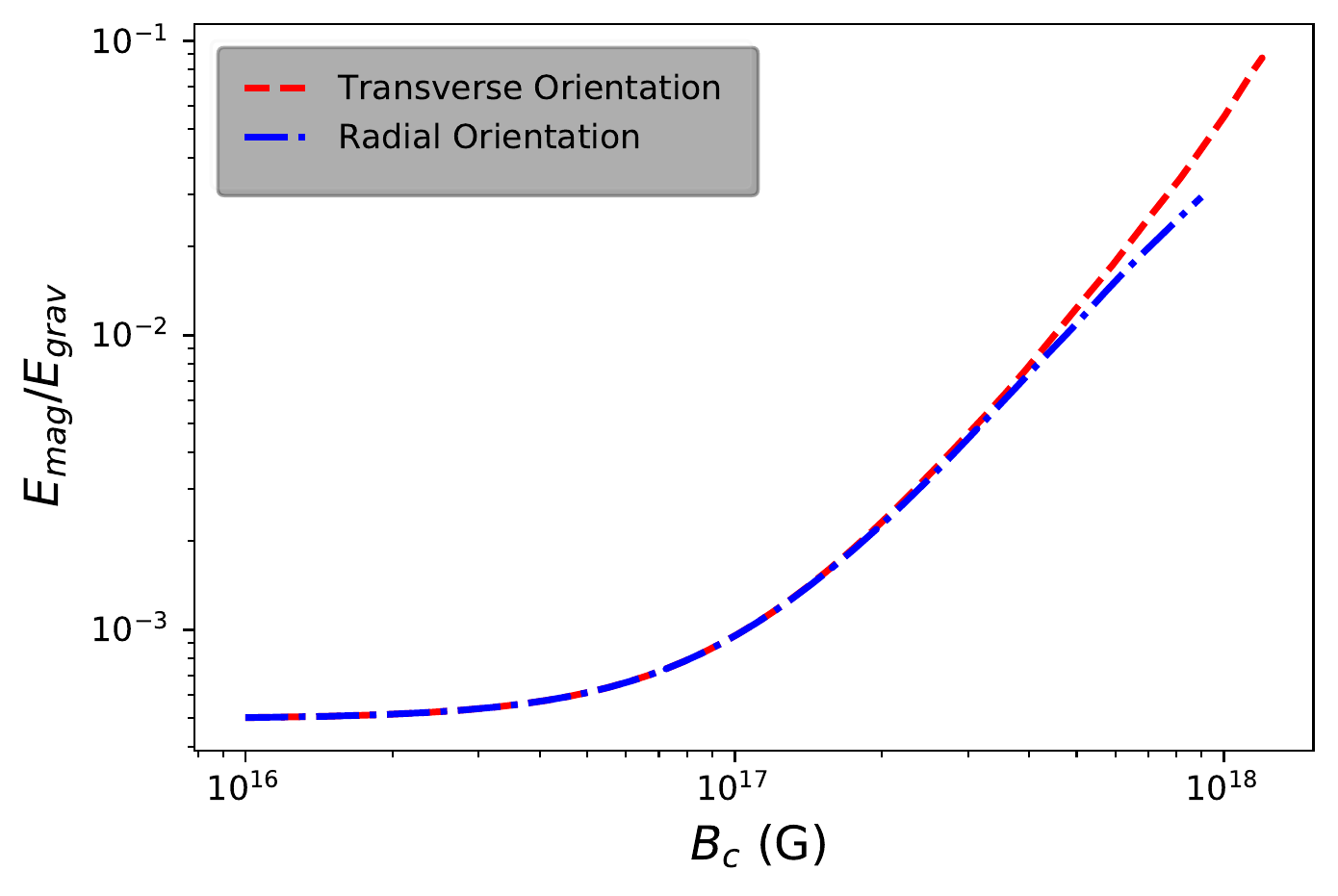} 
\includegraphics[width=0.33\textwidth]{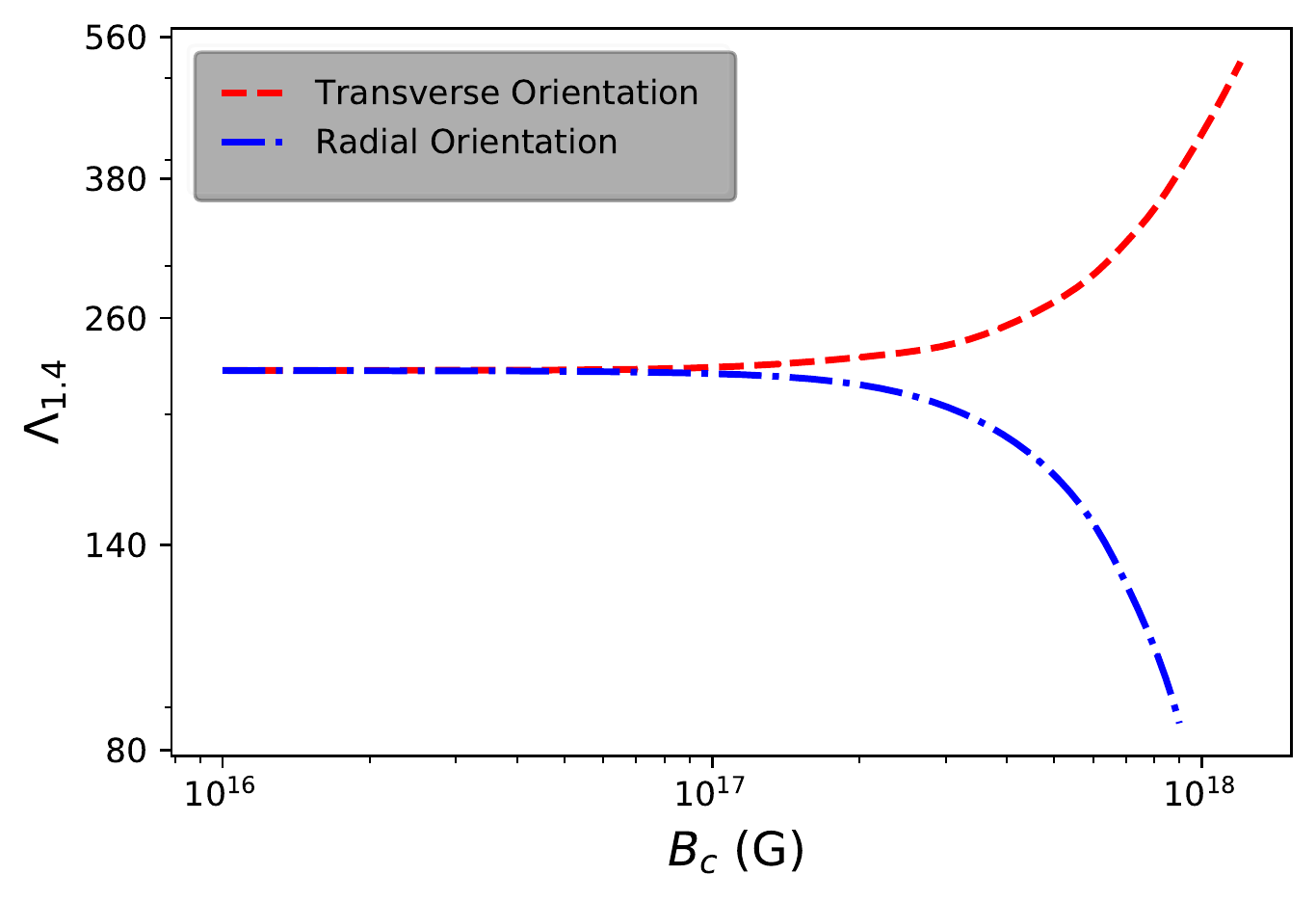} 
\includegraphics[width=0.33\textwidth]{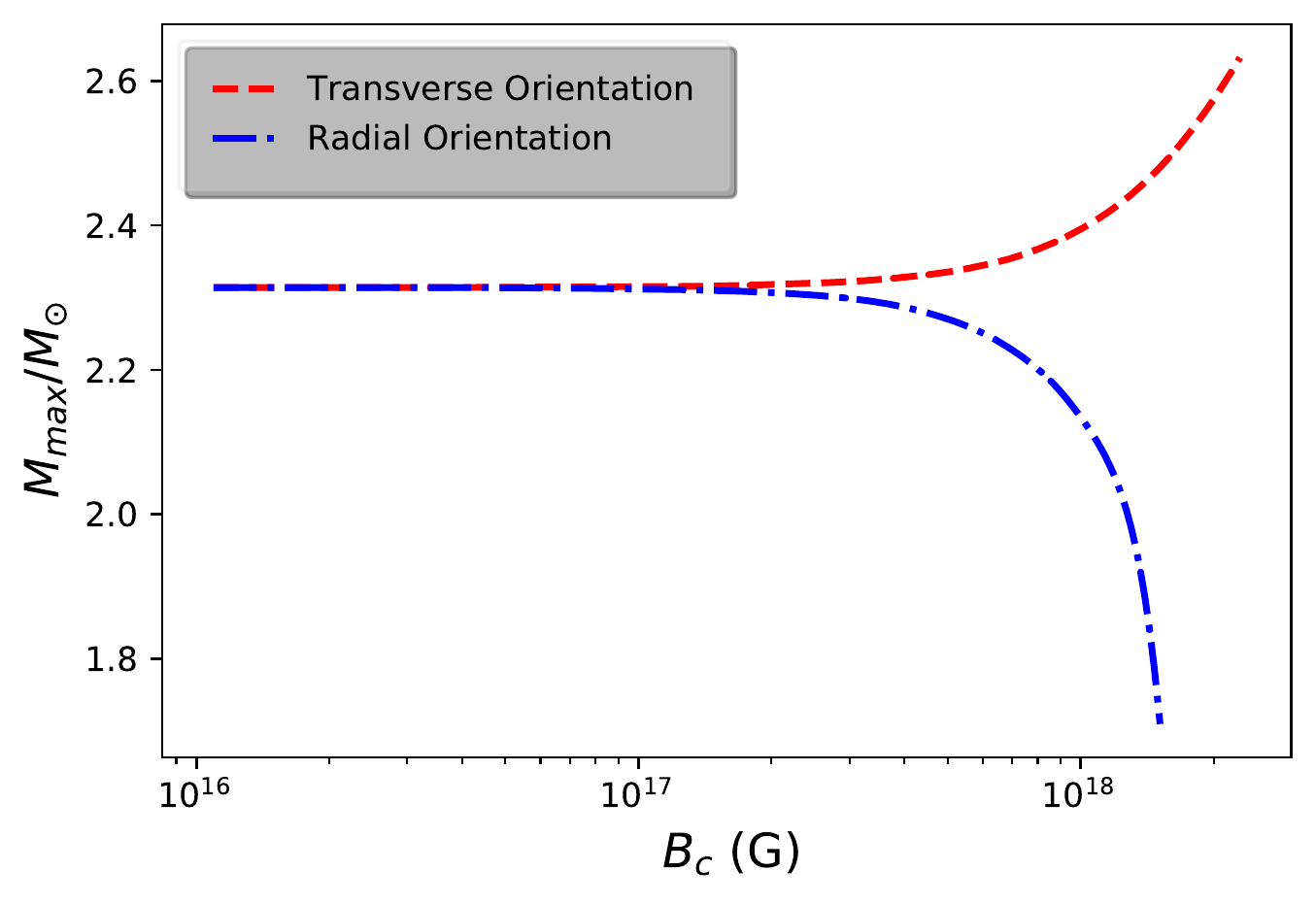}
\includegraphics[width=0.33\textwidth]{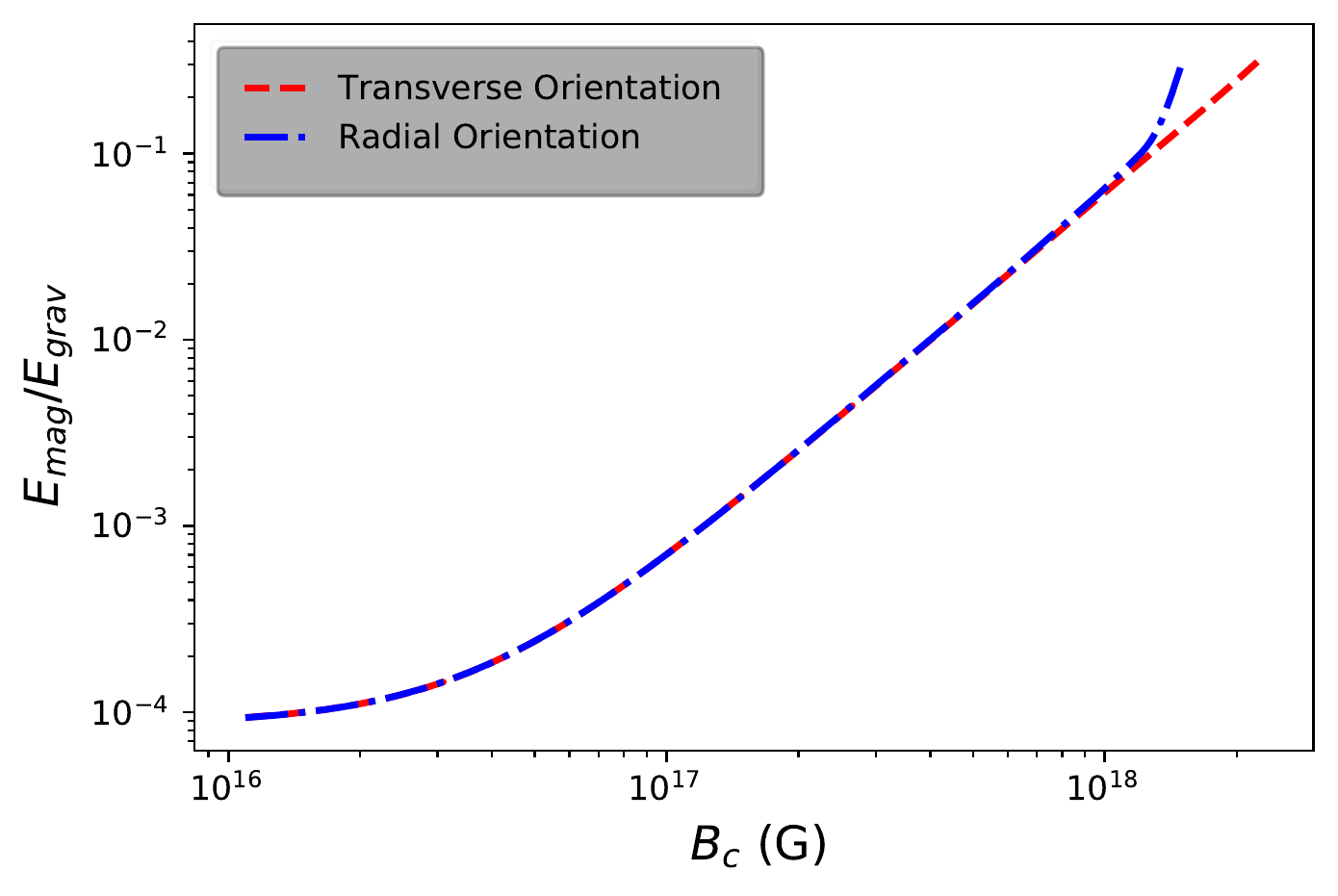} 
\includegraphics[width=0.33\textwidth]{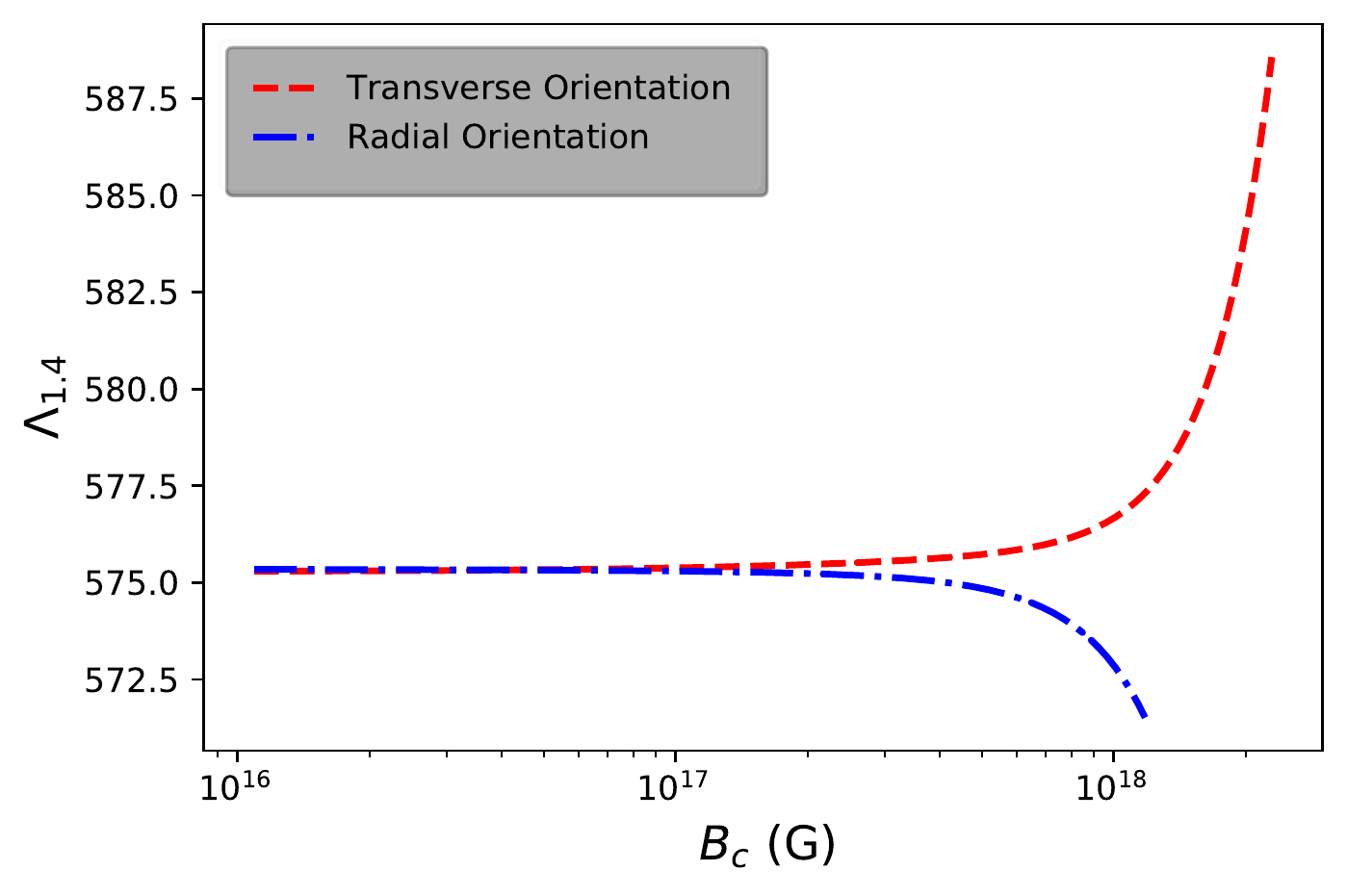} 
\caption{Variation of (i) stellar maximum mass $M_{max}$ ~(left panels), (ii) ratio of magnetic energy $E_{mag}$ to gravitational energy $E_{grav}$~(middle panels) and (iii) tidal deformability ($\Lambda_{1.4}$) for $1.4~M_\odot$ star~(right panels) with central magnetic field strength $B_c$.  In the upper and lower panels, we present cases for NSs and SQSs, respectively. } \label{massmag}
\end{figure*}

%%%%%%%%%%%%%%%%%%%%%%%%%%%%%%%%%%%%%%%%%%%%%%%%%%%%%%%%%%%%%%%%%%%%%%%%%%%%%%%%%%%%%%%%%%%%%%%%%%%%%%%%%%%%%%%%%%%

%%%%%%%%%%%%%%%%%%%%%%%%%%%%%%%%%%%%%%%%%%%%%%%%%%%%%%%%%%%%%%%%%%%%%%%%%%%%%%%%%%%%%%%%%%%%%%%%%%%%%%%%%%%%%%%%%%%

\begin{figure*}[!htpb]
\centering 
\includegraphics[width=0.45\textwidth]{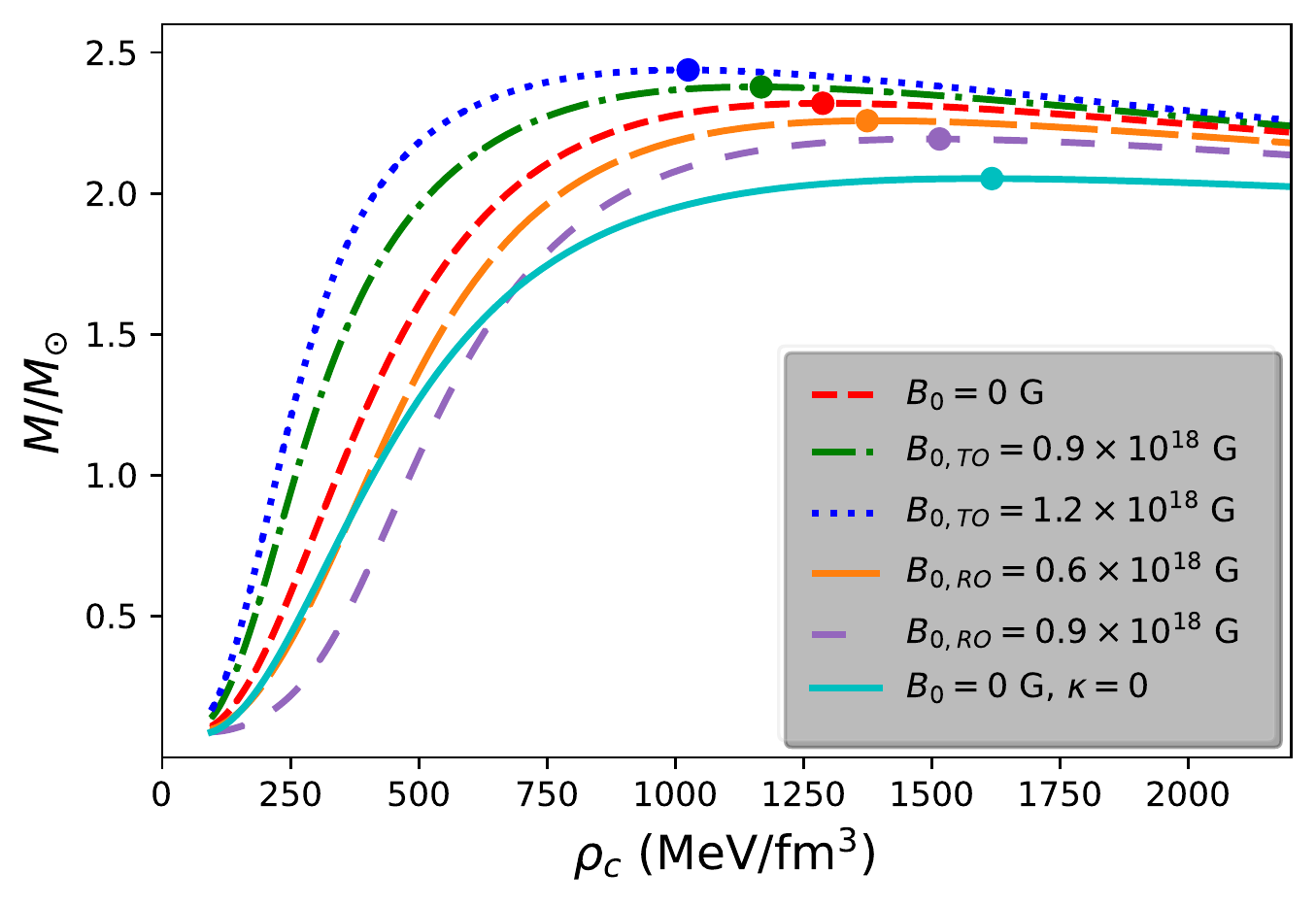} 
\includegraphics[width=0.45\textwidth]{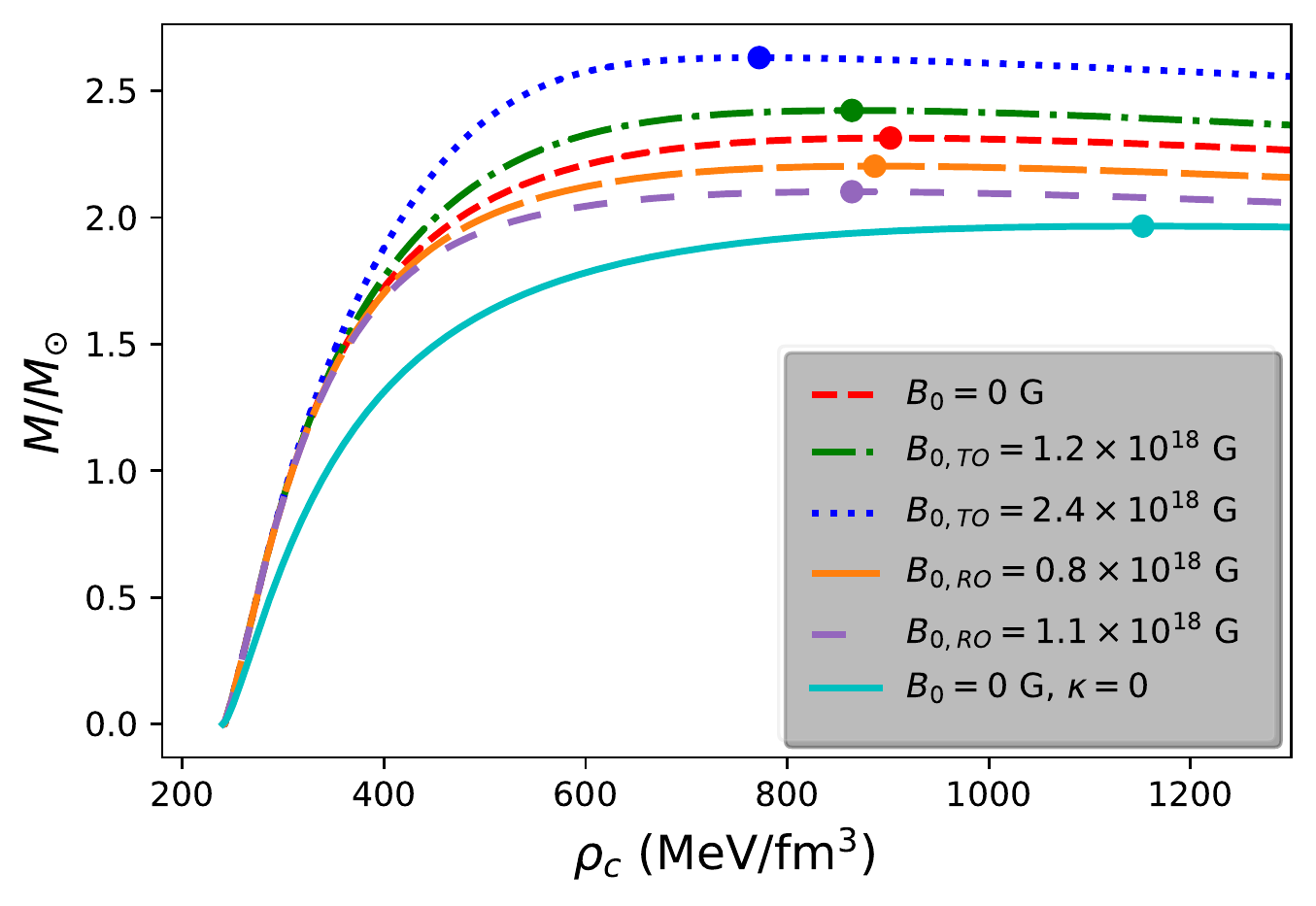}
\caption{$M/M_\odot$ as a function of the matter
  central matter density $\rho_c$ for NSs and SQSs in the left and
  right panels, respectively. Solid circles represent the maximum mass
  stars of each stellar sequence.} \label{masscden}
\end{figure*}

%%%%%%%%%%%%%%%%%%%%%%%%%%%%%%%%%%%%%%%%%%%%%%%%%%%%%%%%%%%%%%%%%%%%%%%%%%%%%%%%%%%%%%%%%%%%%%%%%%%%%%%%%%%%%%%%%%%

The profiles for the matter density $\rho$ from the center to the surface
with the normalized radial coordinate $r/R$, where $R$ is the stellar
radius, are shown for NS and SQS in the left and right upper panels of
Figure~\ref{pressure}, respectively,  for different parametric
values of $B_0$. The corresponding profiles for the radial matter
pressure $p_r$ are featured in the left and right middle panels of
Figure~\ref{pressure} for NSs and SQSs, respectively. Similarly, the
variations of the tangential matter pressure $p_t$ is shown in the
left and right bottom panels of Figure~\ref{pressure}.  Clearly,
Figure~\ref{pressure} shows that $\rho$, $p_r$ and $p_t$ have finite
maximum values at the core of the stellar system and then decrease
gradually to attain their respective minimum values at the surface,
which ensures physical stability of the achieved
solutions. Figure~\ref{pressure} also confirms that the present model
is free from any sort of singularities at the core of the
system. Figure~\ref{EOS} shows the effects of the magnetic field on
the EOSs of the compact stellar objects. In the upper and lower panels
of Figure~\ref{EOS} we present the variations of the system parallel
pressure $p_\parallel$ and transverse pressure $p_\bot$ with the
system density~$(\widetilde{\rho})$, normalized to the system central
density~$(\widetilde{\rho}_c)$, for NSs and SQSs,
respectively. Interestingly, our study reveals that as the magnetic
field strength increases, the system pressures of the compact stars
gradually become stiffer for RO, whereas they gradually become softer
for TO, which is evident in Figure~\ref{EOS}. Importantly, at the
center for each of the cases, $p_\parallel$ and $p_\bot$ have the same
value which ensures zero anisotropy at the stellar core. The variation
of anisotropy, due to both the local anisotropy of the fluid and the
presence of a strong magnetic field, is shown in
Figure~\ref{anisotropy}. Importantly, in the upper and lower panels of
Figure~\ref{anisotropy} one sees that the anisotropy at the center of
both NSs and SQSs is zero, which ensures hydrodynamic stability of the system via the balance of the forces. It is worth mentioning that as
long as the anisotropy is considered only due to the presence of a
strong magnetic field, the anisotropic force shows attractive nature
if the field is in TO, whereas the same force is repulsive for RO
fields. Hence, within the stars, the slopes of the anisotropy profiles
gradually increase for TO as $B_0$ increases, whereas they gradually
decrease with increasing $B_0$ for RO magnetic
fields. Furthermore, the profiles for the density dependent magnetic fields inside NSs and SQSs are featured in the upper and lower panels of Figure.~\ref{mag_field}, respectively,  which show that the magnetic field is maximum at the core of these stars and decreases
monotonically throughout their interiors to reach their minimum values
at the surface.

To shed light on the widely unknown hadronic EOS in the high-density
regime it is important to study the mass-radius relationship of
compact stars, which allows one to rule out or support existing models
of hadronic EOSs. In the present work, the study of the mass-radius
relation is used to analyse the effects of strong magnetic fields,
their orientations and anisotropy on compact stellar configurations
and to control their properties. In the upper and lower left panels of
Figure~\ref{MR}, we show mass-radius relations for NSs and SQSs,
respectively, for a range of different $B_0$ values. The mass-radius
relations for NSs and SQSs due to varying $\kappa$ values are shown in the upper and lower middle panels of Figure.~\ref{MR},  respectively. The upper and lower right panels of Figure~\ref{MR} show
the mass-radius relations due to different parametric choices of
$\eta$ and $\gamma$ for NSs and SQSs, respectively.  From the upper
left panel of Figure~\ref{MR} one sees that for TO magnetic fields and
$B_0=1.2\times{10}^{18}$~G the maximum mass and corresponding radius of the NSs increase by 5.09\% and 12.09\%,  respectively, compared to the anisotropic but non-magnetized case. 
The maximum mass and associated radius increase to 18.75\% and 17.95\%,  respectively,  in comparison to the isotropic
non-magnetized case.

However, for the RO case with $B_0=0.9\times{10}^{18}$~G the maximum
mass and associated radius of NSs decrease by 5.47\% and 9.49\%,
respectively, compared to the anisotropic but non-magnetized case. But
compared to their values in the isotropic non-magnetized case, the
maximum mass increases by 6.82\% while the corresponding radius
decreases by 4.75\%. the upper middle, upper right, lower middle
and lower right panels of Figure~\ref{MR} show that the maximum mass
and its corresponding radius increase gradually for both NSs and SQSs
when the values of $\kappa$, $\eta$ and $\gamma$ gradually
increase. The lower left panel of Figure~\ref{MR} shows that for
$B_0=2.4\times{10}^{18}$~G and the TO field, the maximum mass and the
corresponding radius increase by 13.74\% and 2.20\%, respectively,
compared to anisotropic but non-magnetized SQSs. On the other hand,
these values increase by 8.31\% and 9.48\%, respectively, compared
to the isotropic and non-magnetized case. For the RO case, however,
for $B_0=1.1\times{10}^{18}$~G the maximum mass of SQSs and the
corresponding radius decrease by 9.16\% and 1.62\%, respectively,
compared to the anisotropic but non-magnetized case. With respect to
the isotropic and non-magnetized SQSs, these values decrease by
6.92\% and 5.39\%, respectively.  

In Figure~\ref{tidal} we present the compatibility of our model with respect to 
the tidal deformability ($\Lambda$) for both NSs and SQSs, constrained from the observation of GW emission related to GW170817 event, detected by the LIGO/Virgo Collaboration (LVC)~\citep{Abbott2017,Abbott2018,Abbott2019}. The investigation by LVC sets an upper limit of $\Lambda$ associated with $1.4M_\odot$ pulsars 
($\Lambda_{1.4}$) by~\cite{Abbott2017} which is given as $\Lambda_{1.4}<800$ for the low-spin cases. 
In the upper and lower panels of Figure~\ref{tidal}, we show the variation of $\Lambda$ with 
respect to $M$ for both NSs and SQSs, respectively, for various $B_0$ and $\kappa=0.5$.  Evidently, 
for both the cases, $\Lambda_{1.4}$ lies well below its critical value, which confirms the physical validity 
of the assumed EOSs, viz., SLy and MIT bag model EOSs for NSs and SQSs, respectively.  
In Figure~\ref{tidal}, one may also notice that $\Lambda$ increases with increasing $B_0$ for TO, 
whereas it decreases with increasing $B_0$ for RO.

In Figure~\ref{massmag} we present the effect of magnetic field orientation on the physical properties of both the stars, such as $M_{max}$, the ratio of magnetic to gravitational energies and tidal deformability for $1.4~M_\odot$ stars.  Although the masses of NSs and SQSs increase or decrease
  (based on the orientation of the magnetic field) compared to their
  values in the anisotropic and non-magnetized cases, they change
  asymmetrically, which is shown in the upper left and lower left
  panels of Figure~\ref{massmag}.  From the upper left panel of Figure~\ref{massmag}, one sees
  that for $B_0={10}^{18}$~G and $\kappa=0.5$ the maximum masses of
  NSs are $2.39~M_\odot$ for the TO case and $2.17~M_\odot$ for the RO
  case, which leads to a 10.38\% asymmetry in the masses. Similarly,
  the lower left panel of Figure~\ref{massmag} reveals that for
  $B_0=1.5\times{10}^{18}$~G and $\kappa=0.5$, the maximum masses of
  SQSs are $2.48~M_\odot$ and $1.73~M_\odot$ for TO and RO cases,
  respectively, which leads to $42.91\%$ asymmetry in the masses.
  Evidently, for both the stars, as the central magnetic field $B_c$ increases, the
  effects of magnetic field orientations via mass-asymmetry become
  larger gradually, which corroborate the conclusion of~\cite{Chu2014}
  that orientations of the magnetic field have a significant effect on
  the maximum mass of magnetized compact stars. Further, the left upper and lower panels of Figure~\ref{massmag} exhibit that for $B_c<10^{17}$~G, the anisotropic compact stars are not sensitive to the present magnetic field strength and their orientations within the stars.  Again, note that \cite{Sinha2013} showed for the magnetic field strength less than $3 \times 10^{18}$~G, the effects of Landau quantization are not considerable within the magnetized compact stars.  We therefore choose to constrain $B_c$ in our work as $10^{17}$~G$<B_c<3 \times 10^{18}$~G. In the right upper and lower panels of Figure~\ref{massmag}, we also show the effect of magnetic field orientation on $\Lambda_{1.4}$,
  which increases with increasing $B_c$ for the TO case and decreases with
  increasing $B_c$ for the RO case.  \cite{Chu2021} found the
  same dependency of $\Lambda_{1.4}$ on the magnetic field
  orientations which confirms our results in the case of anisotropic
  magnetized compact stars. Hence, through this work, we explore that
  for anisotropic magnetized stars, anisotropy, magnetic field
  strength and orientations of the magnetic field have a significant
  effect on $\Lambda_{1.4}$. Hence, RO and TO cases of the magnetic field play a significant role in magnetized stellar configurations. 
  
\cite{Chandrasekhar1953} found that in the case of magnetized relativistic stars, instead of $\Gamma>\frac{4}{3}$ the system may be dynamically unstable due to a sufficiently strong internal magnetic field, which may induce dynamical instability in compact stars. They found that in magnetized stars, the necessary condition to achieve a stellar equilibrium configuration is $\mid E_{grav}\mid > E_{mag}$, where $E_{grav}$ and $E_{mag}$ are the gravitational potential energy and magnetic energy, respectively. In the middle upper and lower 
panels of Figure~\ref{massmag} we show that for both NS and SQS, respectively, $E_{grav}$ dominates significantly over $E_{mag}$ for both orientations of the magnetic field, which confirms the dynamic stability of these magnetized compact stellar objects. Further to discuss the stability of a spherically symmetric static stellar structure, the model must be consistent with the condition $\mathrm{d}M/\mathrm{d}\rho_c>0$, say, up to the maximum mass~\citep{Harrison1965}. From the left and right panels of Figure~\ref{masscden}, it is evident that both NSs and SQSs fulfill this stability criterion.  

\begin{figure}
\centering 
\includegraphics[width=0.45\textwidth]{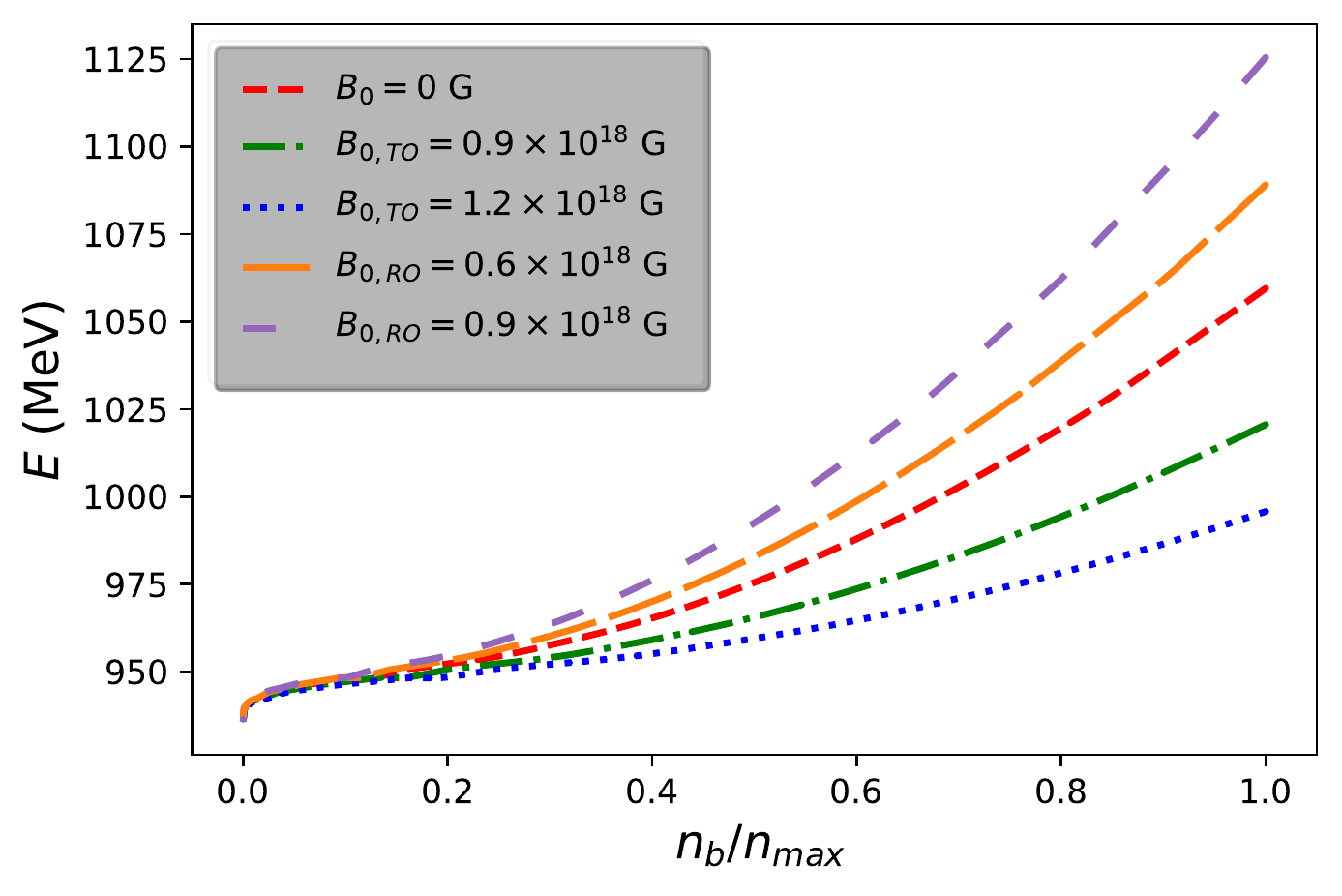} 
\includegraphics[width=0.45\textwidth]{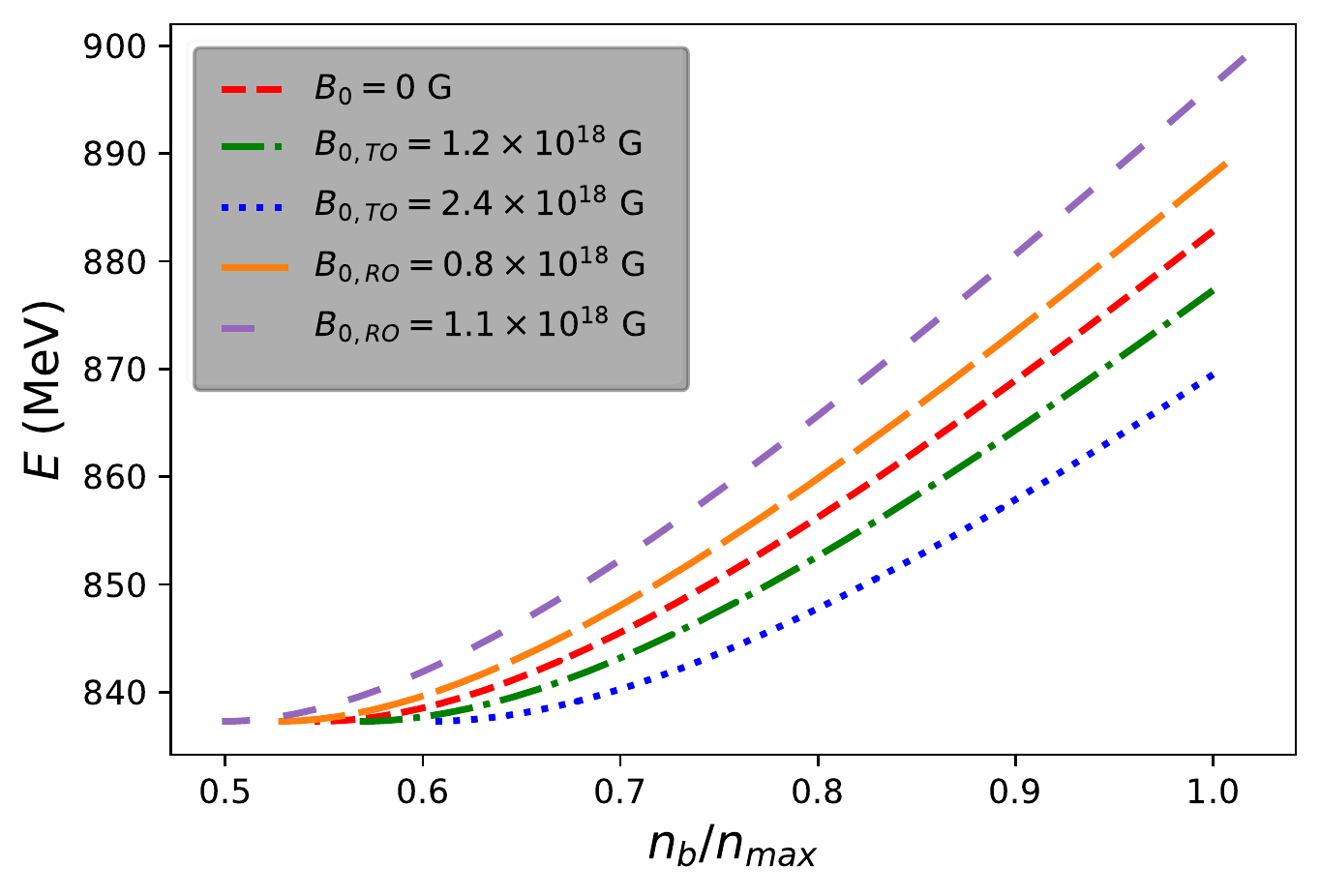} 
\caption{Variation of energy per baryon ($E$) with the ratio of baryon number density ($n_b$) to the maximum value of $n_b$ ($n_{max}$) for
various $B_0$ and $\kappa=0.5$.  In the upper and lower panels we present cases for NSs and SQSs, respectively. } \label{energy}
\end{figure}

Further, we check the absolutely stable condition for both the EOSs of NSs and SQSs.  
We find the minimum energy per baryon for SQM is less than 930 MeV for the chosen values of 
$B_0$ in the case of MIT bag model EOS.  On the other hand,  the minimum energy per baryon for NS matter described by SLy EOS is greater than 930 MeV for the chosen values of $B_0$.  In Figure~\ref{energy}, we show the variation of the energy per baryon with the ratio of baryon number density ($n_b$) to its maximum value ($n_{max}$), which also confirms that SQM may be the true ground state for strong interactions. 

\begin{figure}[!htpb]
\centering 
\includegraphics[width=0.45\textwidth]{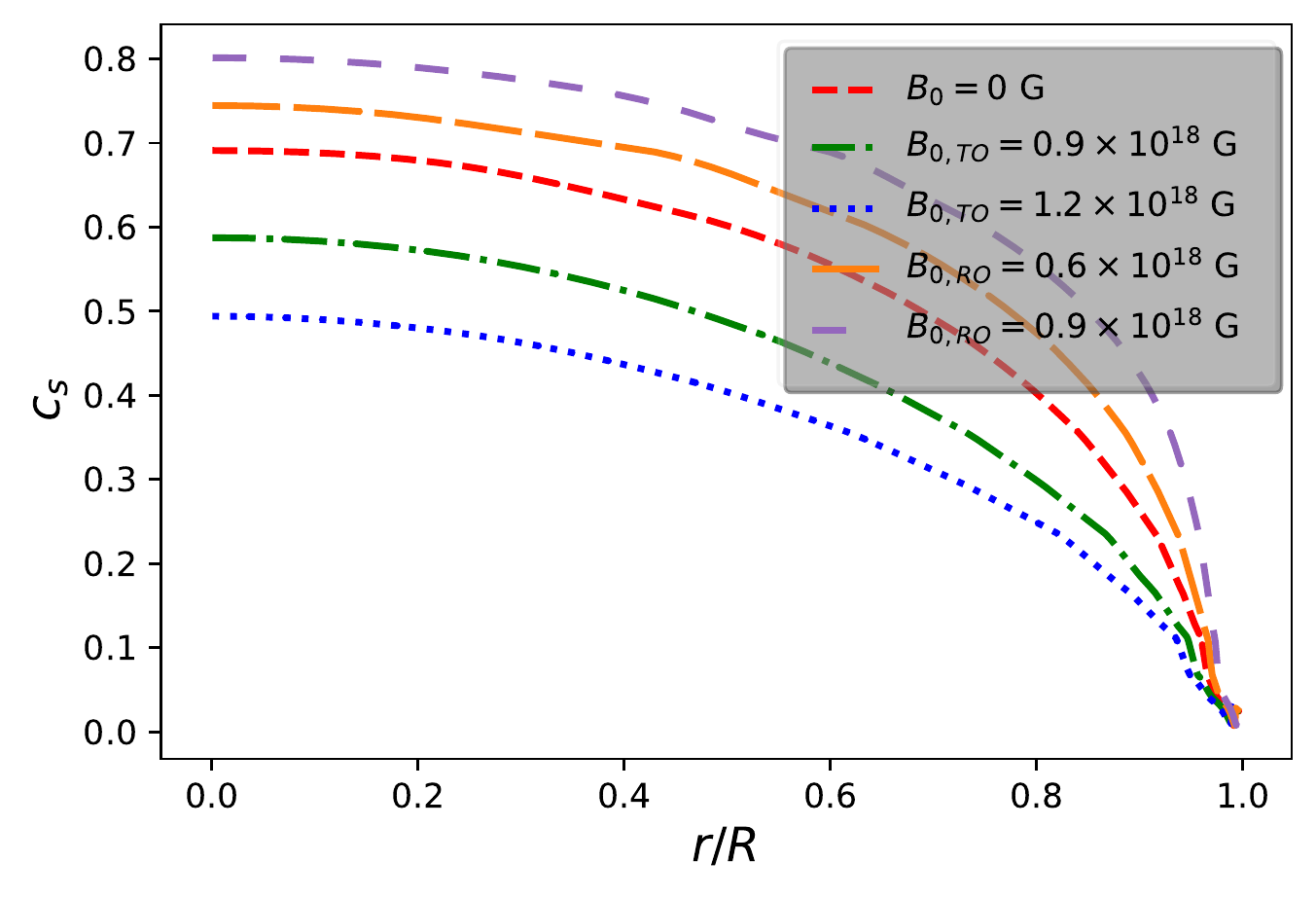} 
\caption{Variation of sound speed $(c_s)$ with
  radial coordinate $r/R$ for $2.01\pm0.04~M_\odot$~\citep{Antoniadis2013} NS candidate PSR~J0348+0432.} \label{soundvel}
\end{figure}

We also examine the sound speed ($c_s$) for all the cases due to NS.  We find that the sound speeds for NS are well within the causality limit for different $B_0$ as shown in Figure~\ref{soundvel}. Since we use MIT bag model EOS to describe SQM distribution within SQS, $c_s$ for
SQS is always given by $c_s=\sqrt{d{p_r}/d\rho} =\sqrt{1/3} \sim 0.58$.
Therefore, in this work, the causality condition does not violate for any chosen EOSs with or without 
strong magnetic fields.

%%%%%%%%%%%%%%%%%%%%%%%%%%%%%%%%%%%%%%%%%%%%%%%%%%%%%%%%%%%%%%%%%%%%%%%%%%%%%%%%%%%%%%%%%%%%%%%%%%%%%%%%%%%%%%%%%%%

%%%%%%%%%%%%%%%%%%%%%%%%%%%%%%%%%%%%%%%%%%%%%%%%%%%%%%%%%%%%%%%%%%%%%%%%%%%%%%%%%%%%%%%%%%%%%%%%%%%%%%%%%%%%%%%%%%%

%%%%%%%%%%%%%%%%%%%%%%%%%%%%%%%%%%%%%%%%%%%%%%%%%%%%%%%%%%%%%%%%%%%%%%%%%%%%%%%%%%%%%%%%%%%%%%%%%%%%%%%%%%%%%%%%%%%

%%%%%%%%%%%%%%%%%%%%%%%%%%%%%%%%%%%%%%%%%%%%%%%%%%%%%%%%%%%%%%%%%%%%%%%%%%%%%%%%%%%%%%%%%%%%%%%%%%%%%%%%%%%%%%%%%%%

%%%%%%%%%%%%%%%%%%%%%%%%%%%%%%%%%%%%%%%%%%%%%%%%%%%%%%%%%%%%%%%%%%%%%%%%%%%%%%%%%%%%%%%%%%%%%%%%%%%%%%%%%%%%%%%%%%%

\begin{figure}[!htpb]
\centering 
\includegraphics[width=0.45\textwidth]{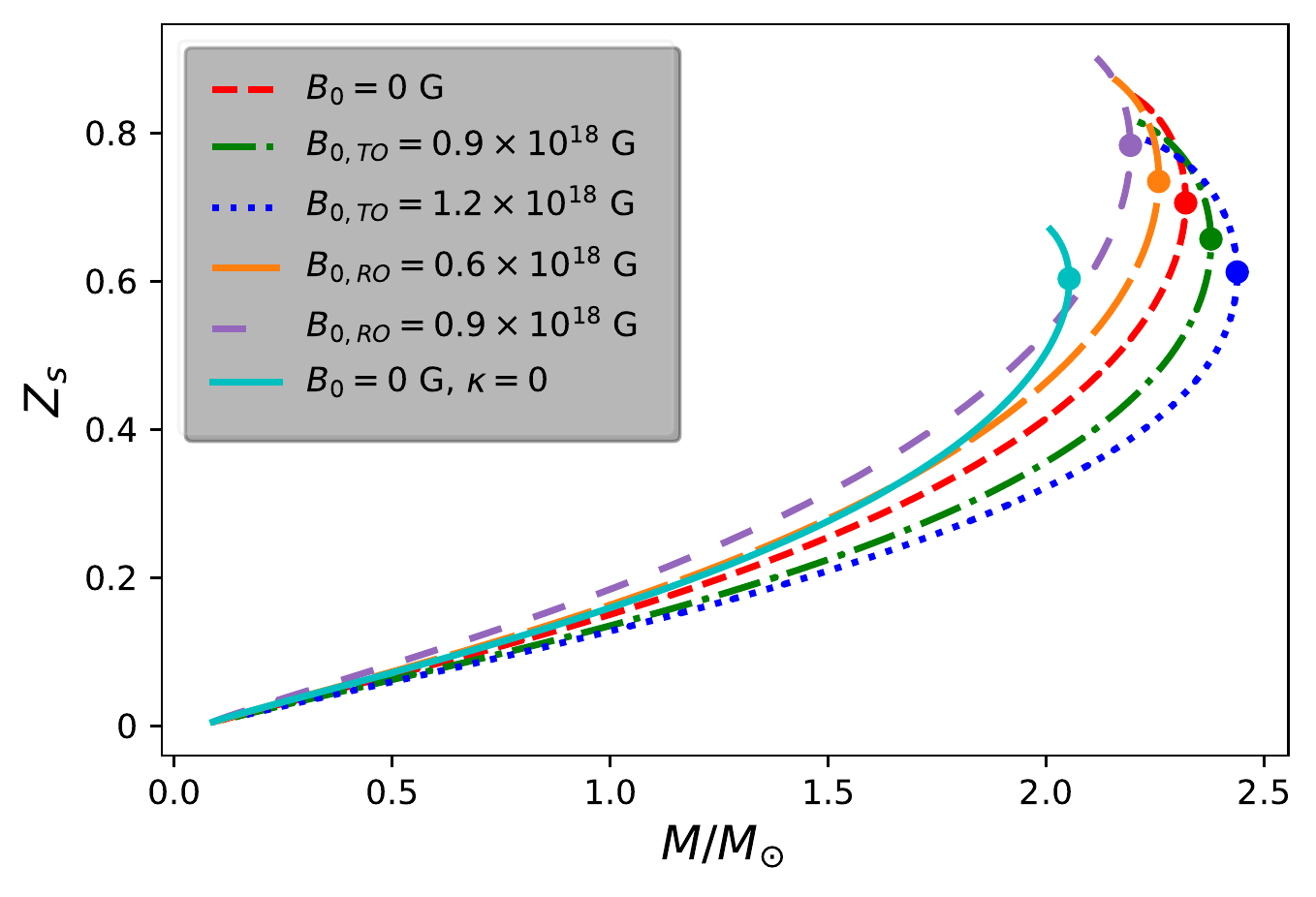} 
\includegraphics[width=0.45\textwidth]{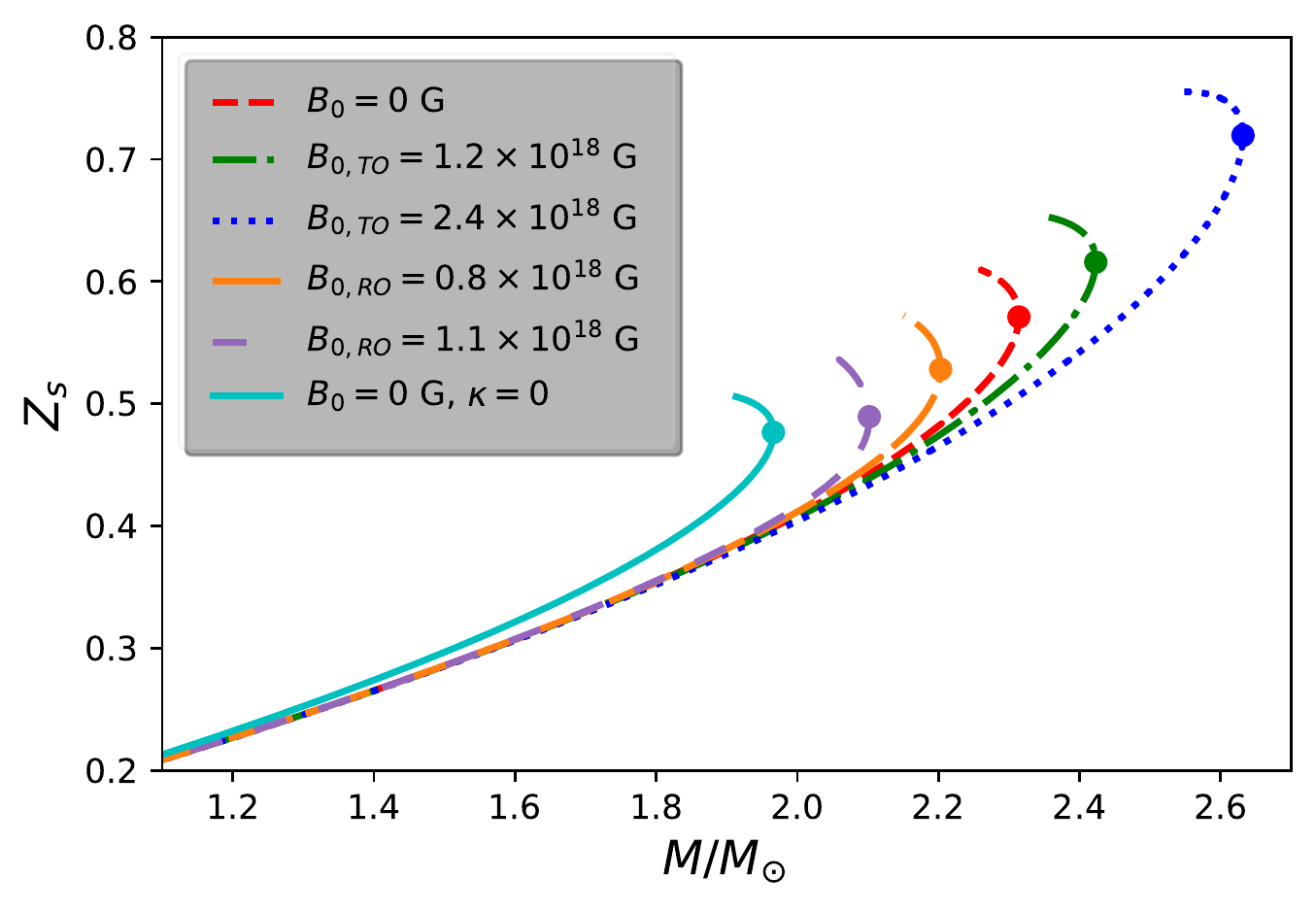}
	\caption{Variation of surface redshift $(Z_s)$ with $M/M_{\odot}$ for NSs (upper panel) 
	and SQSs (lower panel).} \label{Redshift}
\end{figure}

%%%%%%%%%%%%%%%%%%%%%%%%%%%%%%%%%%%%%%%%%%%%%%%%%%%%%%%%%%%%%%%%%%%%%%%%%%%%%%%%%%%%%%%%%%%%%%%%%%%%%%%%%%%%%%%%%%%

For a better understanding of the effects due to anisotropy, the role
of magnetic fields and their orientations, in Tables~\ref{Table 1},
\ref{Table 2}, \ref{Table 3} and \ref{Table 4} we present numerical
values of different stellar properties, viz., maximum mass, the
corresponding radius, $B_c$, $\widetilde{\rho}_c$, $\widetilde{p}_c$,
Buchdahl ratio, surface redshift ($Z_s$), the ratio of magnetic
energy ($E_{mag}$) to gravitational energy ($E_{grav}$) and $\Lambda_{1.4}$ for NSs and
SQSs. One sees from Tables~\ref{Table 1} and \ref{Table 3} that for TO
magnetic fields the maximum mass and the corresponding radius increase with increasing values of $B_0$, whereas for RO magnetic fields the
maximum mass and the corresponding radius decrease gradually with
increasing values of $B_0$. On the other hand, Tables~\ref{Table 2}
and \ref{Table 4} show that the mass and the radius of a star increases
if the strength of the anisotropy, $\kappa$, is increased. For all
cases, the value of $2M/R$ lies well below the critical value of
$8/9$. In Tables~\ref{Table 1}-\ref{Table 4} we also demonstrate that
 in each case, $\mid E_{grav} \mid$ is significantly higher than
$E_{mag}$, which confirms stability of these
stars~\citep{Chandrasekhar1953}. In Figure~\ref{Redshift} we show the
variation of the surface redshift with mass, for TO and RO magnetic
fields and different values of $B_0$. Numerical values of the surface
redshifts of maximum-mass stars for different cases are listed in
Tables~\ref{Table 1}-\ref{Table 4}.  We also show in Tables~\ref{Table 1}-\ref{Table 4} that in each case tidal deformability for $1.4~M_\odot$ star is well bellow the critical value $\Lambda_{1.4,crit}=800$.

\begin{figure}[!htpb]
\centering
\includegraphics[width=0.45\textwidth]{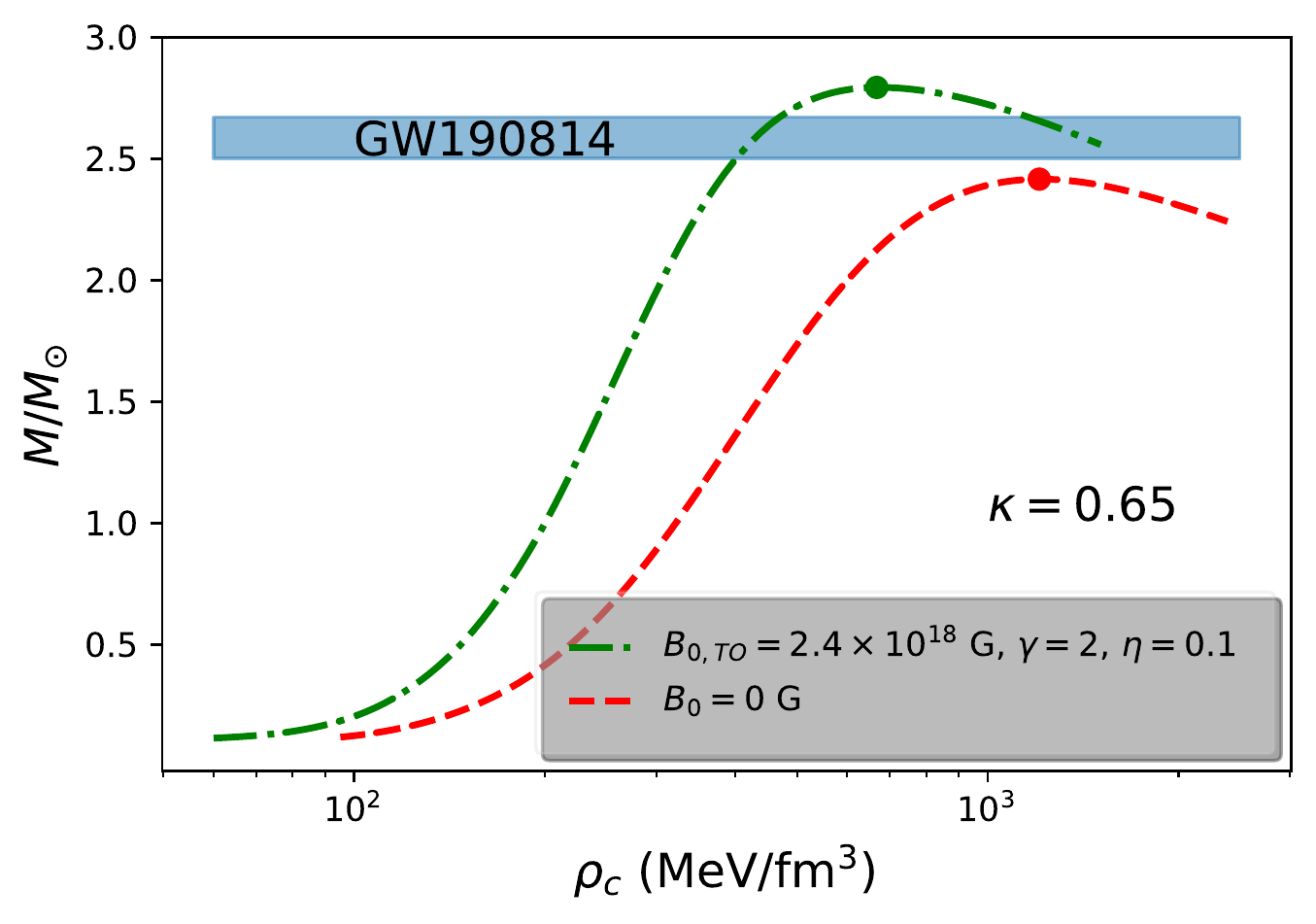}
\caption{Variation of the total mass of the NSs in the
  units of Solar mass $(M/M_\odot)$ with the matter central density $\rho_c$. Here,
  we choose $\kappa=0.65$.} \label{MRNS_massgap}
\end{figure}

From the mass-radius relations shown in Figure~\ref{MR} and the data
provided in Tables~\ref{Table 1}-\ref{Table 4}, the following general
conclusions can be drawn.  As $B_0$ increases in compact stars with TO
magnetic fields, the anisotropic and magnetized stellar objects become
more massive and larger due to the gradual increase of the repulsive,
effectively anisotropic and Lorentz forces. The stars become also more
massive and larger if the strength of the anisotropy, $\kappa$,
increases. On the contrary, the stars with RO magnetic fields become
less massive and smaller in size for gradually increasing values of
the magnetic field, since the corresponding effective anisotropic
force simultaneously decreases. Interestingly, we point out that
anisotropic, magnetized compact stars can have masses that are in the
mass range $2.50-2.67~M_\odot$ deduced for the lighter companion in
the binary compact-object coalescence event GW190814, observed
recently by LIGO and Virgo \citep{Abbott2020}, as shown in
Figure~\ref{MRNS_massgap}. With the appropriate choice of the physical parameters, such as $B_0=2.4\times {10}^{18}$~G, $\eta=0.1$, $\gamma=2$ and $\kappa=0.65$, we find that the maximum possible mass of a NS is $\sim 2.79~M_\odot$,  which comfortably accommodates the anomalously high mass of the lighter object associated with GW190814. We note that by considering rotation of the anisotropic and magnetized stars in a future study, the maximum mass of a NS will be pushed to even higher values.

%%%%%%%%%%%%%%%%%%%%%%%%%%%%%%%%%%%%%%%%%%%%%%%%%%%%%%%%%%%%%%%%%%%%%%%%%%%%%%%%%%%%%%%%%%%%%%%%%%%%%%%%%%%%%%%%%%%

\begin{table*}[htbp!]
  \centering
    \caption{Numerical values of physical parameters for the NSs with
      $\kappa = 0.5$ for different values of $B_0$}
\label{Table 1}
    %\begin{ruledtabular}
        \begin{tabular}{cccccccccccccccccccc}
%%%%%%%%%%%%%%%%%%%%%%%%%%%%%%%%%%%%%%%%%%%%%%%%%%%%%%%%%%%%%%%%%%%%%%%%%%%%%%%%%%
\hline\hline Orientation & Value & Value of & Corresponding & Central
& Central & Central & & Surface & \\ of Magnetic & of $B_0$ & Maximum
& Predicted & Magnetic field & Density & Pressure & $\frac{2M}{R}$ &
Redshift & $\frac{E_{mag}}{E_{grav}}$ & $\Lambda_{1.4}$\\ field & (Gauss) & Mass
$(M_\odot)$ & Radius (km) & $B_c$ (Gauss) & $\widetilde{\rho}_c~(\rm
g/{cm}^3)$ & $\widetilde{p}_c~(\rm dyne/{cm}^2)$ & & $(Z_s)$ & &
\\ \hline \vspace{-0.2cm}
%%%%%%%%%%%%%%%%%%%%%%%%%%%%%%%%%%%%%%%%%%%%%%%%%%%%%%%%%%%%%%%%%%%%%%%%%%%%%%%%%%%
\multirow{2}{*}{TO} & $1.2\times {10}^{18}$ & $2.438$ & $11.685$ & $1.2\times {10}^{18}$ & $1.891\times {10}^{15}$ & $6.316\times {10}^{35}$ & $0.62$ & $0.61$ & $0.088$ & $523.17$  \\ 
\
                            & $0.9 \times {10}^{18}$ & $2.378$ & $11.030$ & $0.9\times {10}^{18}$ & $2.117\times {10}^{15}$ & $7.838\times {10}^{35}$ & $0.64$ & $0.66$ & $0.043$ & $388.44$ \\ 
\
$B=0$ & - & $2.320$ & $10.425$ & - & $2.295\times {10}^{15}$ & $9.120\times {10}^{35}$ & $0.66$ & $0.71$ & - & $225.90$\\ 
\
\multirow{2}{*}{RO}    & $0.6 \times {10}^{18}$ & $2.258$ & $9.975$ & $0.6\times {10}^{18}$ & $2.465\times {10}^{15}$ & $1.020\times {10}^{36}$ & $0.67$ & $0.73$ & $0.015$ & $149.11$ \\ 
\
                            & $0.9 \times {10}^{18}$ & $2.193$ & $9.436$ & $0.9\times {10}^{18}$ & $2.736\times {10}^{15}$ & $1.209\times {10}^{36}$ & $0.69$ & $0.78$ & $0.029$ & $86.29$\\
  \hline\hline
  \end{tabular}
   % \end{ruledtabular}
    \end{table*}

%%%%%%%%%%%%%%%%%%%%%%%%%%%%%%%%%%%%%%%%%%%%%%%%%%%%%%%%%%%%%%%%%%%%%%%%%%%%%%%%%%%%%%%%%%%%%%%%%%%%%%%%%%%%%%%%%%%

%%%%%%%%%%%%%%%%%%%%%%%%%%%%%%%%%%%%%%%%%%%%%%%%%%%%%%%%%%%%%%%%%%%%%%%%%%%%%%%%%%%%%%%%%%%%%%%%%%%%%%%%%%%%%%%%%%%

\begin{table*}[htbp!]
  \centering
    \caption{Numerical values of physical parameters for the NSs with $B_0 = 0.9\times{10}^{18}$ for different values of $\kappa$} \label{Table 2}
    %\begin{ruledtabular}
        \begin{tabular}{cccccccccccccccccc}
%%%%%%%%%%%%%%%%%%%%%%%%%%%%%%%%%%%%%%%%%%%%%%%%%%%%%%%%%%%%%%%%%%%%%%%%%%%%%%%%%%
\hline\hline Value & Value of & Corresponding & Central & Central & Central &  & Surface & &  \\
of $\kappa$ & Maximum & Predicted & Magnetic field & Density & Pressure & $\frac{2M}{R}$ & Redshift & $\frac{E_{mag}}{E_{grav}}$ & $\Lambda_{1.4}$\\
       &  Mass $(M_\odot)$ & Radius~(km) & $B_c$ (Gauss) & $\widetilde{\rho}_c~(\rm g/{cm}^3)$ & $\widetilde{p}_c~(\rm dyne/{cm}^2)$ &   &  $(Z_s)$ & & \\ 
\hline \vspace{-0.2cm} 
%%%%%%%%%%%%%%%%%%%%%%%%%%%%%%%%%%%%%%%%%%%%%%%%%%%%%%%%%%%%%%%%%%%%%%%%%%%%%%%%%%%
 $0.15$ & $2.179$ & $10.698$ & $9.000\times {10}^{17}$ & $2.406\times {10}^{15}$ & $9.715\times {10}^{35}$ & $0.60$ & $0.58$ & $0.0428$ & $423.41$ \\ 
\
 $0.30$ & $2.260$ & $10.860$ & $9.000\times {10}^{17}$ & $2.257\times {10}^{15}$ & $8.880\times {10}^{35}$ & $0.61$ & $0.61$ & $0.0431$ & $392.41$ \\ 
\
 $0.45$ & $2.347$ & $11.031$ & $9.000\times {10}^{17}$ & $2.117\times {10}^{15}$ & $7.838\times {10}^{35}$ & $0.63$ & $0.64$ & $0.0434$ & $385.45$ \\ 
\
 $0.60$ & $2.443$ & $11.210$ & $9.000\times {10}^{17}$ & $1.986\times {10}^{15}$ & $6.257\times {10}^{35}$ & $0.64$ & $0.67$ & $0.0438$ & $395.52$\\
 \hline\hline
  \end{tabular}
    %\end{ruledtabular}
    \end{table*}

%%%%%%%%%%%%%%%%%%%%%%%%%%%%%%%%%%%%%%%%%%%%%%%%%%%%%%%%%%%%%%%%%%%%%%%%%%%%%%%%%%%%%%%%%%%%%%%%%%%%%%%%%%%%%%%%%%%

%%%%%%%%%%%%%%%%%%%%%%%%%%%%%%%%%%%%%%%%%%%%%%%%%%%%%%%%%%%%%%%%%%%%%%%%%%%%%%%%%%%%%%%%%%%%%%%%%%%%%%%%%%%%%%%%%%%

\begin{table*}[htbp!]
  \centering
    \caption{Numerical values of physical parameters for the SQSs with $\kappa = 0.5$ and $\mathcal{B} = 60~\mathrm{MeV/{fm^3}}$ for different values of $B_0$ } \label{Table 3}
    %\begin{ruledtabular}
        \begin{tabular}{cccccccccccccccccc}
%%%%%%%%%%%%%%%%%%%%%%%%%%%%%%%%%%%%%%%%%%%%%%%%%%%%%%%%%%%%%%%%%%%%%%%%%%%%%%%%%%
\hline\hline
Orientation & Value & Value of & Corresponding & Central & Central & Central &  & Surface & &  \\
of Magnetic & of $B_0$ & Maximum & Predicted & Magnetic field & Density & Pressure & $\frac{2M}{R}$ & Redshift & $\frac{E_{mag}}{E_{grav}}$  & $\Lambda_{1.4}$\\
 field & (Gauss)  &  Mass $(M_\odot)$ & Radius~(km) & $B_c$ (Gauss) & $\widetilde{\rho}_c~(\rm g/{cm}^3)$ & $\widetilde{p}_c~(\rm dyne/{cm}^2)$ &   &  $(Z_s)$ & & \\ 
\hline \vspace{-0.2cm} 
%%%%%%%%%%%%%%%%%%%%%%%%%%%%%%%%%%%%%%%%%%%%%%%%%%%%%%%%%%%%%%%%%%%%%%%%%%%%%%%%%%%
\multirow{2}{*}{TO} & $2.4 \times {10}^{18}$ & $2.632$ & $11.730$ & $2.296\times {10}^{18}$ & $1.610\times {10}^{15}$ & $4.939\times {10}^{35}$ & $0.66$ & $0.72$ & $0.33$ & $587.91$ \\ 
\
                            & $1.2 \times {10}^{18}$ & $2.423$ & $11.585$ & $1.185\times {10}^{18}$ & $1.603\times {10}^{15}$ & $3.892\times {10}^{35}$ & $0.62$ & $0.62$ & $0.09$ & $577.33$\\ 
\
$B=0$ & - & $2.314$ & $11.477$ & - & $1.609\times {10}^{15}$ & $3.537\times {10}^{35}$ & $0.59$ & $0.57$ & -  & $575.32$\\ 
\
\multirow{2}{*}{RO}    & $0.8 \times {10}^{18}$ & $2.203$ & $11.367$ & $7.931\times {10}^{17}$ & $1.609\times {10}^{15}$ & $3.204\times {10}^{35}$ & $0.57$ & $0.53$ & $0.04$ & $573.98$ \\ 
\
                            & $1.1 \times {10}^{18}$ & $2.102$ & $11.291$ & $1.086\times {10}^{18}$ & $1.593\times {10}^{15}$ & $2.864\times {10}^{35}$ & $0.55$ & $0.49$ & $0.08$ & $572.28$  \\
                            \hline\hline
  \end{tabular}
    %\end{ruledtabular}
    \end{table*}

%%%%%%%%%%%%%%%%%%%%%%%%%%%%%%%%%%%%%%%%%%%%%%%%%%%%%%%%%%%%%%%%%%%%%%%%%%%%%%%%%%%%%%%%%%%%%%%%%%%%%%%%%%%%%%%%%%%

%%%%%%%%%%%%%%%%%%%%%%%%%%%%%%%%%%%%%%%%%%%%%%%%%%%%%%%%%%%%%%%%%%%%%%%%%%%%%%%%%%%%%%%%%%%%%%%%%%%%%%%%%%%%%%%%%%%

\begin{table*}
  \centering
    \caption{Numerical values of physical parameters for the SQSs with
      $B_0 = 2.4\times{10}^{18}$ G and $\mathcal{B} =
      60~\mathrm{MeV/{fm^3}}$, for different values of $\kappa$
    } \label{Table 4}
    %\scalebox{0.8}{
    \begin{ruledtabular}
        \begin{tabular}{cccccccccccccccccc}
%%%%%%%%%%%%%%%%%%%%%%%%%%%%%%%%%%%%%%%%%%%%%%%%%%%%%%%%%%%%%%%%%%%%%%%%%%%%%%%%%%
 Value & Value of & Corresponding & Central & Central & Central &  & Surface &  &\\
of $\kappa$ & Maximum & Predicted & Magnetic field & Density & Pressure & $\frac{2M}{R}$ & Redshift & $\frac{E_{mag}}{E_{grav}}$ & $\Lambda_{1.4}$\\
       &  Mass $(M_\odot)$ & Radius~(km) & $B_c$ (Gauss) & $\widetilde{\rho}_c~(\rm g/{cm}^3)$ & $\widetilde{p}_c~(\rm dyne/{cm}^2)$ &   &  $(Z_s)$ & & \\ 
\hline \vspace{-0.2cm} 
%%%%%%%%%%%%%%%%%%%%%%%%%%%%%%%%%%%%%%%%%%%%%%%%%%%%%%%%%%%%%%%%%%%%%%%%%%%%%%%%%%%
 $0.15$ & $2.365$ & $11.297$ & $2.381\times {10}^{18}$ & $1.853\times {10}^{15}$ & $5.775\times {10}^{35}$ & $0.62$ & $0.62$ & $0.3280$ & $478.50$\\ 
\
 $0.30$ & $2.473$ & $11.474$ & $2.357\times {10}^{18}$ & $1.748\times {10}^{15}$ & $5.430\times {10}^{35}$ & $0.64$ & $0.66$ & $0.3285$ & $520.49$ \\ 
\
 $0.45$ & $2.590$ & $11.673$ & $2.308\times {10}^{18}$ & $1.631\times {10}^{15}$ & $5.017\times {10}^{35}$ & $0.65$ & $0.70$ & $0.3302$ & $567.95$ \\ 
\
 $0.60$ & $2.718$ & $11.881$ & $2.238\times {10}^{18}$ & $1.530\times {10}^{15}$ & $4.629\times {10}^{35}$ & $0.67$ & $0.75$ & $0.3319$ & $629.68$
  \end{tabular}
    \end{ruledtabular}%}
    \end{table*}

%%%%%%%%%%%%%%%%%%%%%%%%%%%%%%%%%%%%%%%%%%%%%%%%%%%%%%%%%%%%%%%%%%%%%%%%%%%%%%%%%%%%%%%%%%%%%%%%%%%%%%%%%%%%%%%%%%%

\section{Prospects of future work on white dwarfs}\label{secIII}

\begin{figure}[!htpb]
\centering
\includegraphics[width=0.45\textwidth]{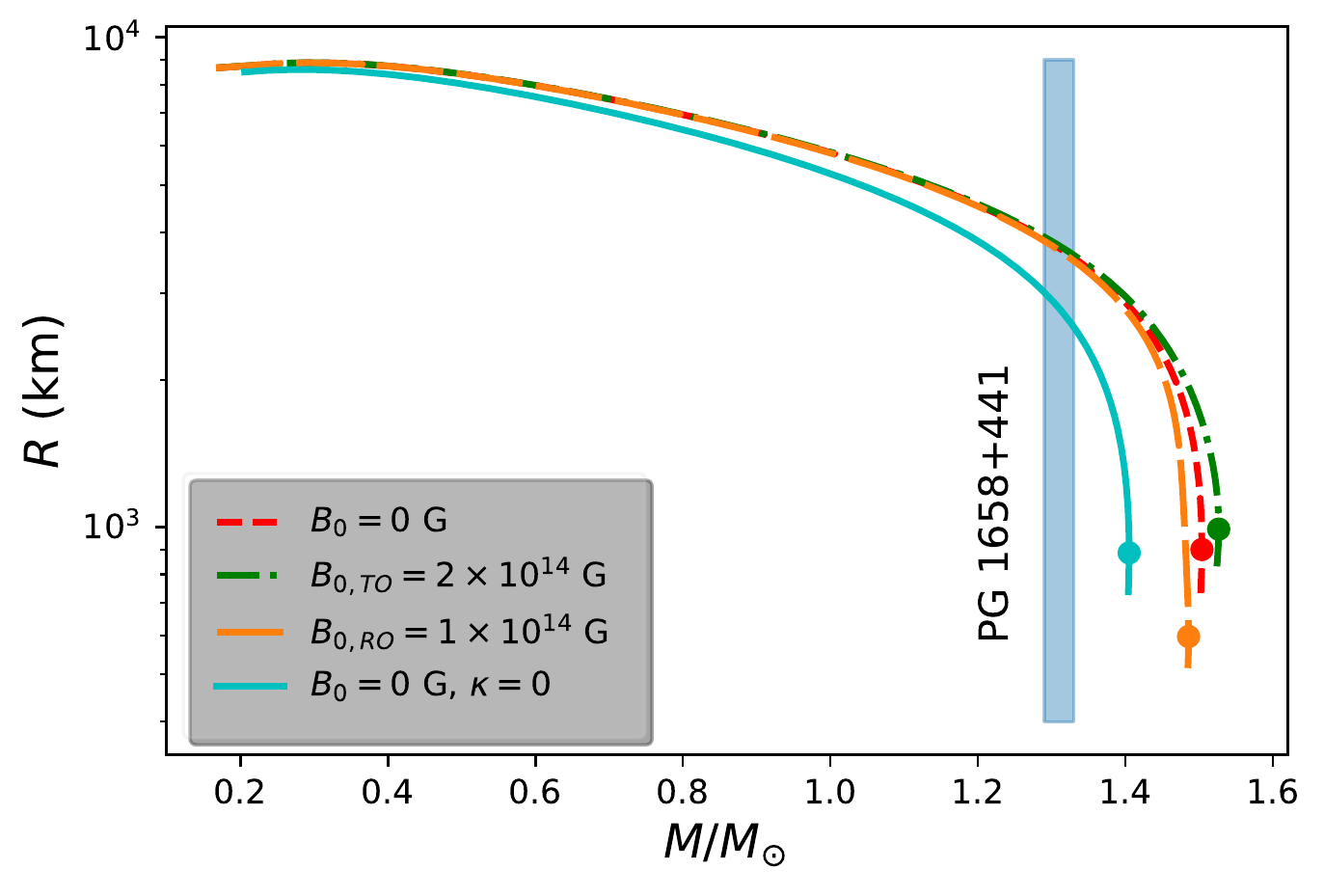}
\caption{ Mass-radius relations of white dwarfs for
    different magnetic field strengths. Solid circles represent the
  maximum mass star of each stellar sequence.} \label{MR_WDNSQS}
\end{figure}

In their recent study,~\cite{Chowdhury2019} attempted to explain white
dwarfs (WDs) based on an anisotropic spherically-symmetric model in
the framework of modified gravity theory and indicated the possible
existence of super-Chandrasekhar WDs beyond the standard Chandrasekhar
mass limit. It is worth mentioning that during the last few years
Mukhopadhyay and
collaborators~\citep{Das2012,Das2013,Vishal2014,Das2015a,Subramanian2015,Mukhopadhyay2016,Mukhopadhyay2017,Bhattacharya2018,Kalita2018,Kalita2019}
through their series of important works have established the possible
existence of highly super-Chandrasekhar mass WDs. They found that the
presence of a strong magnetic field leads to super-Chandrasekhar WDs
which suitably served as a progenitor of the peculiar overluminous
type Ia supernovae (SNeIa). Although in this article, we mainly focus
on ultra-dense compact stars it will be interesting to investigate the
effects of anisotropy, strong magnetic fields and the orientation of
the magnetic field on WDs. The upper limit of density for these highly
magnetized WDs (B-WDs) is constrained by the effects of electron
capture and pycnonuclear fusion reactions~\citep{Otoniel2019}.  In
Figure~\ref{MR_WDNSQS} we demonstrate that inclusion of anisotropy and
TO of the magnetic field increases the maximum mass of B-WDs compared
to the non-magnetized isotropic case, whereas in the case of a RO of
the magnetic field the maximum mass of B-WDs drops compared to the
non-magnetized anisotropic case. Although, we take the opportunity to
discuss whether the present model can suitably explain anisotropic
B-WDs, the understanding of their properties requires a detailed
discussion which is beyond the scope of this study. We are going to
report it in our next work, which is in progress.

%%%%%%%%%%%%%%%%%%%%%%%%%%%%%%%%%%%%%%%%%%%%%%%%%%%%%%%%%%%%%%%%%%%%%%%%%%%%%%%%%%%%%%%%%%%%%%%%%%%%%%%%%%%%%%%%%%%

%%%%%%%%%%%%%%%%%%%%%%%%%%%%%%%%%%%%%%%%%%%%%%%%%%%%%%%%%%%%%%%%%%%%%%%%%%%%%%%%%%%%%%%%%%%%%%%%%%%%%%%%%%%%%%%%%%%

%%%%%%%%%%%%%%%%%%%%%%%%%%%%%%%%%%%%%%%%%%%%%%%%%%%%%%%%%%%%%%%%%%%%%%%%%%%%%%%%%%%%%%%%%%%%%%%%%%%%%%%%%%%%%%%%%%%

\section{Conclusion}\label{secIV}

In this work, we study the combined effects of anisotropy, strong
magnetic fields and their orientations on the properties of
spherically symmetric compact objects, viz., NSs and SQSs. Our study
reveals that in a magnetized compact stellar object the magnetic field
strength, the orientation of the magnetic field and the anisotropy
influence the EOS of the system by modifying both the matter and
pressures of the system. Although in the present study we consider
spherically symmetric compact objects, one may point out that the
occurrence of anisotropy may push the system toward non-spherical
symmetry. However, it is already known that for a toroidally dominated
field magnetized stars exhibit negligible deviations from spherical
symmetry~\citep{Das2015b,Subramanian2015,Kalita2019}. For example, the
maximum value of anisotropy in the case of TO magnetic fields with
$B_0=1.2\times{10}^{18}$~G in a NS is $\sim80\%$ lower than $p_c$ (see the upper panel of Figure~\ref{anisotropy}), and for $B_0=2.4\times{10}^{18}$~G in a SQS it is $\sim81\%$ lower than $p_c$ (see the lower panel of Figure~\ref{anisotropy}).  This indicates that treating magnetized anisotropic stars as spherically symmetric objects has no considerable influence on the %physical properties and 
geometry of the stellar configurations.

~\cite{Ferrer2010} showed that inclusion of a strong magnetic field
invites anisotropy within the system due to the distinction between
parallel and transverse pressures. However, it is also important
for such anisotropic and magnetized stars that they are consistent
with the TOV equations throughout the stars, which ensures hydrostatic
stability of the systems via equilibrium of forces. Nevertheless, to
the best of our knowledge, prior to this study, the issue of non-zero
value of anisotropic force in the center of highly magnetized compact
stars mostly remained unnoticed. One may easily check via the TOV
equations (see Eq.~\ref{1.11}) that the non-zero value of anisotropy
at the center leads to instability due to non-equilibrium of the
forces. On the other hand, the anisotropy which originates due to the
presence of the strong magnetic field via the distinction between the
parallel and transverse pressures, which is $\sim \mid B^2 \mid$,
cannot be zero at the center due to the maximum finite value of the
magnetic field at the stellar core. Interestingly, we find that this
situation can be taken care of by considering the anisotropic effect
due to both the local anisotropy of the fluid and the presence of the
strong magnetic field.

Of course, the present study offers only a simplified treatment of the
complex structure of magnetised compact stellar configurations. But
through our work, we are able to demonstrate that to study magnetised
compact stars, it is essential to consider the effective anisotropies
of both the fluid and the magnetic field, which have a significant
influence on the properties of compact stars. Based on the orientation
of the magnetic field, the maximum mass of static magnetized compact
stars may be enhanced or reduced, which resolves the long-standing
issue whether or not the mass of the system increases or decreases due to the presence of a strong magnetic field. Importantly, the present
study has also explored the magneto-hydrostatic stability of the
system, which demonstrates the physical validity of this model.

\section{acknowledgments}
Research of D. D. is funded by the C. V. Raman Postdoctoral Fellowship
(Reg. No. R(IA)CVR-PDF/2020/222) from Department of Physics, Indian
Institute of Science. F. W. is supported through the U.S.\ National
Science Foundation under Grants PHY-1714068 and PHY-2012152. 
This work is partly supported by a fund of Department
of Science and Technology (DST-SERB) with research Grant
No. DSTO/PPH/BMP/1946 (EMR/2017/001226).  
One of the authors (D. D.) is thankful to Surajit Kalita of IISc for his
pertinent suggestions which helped to upgrade the paper.  We are also thankful to the referee for their pertinent comments, which helped us in upgrading our work substantially.  All computations were
performed in open source softwares, and the authors are sincerely
thankful to the open source community.

\end{document}